\documentclass[]{aa}

\usepackage{placeins}
\usepackage{xtab,booktabs}
\usepackage{ltablex}
\usepackage{changepage}
\usepackage{graphicx,lscape}
\usepackage{subfig}
\usepackage{hyperref}
\usepackage{txfonts,longtable}
\usepackage{graphicx,rotating}
\usepackage{subfig}
\usepackage{txfonts}
\usepackage[T1]{fontenc}
\usepackage{epsfig}
\usepackage{natbib}
\usepackage{color}
\usepackage{natbib}
\usepackage{amssymb}  
\usepackage{tabularx}
\usepackage{comment}
\usepackage{stackengine}

\DeclareGraphicsExtensions{.pdf,.png,.jpg,.ps,.tif}
\bibpunct{(}{)}{;}{a}{}{,} % to follow the A&A style
\begin{document} 

  \title{Star-forming and gas-rich brightest cluster galaxies at $z\sim0.4$ in the Kilo-Degree Survey\fnmsep\thanks{Spectra in Fig.~\ref{fig:BCG_spectra} are available in electronic form at the CDS via anonymous ftp to cdsarc.u-strasbg.fr (130.79.128.5) or via http://cdsweb.u-strasbg.fr/cgi-bin/qcat?J/A+A/}}
  
  %{ Tables n are only available in electronic form at the CDS via anonymous ftp to cdsarc.u-strasbg.fr (130.79.128.5) or via http://cdsweb.u-strasbg.fr/cgi-bin/qcat?J/A+A/}
 %  \title{Molecular gas in three KiDS brightest cluster galaxies at $z\sim0.4$}

   \author{G. Castignani
          \inst{1,2}\fnmsep\thanks{e-mail: gianluca.castignani@unibo.it}
          \and
          M. Radovich\inst{3}
        \and
          F. Combes\inst{4,5}
          \and
          P. Salom\'e\inst{4}
          \and
          M. Maturi\inst{6,7}
          \and
          L. Moscardini\inst{1,2,8}
          \and
          S. Bardelli\inst{2}
          \and
          C. Giocoli\inst{2,8}
          \and
          G. Lesci\inst{1,2}
          \and
          F. Marulli\inst{1,2,8}
          \and
          E. Puddu\inst{9}
          \and
          M. Sereno\inst{2,8}          
          %\and
          %... KiDS people ...
          }

   \institute{Dipartimento di Fisica e Astronomia ''Augusto Righi'', Alma Mater Studiorum Università di Bologna, Via Gobetti 93/2, I-40129 Bologna, Italy   
   \and
   INAF - Osservatorio  di  Astrofisica  e  Scienza  dello  Spazio  di  Bologna,  via  Gobetti  93/3,  I-40129,  Bologna,  Italy
   \and
   INAF - Osservatorio Astronomico di Padova, vicolo dell'Osservatorio 5, I-35122 Padova, Italy
   \and
   Observatoire de Paris, LERMA, CNRS, Sorbonne University, PSL Research Universty, 75014 Paris, France 
   \and
   Coll\`{e}ge de France, 11 Place Marcelin Berthelot, 75231 Paris, France
   \and
   Center for Astronomy - University of Heidelberg, Albert-Ueberle-Stra{\ss}e 2, 69120 Heidelberg, Germany 
   \and Institute of Theoretical Physics - University of Heidelberg, Albert-Ueberle-Stra{\ss}e 2, 69120 Heidelberg, Germany 
   \and INFN - Sezione di Bologna, viale Berti-Pichat 6/2, I-40127 Bologna, Italy
   \and
   INAF - Osservatorio di Capodimonte, Salita Moiariello 16, 80131 Napoli, Italy      
        }
\date{Received: March 31, 2022; Accepted July 25, 2022}
% \abstract{}{}{}{}{} 
% 5 {} token are mandatory

  \abstract
   {Brightest cluster galaxies (BCGs) are { typically massive} ellipticals at the centers of clusters. They are believed to experience strong environmental processing, {and} their mass assembly and star formation history are  still debated. We have selected three star-forming BCGs in the equatorial field of the Kilo-Degree Survey (KiDS). They are    KiDS~0920 ($z=0.3216$), KiDS~1220 ($z=0.3886$), and KiDS~1444 ($z=0.4417$). We have observed them with the IRAM 30m telescope in the first three CO transitions. We remarkably detected all BCGs at high signal-to-noise ratio, ${\rm S/N}\simeq(3.8-10.2)$, for a total of seven detected lines out of eight, corresponding to a success rate of {88$\%$.} This allows us to double the number of distant BCGs with clear detections in{ at least} two CO lines. 
   We then combined our observations with available stellar, star formation, and dust properties of the BCGs and compared them with a sample of $\sim100$ distant cluster galaxies with observations in CO. Our analysis yields large molecular gas reservoirs $M_{H_2}\simeq(0.5-1.4)\times10^{11}~M_\odot$, high excitation ratios $r_{31}= L^{\prime}_{\rm CO(3\rightarrow2)}/L^{\prime}_{\rm CO(1\rightarrow0)}\simeq(0.1-0.3)$, long depletion times $\tau_{\rm dep}\simeq(2-4)$~Gyr, and high $M_{H_2}/M_{\rm dust}\simeq(170-300)$ for the three targeted BCGs. The excitation ratio $r_{31}$ of intermediate-$z$ BCGs, including RX1532 and M1932 from previous studies, {appears to be well correlated  with the star formation rate and efficiency,} which suggests that excited gas is found only in highly star-forming and cool-core BCGs. By performing color-magnitude plots and a red-sequence modeling, we find that recent bursts of star formation are needed to explain the fact that the BCGs are measurably bluer than photometrically selected cluster members.
   To explain the global observed phenomenology, we suggest that a substantial amount of the molecular gas has been accreted by the KiDS BCGs but still not efficiently converted into stars. KiDS~1220 also shows a double-horn emission in  CO(3$\rightarrow$2), which implies a low gas concentration. The modeling of the spectrum yields an extended molecular gas reservoir of $\sim$9~kpc, which is reminiscent of the mature extended-disk phase observed in some local BCGs.}
   
   \maketitle

%-------------------------Fig BCG IMAGES  ---------------
\begin{figure*}[th!]\centering
\captionsetup[subfigure]{labelformat=empty}
%%%%%%%%%%%%%%%%%%%%%%%%%%%%%%%%%%%%%%%%%%%%5
\subfloat[]{\hspace{0.cm}\includegraphics[trim={0.5cm 0cm 3cm 
0cm},clip,width=0.32\textwidth,clip=true]{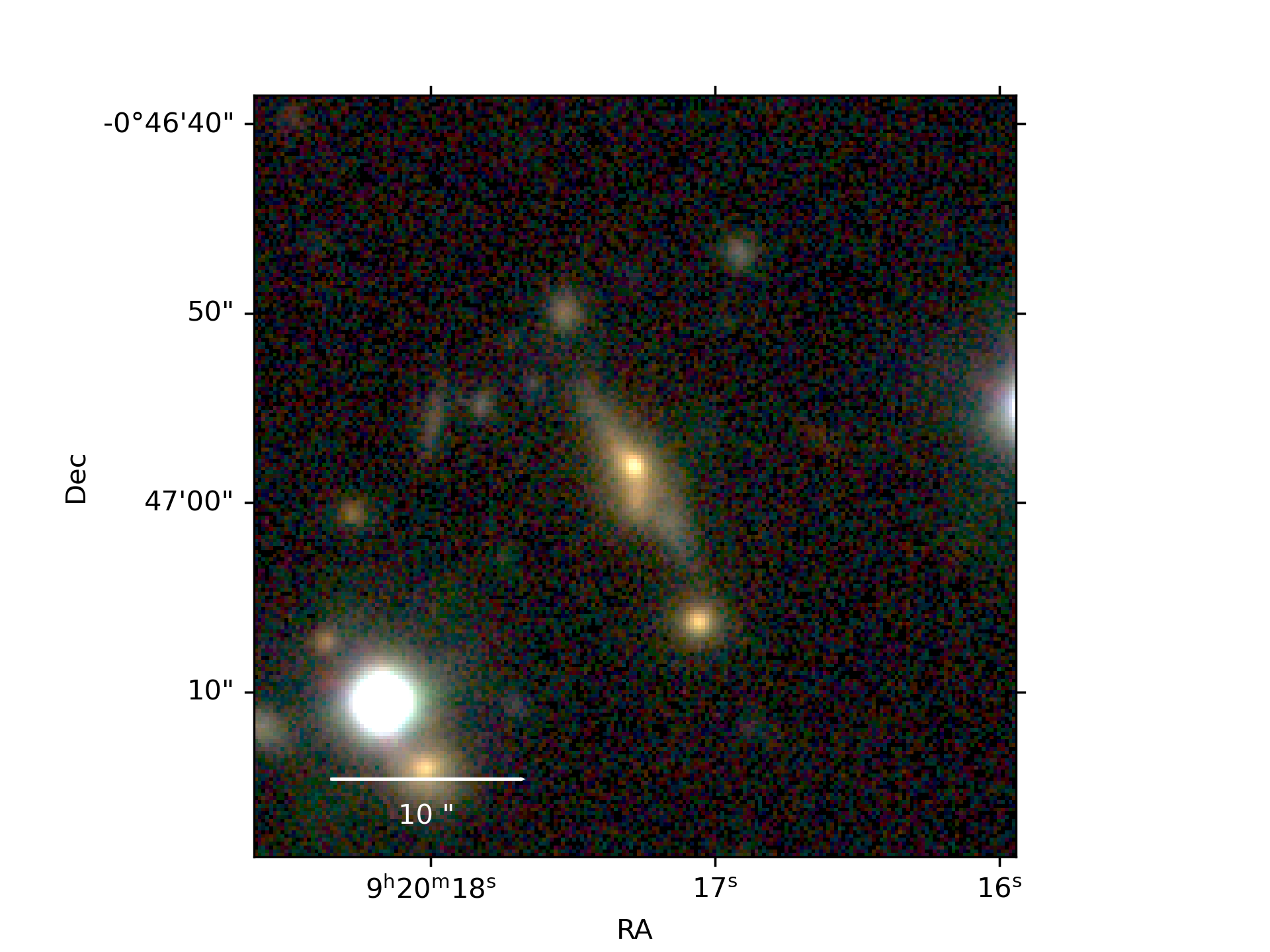}}
\hspace{0.1cm}\subfloat[]{\hspace{0.cm}\includegraphics[trim={0.5cm 0cm 3cm 
0cm},clip,width=0.32\textwidth,clip=true]{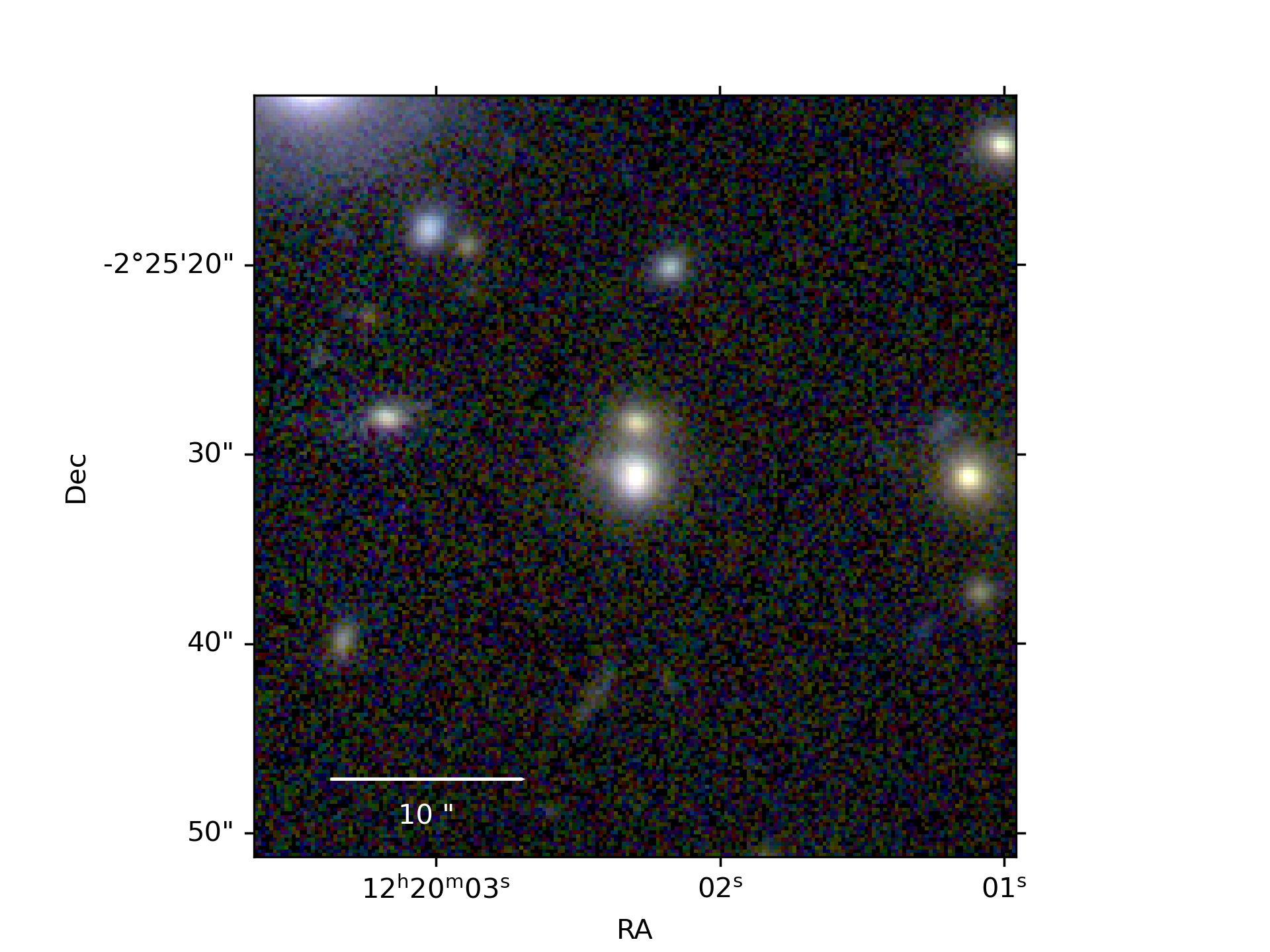}}
\hspace{0.1cm}\subfloat[]{\hspace{0.cm}\includegraphics[trim={0.5cm 0cm 3cm 
0cm},clip,width=0.32\textwidth,clip=true]{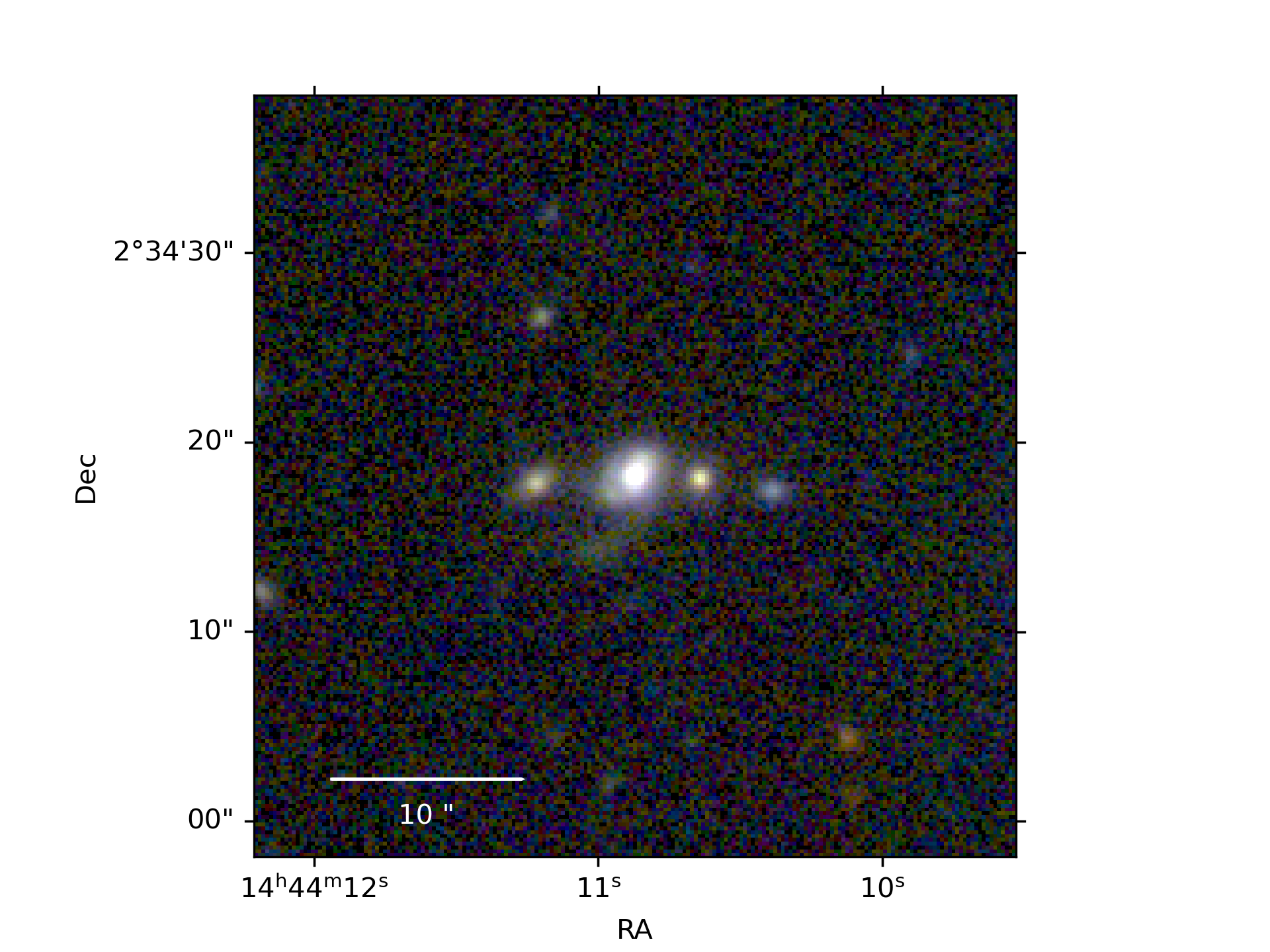}}
%%%%%%%%%%%%%%%%%%%%%%%%%%%
\caption{Color composite $40^{\prime\prime}\times40^{\prime\prime}$ KiDS images (red~=~\textsf{i} band; green~=~\textsf{r} band; blue~=~\textsf{g} band) with OmegaCAM at VST 
centered around the BCGs KiDS~0920 (left), KiDS~1220 (center), and KiDS~1444 (right). North is up, and  east is to the left.}\label{fig:BCG_images}
\end{figure*}
%%%%%%%%%%%%%%%%%%%%%%%%%%%%%%%%%%%%%%%%%%%%5

%
%-------------------------------------------------------------------
\section{Introduction}\label{sec:introduction}
Brightest cluster galaxies (BCGs) are among the most massive galaxies in the Universe, are typically massive ellipticals of cD type, {and are often} radio galaxies \citep{Zirbel1996}. There exists a strong, albeit still debated, coevolution between BCGs, their cluster environments \citep{Lauer2014}, and the accretion processes onto the supermassive black holes residing at galaxy centers \citep{Fabian2012}. 
%, which ultimately launch powerful large-scale radio jets

Because of their exceptional masses and luminosities,  BCGs are believed to evolve via phenomena such as dynamical friction \citep{White1976}, galactic cannibalism  \citep{Hausman_Ostriker1978}, and interactions with the intracluster medium \citep[ICM;][]{Stott2012}. {Active galactic nucleus (AGN) feedback-regulated cooling in the ICM is another mechanism commonly invoked to explain the evolution of BCGs and in particular the fact that large reservoirs of molecular gas are rarely observed in BCGs \citep[e.g.,][]{Edge2001,Salome_Combes2003}. While condensation of the ICM occurs in the central and low-entropy regions of the cluster, the energy injected into the ICM via radio jets propagating from the BCGs themselves may prevent the formation of the strong ($\sim({\rm 100-1000})~M_\odot~{\rm yr}^{-1}$) cooling flows that were theoretically predicted \citep[e.g.,][]{Fabian1994,McNamara2000,Peterson2003,Peterson_Fabian2006,McNamara_Nulsen2012}.}

Most of the BCG stellar mass is assembled at high redshifts ($z\sim3-5$) {from} smaller sources that are later swallowed by the BCG \citep{DeLucia_Blaizot2007}. Indeed, BCGs have doubled their stellar mass since $z\sim1$ \citep{Lidman2012}, which is consistent with evolution via the dry accretion of satellites \citep{Collins2009,Stott2011,Lavoie2016}. However, recent studies have proposed that high star formation is present in distant BCGs \citep{McDonald2016,Webb2015a,Webb2015b,Bonaventura2017}, a scenario where star formation is fed by rapid gas deposition, possibly regulated by cooling flows \citep{Salome_Combes2003,Olivares2019}. Such results imply that at least some distant BCGs should experience strong environment-driven gas processing and quenching mechanisms, such as strangulation, ram pressure stripping, and galaxy harassment \citep[e.g.,][]{Larson1980,Moore1999}, which are still, however, substantially unexplored for distant BCGs.

In support of the latter scenario, some studies report growing evidence for large reservoirs of molecular gas ($\gtrsim10^{10}~M_\odot$) feeding the star formation in some intermediate- to high-redshift BCGs \citep{McDonald2014,Russell2017,Castignani2020a,Damato2020,Dunne2021,OSullivan2021}. 
However, a number of other distant star-forming BCGs have only robust upper limits on the order of $\sim10^{10}~M_\odot$ in their $H_2$ gas masses \citep{Castignani2020a,Castignani2020b,Castignani2020c}. In \citet{Castignani2020c}, in particular, we found gas-rich companions around a distant BCG at $z=1.7$, which is instead gas-poor.

These results suggest that gas-rich BCGs in the distant Universe (${z\sim0.3-2}$) belong to a rare population of star-forming cluster ellipticals that experience gas deposition and a subsequent conversion of gas{ into star} formation. As suggested in \citet{Castignani2020a} for distant BCGs of the Cluster Lensing And Supernova survey with {\it Hubble} (CLASH) sample \citep{Postman2012}, this is an environment-driven process. The detection of large molecular gas reservoirs is likely due to the strong cooling of the ICM at the cluster cores, which ultimately favors condensation into molecular gas, similar to what has been observed in some local BCGs \citep{Salome2006,Tremblay2016}.

To understand the effect of the cluster environments on  the mass assembly and star formation of distant BCGs, it is essential to start building a sample of star-forming BCGs detected at different molecular gas transitions, which probe different gas densities, and possibly different degrees of environmental processing.  However, this is challenging as it requires a large sample of{ well-defined} clusters and a robust characterization of their BCGs  at 
{multiple wavelengths} to enable the stellar and star formation properties to be accurately inferred.

To this aim, in this work we report the study of three star-forming BCGs that we selected from the multiwavelength wide-field Kilo Degree Survey \citep[KiDS;][]{deJong2017,Kuijken2019} and that we observed in the first three CO transitions with the {Institut de Radioastronomie Millimétrique (IRAM)} 30m telescope. The present study is part of a larger campaign targeting galaxies in and around distant clusters in CO, mainly BCGs \citep{Castignani2018,Castignani2019,Castignani2020a,Castignani2020b,Castignani2020c,Castignani2020d,Castignani2022a}, with the final goal of evaluating the impact of dense megaparsec-scale environments in processing the galaxies' gas reservoirs.

The paper is structured as follows. In Sect.~\ref{sec:BCGsample} we describe the BCGs that are the subject of this work, in Sect.~\ref{sec:observations_and_data_reduction} we describe the molecular gas observations and data analysis, {in Sect.~\ref{sec:comparison_sample} we describe our comparison samples,} in Sect.~\ref{sec:results} we present the results, and in Sect.~\ref{sec:conclusions} we summarize them and draw our conclusions. Stellar mass and star formation rate (SFR) estimates reported in this work for the three targeted BCGs rely on the \citet{Chabrier2003} initial mass function (IMF). Magnitudes are reported in the AB system. Throughout this work we adopt a flat {$\Lambda$-cold dark matter} ($\Lambda$CDM) cosmology with matter density $\Omega_{\rm m} = 0.30$, dark energy density $\Omega_{\Lambda} = 0.70$, and Hubble constant $h=H_0/(100\, \rm km\,s^{-1}\,Mpc^{-1}) = 0.70$.

\begin{figure}[]\centering
\captionsetup[subfigure]{labelformat=empty}
%%%%%%%%%%%%%%%%%%%%%%%%%%%%%%%%%%%%%%%%%%%%5
\subfloat[KiDS~0920]{\hspace{0.cm}\includegraphics[trim={0cm 0cm 0cm 
0cm},clip,width=0.5\textwidth,clip=true]{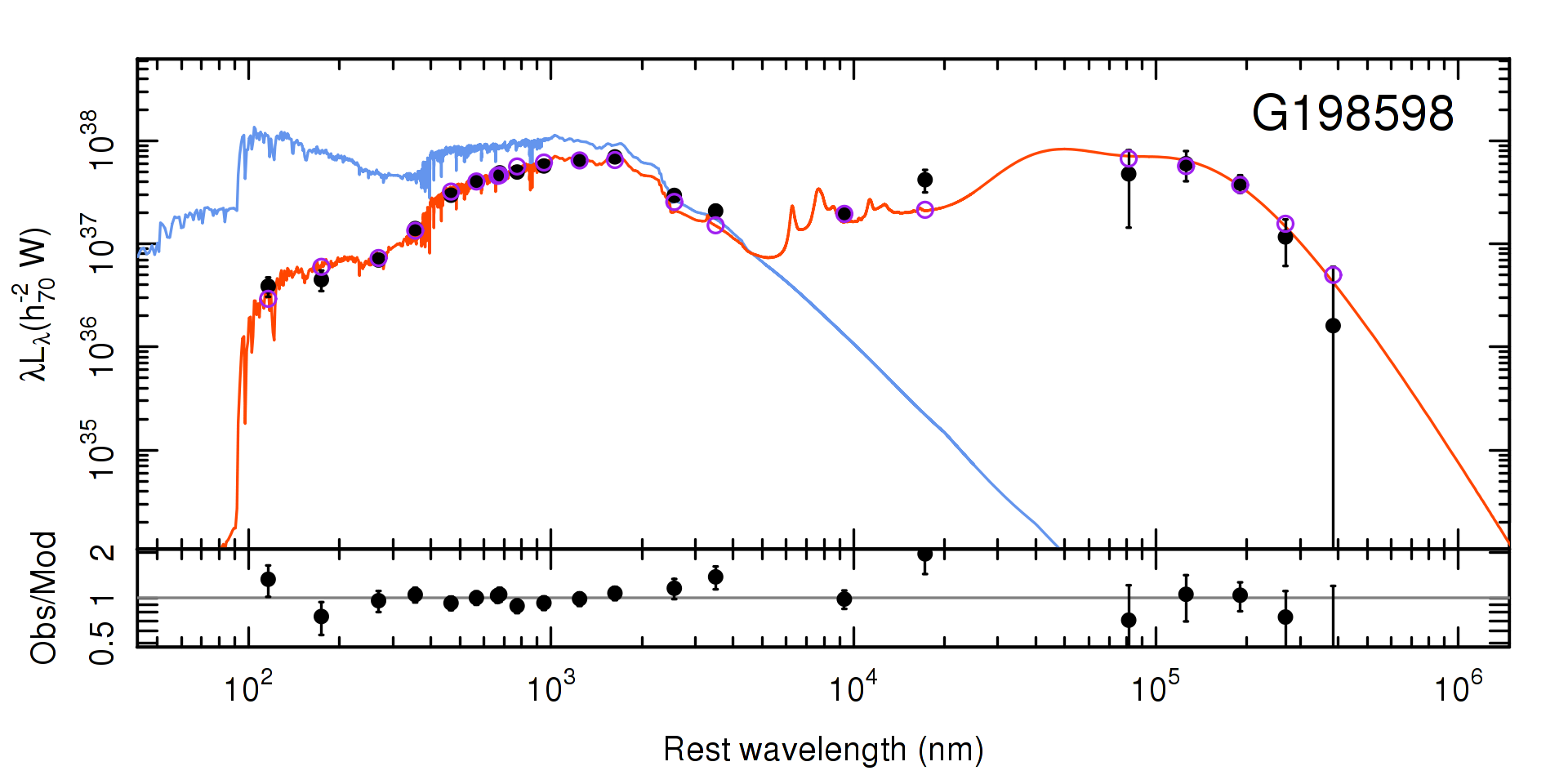}}
\hspace{0.2cm}\subfloat[KiDS~1220]{\hspace{0.cm}\includegraphics[trim={0cm 0cm 0cm 
0cm},clip,width=0.5\textwidth,clip=true]{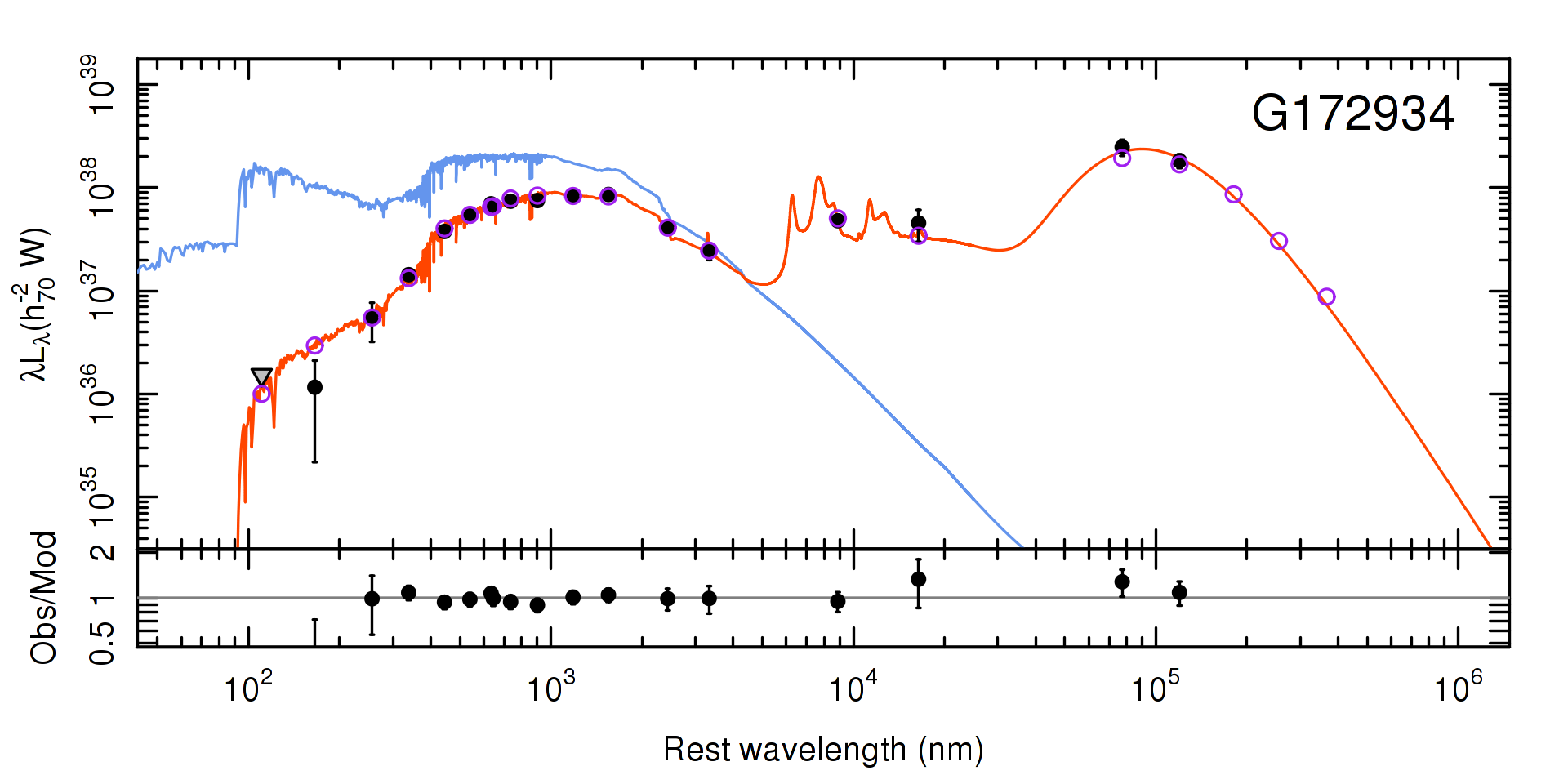}}
\hspace{0.2cm}\subfloat[KiDS~1444]{\hspace{0.cm}\includegraphics[trim={0cm 0cm 0cm 
0cm},clip,width=0.5\textwidth,clip=true]{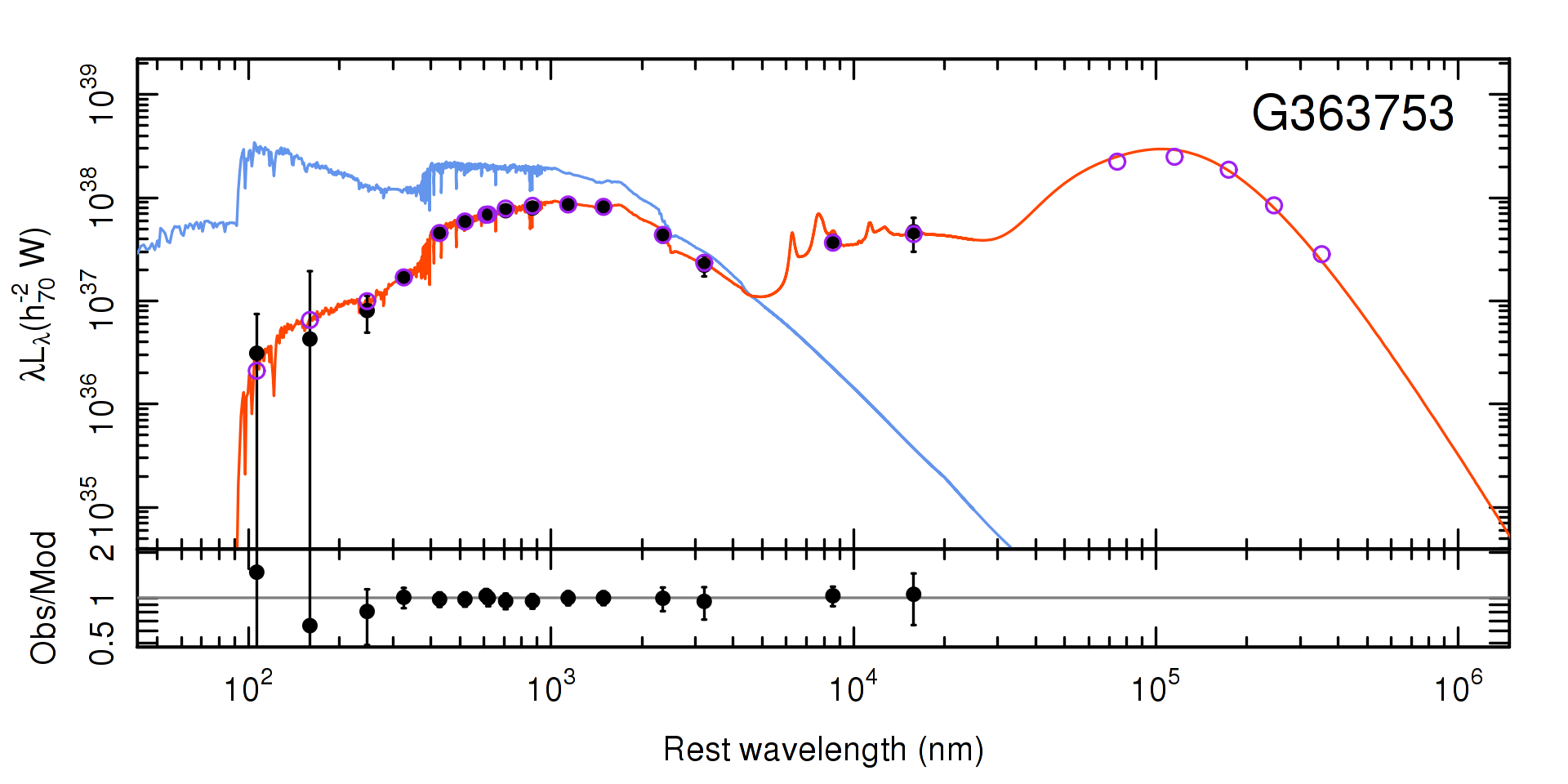}}
%%%%%%%%%%%%%%%%%%%%%%%%%%%
\caption{Ultraviolet to far-infrared SEDs of the three BCGs of this work taken from GAMA DR3 \citep{Driver2018}. Dust-attenuated (red curves) and dust-unattenuated (blue curves) MAGPHYS fits are overplotted. Filled black dots and open circles are the observed and model photometry, respectively. The triangle for KiDS~1220 is a GALEX far-ultraviolet upper limit. Bottom panels show the residuals.}\label{fig:BCG_SEDs}
\end{figure}
%%%%%%%%%%%%%%%%%%%%%%%%%%%%%%%%%%%%%%%%%%%%5

%-------------- Table: source properties-------------------
\begin{table*}[]\centering
\begin{adjustwidth}{-1cm}{}
{\small
\begin{center}
\begin{tabular}{ccccccccccc}
\hline\hline
Galaxy  & R.A. & Dec. & $z_{spec}$ &  $\log(M_\star/M_\odot)$ & $\log(L_{\rm dust}/L_\odot)$ & $\log(M_{\rm dust}/M_\odot)$ & SFR & sSFR & sSFR$_{\rm MS}$ & $\log(M_{200}/M_{\odot})$  \\
   & (hh:mm:ss.s) & (dd:mm:ss.s) &  &   & &  & ($M_\odot$/yr) &  (Gyr$^{-1}$) & (Gyr$^{-1}$) & \\ 
  (1) & (2) & (3) & (4) & (5) & (6) & (7) & (8) & (9) & (10) & (11) \\
 \hline
%A209  &     01:31:52.53 &  -13:36:40.5   & 0.206 & $2.00\pm0.16$ & $<0.12$~(UV) & $<6.0\times10^{-4}$ & 0.04 & $106\pm27$\\ %old coord : 01:31:52.57 , -13:36:38.8
KiDS~0920 & 09:20:17.28 & -00:46:58.51 & 0.3216 & $11.35^{+0.09}_{-0.06}$ & $11.46^{+0.07}_{-0.07}$  & $8.46^{+0.24}_{-0.21}$ &  $22.2^{+3.9}_{-4.2}$  & $0.10^{+0.04}_{-0.02}$ & 0.06 &  ${13.9^{+0.1}_{-0.2}}$ \\ %M_SUNH_A = 0.269 (1e+14 Msun/h); SN_cluster = 3.74
KiDS~1220 & 12:20:02.30 & -02:25:31.15 & 0.3886 &  $11.48^{+0.09}_{-0.08}$ & $11.81^{+0.08}_{-0.03}$ & $8.66^{+0.29}_{-0.15}$ &  $32.5^{+10.3}_{-3.5}$ & $0.11^{+0.06}_{-0.02}$ & 0.07 & ${14.2^{+0.1}_{-0.2}}$ \\ %M_SUNH_A = 1.274 (1e+14 Msun/h); SN_cluster = 4.71   
KiDS~1444 & 14:44:10.87 & +02:34:18.23 & 0.4417 & $11.38^{+0.09}_{-0.09}$ & $11.69^{+0.23}_{-0.16}$ & $8.84^{+0.49}_{-0.45}$ & $33.7^{+20.9}_{-10.1}$ &  $0.15^{+0.09}_{-0.05}$ & 0.09 & ${13.8^{+0.1}_{-0.2}}$ \\ %M_SUNH_A = 0.285 (1e+14 Msun/h); SN_cluster = 3.72 
\hline 
\end{tabular}
 \end{center} }
 \end{adjustwidth}
\caption{Properties of our targets. (1) BCG name; (2-3) J2000 equatorial coordinates; (4) spectroscopic redshift from GAMA; (5) stellar mass; (6-7) dust luminosity and mass; (8) SFR; (9) specific SFR; (10) specific SFR at the MS estimated using the \citet{Speagle2014} prescription; (11) richness-based $M_{200}$ mass estimate of the cluster. Columns (5-9) are from MAGPHYS SED fits \citep{Driver2018}.}
\label{tab:BCG_properties}
\end{table*}
%%%%%%%%%%%%%%%%%%%%%%%%%%%%%%%%%%%%%%

%SED fits were performed with the galaxy templates here:
%/home/gcastignani/Radio_Galaxy_IRAM_project/macros/LePhare_macros/SEDlists/CE_NEW_MOD_CHARY_ELBAZ.list
%see /home/gcastignani/LePhare/lephare_dev/sed/GAL/CE_NEW/README/
%but Mstar was assigned using Pegase2 template:
%/home/gcastignani/Radio_Galaxy_IRAM_project/macros/LePhare_macros/SEDlists/PEGASE2_ELL3.list >> here Rana & Basu 1992 IMF is assumed 
%From Kiuchi et al. (2009, 2009ApJ...696.1051K) I read The stellar mass derived with the IMF of Rana & Basu (1992)is∼0.1 dex smaller than that of Salpeter (1955). 
%Also: Madau & Dickinson 2014:  "To rescale stellar masses from Chabrier or Kroupa to Salpeter IMF, we divide by constant factors 0.61 and 0.66, respectively. " 
%%%%%%%%%%%%%%%%%%%%%%%%%%%%%%%%%%%%%5
%SFR are estimated from Ldust (Bruzual & Charlot 2003 model) assuming a Chabrier 2003 IMF.

\section{Three star-forming BCGs at $z\sim0.4$}\label{sec:BCGsample}
\subsection{The selection of the BCGs in KiDS}\label{sec:BCGselection}
KiDS is a wide field imaging survey that {in its final release will cover} around 1,350~deg$^2$ of the sky \citep{deJong2017,Kuijken2019}, equally divided between an equatorial field (KiDS-N) and a southern field (KiDS-S). VLT Survey Telescope (VST) \textsf{ugri} optical observations  and {Visible and Infrared Survey Telescope for Astronomy (VISTA) telescope \textsf{ZYJHK} infrared imaging} enable the detection of $\sim5\times10^{7}$ galaxies down to a limiting AB magnitude  $\textsf{r}\sim25.0$  ($5\sigma$ in $2^{\prime\prime}$ aperture). Optical-infrared photometry enabled the determination of accurate photometric redshifts up to $z\sim1$ \citep{Hildebrandt2021,Benitez2000}, which are complemented by spectroscopic information as KiDS is also covered by Galaxy And Mass Assembly \citep[GAMA;][]{Baldry2010} and Sloan Digital Sky Survey \citep[SDSS;][]{York2000} spectroscopic surveys. 

This rich data set allows the detection of distant clusters in KiDS and the characterization of their galaxy populations. Distant clusters up to $z\sim1$ have been indeed detected at signal-to-noise ratio S/N$>3$ within KiDS \citep{Maturi2019,Bellagamba2019} with the {Adaptive Matched Identifier of Clustered Objects (AMICO)} cluster finder \citep{Bellagamba2018}, which looks for overdensities of galaxies using photometric redshifts and a matched filtering. 

For this work we consider the parent sample of 5,251 BCGs (of which 1,484 { have} spectroscopic redshifts) in the equatorial KIDS-N area, which can be targeted with observational facilities in the northern hemisphere. These BCGs were selected by \citet{Radovich2020} within the Data Release (DR)~3 of KiDS \citep{deJong2017} as the brightest galaxies in the \textsf{r} band within the cluster cores, on the basis of AMICO probabilistic cluster memberships. 

We aim at targeting actively star-forming BCGs, to investigate their molecular gas reservoirs feeding the star formation, and explore their properties in connection with those of the host galaxies and their environments.
We thus limit ourselves to the 684 BCGs in DR3 KiDS-N with spectroscopic redshifts $z>0.3$. { This redshift cut allows us to focus on distant BCGs, and on their evolution, while BCGs at lower redshifts, including those of Abell clusters, have been extensively studied in previous work \citep{Edge2001,Salome_Combes2003,Olivares2019}.}

Similarly to the target selection we did in \citet{Castignani2019} we { further} limit ourselves to those sources with clear counterparts in the all-sky data release of the Wide-field Infrared Survey Explorer \citep[WISE;][]{Wright2010} at 22~$\mu$m in the observer frame (i.e., the W4 channel). These are thus good star-forming BCG candidates in the distant Universe. We found only nine such BCGs with WISE W4 counterparts at S/N$>3$, 
of which six had S/N$>3.3$. Among the latter in this work we consider the three with highest values in the range S/N$\sim4.1-6.5$, while the remaining three will be the subject of a forthcoming paper.

The three intermediate-$z$ BCGs of this work are KiDS~0920 (i.e., G198598 at $z=0.3216$), KiDS~1220 (G172934 at $z=0.3886$), and KiDS~1444 (G363753 at $z=0.4417$). { In} parentheses, we report the ID numbers of the sources from GAMA DR3 \citep{Baldry2018} and the corresponding spectroscopic redshifts.~\footnote{\href{http://www.gama-survey.org/dr3/}{http://www.gama-survey.org/dr3/}} The three sources belong to moderately rich clusters, { which have been detected with  S/N = 3.7, 4.7, and 3.7 \citep{Maturi2019,Bellagamba2019}, while they have intrinsic richness $\lambda_\star=20\pm3$, $33\pm6$, and $19\pm3$, respectively. The corresponding richness-based cluster masses range between $\log(M_{200}/M_\odot)=13.8-14.2$ (see Table~\ref{tab:BCG_properties}). Here $\lambda_\star$ is the intrinsic cluster richness as defined in \citet{Maturi2019}. Based on these properties we infer, on average, a small probability of $\sim0.3\%$ of the three clusters being false positives \citep[see Fig.~12 of][]{Maturi2019}. We additionally searched for our clusters in published catalogs of clusters. KiDS~1220 matches with WHL~J121959.1-022611, which is an optically identified cluster in SDSS, found by both the redMaPPer \citep{Rykoff2014,Rozo2015} and the \citet{Wen2012} cluster finding algorithms. On the other hand, our search did not produce positive results when looking in catalogs of X-rays and  Sunyaev-Zel'dovich clusters.
This is not surprising as our considered clusters have relatively low masses and are at intermediate redshifts.}

%eROSITA Final Equatorial-Depth Survey \citep{Liu2022}, 
%\bibitem[Liu et al.(2022)]{Liu2022} Liu, A., Bulbul, E., Ghirardini, V. et al. 2022, A\&A, 661, 2

Figure~\ref{fig:BCG_images} displays the color composite  images for the three KiDS BCGs. The images show that our targets are  massive elliptical galaxies,  bulge-dominated, but they also show complex and disturbed morphology. Indeed, KiDS~0920 shows an elongated reddish emission, possibly associated with the stellar disk, along the northeast to southwest direction. Similarly, KiDS~1444 shows disturbed morphology, with evidence of a faint elongated tail to the south. Both KiDS~1220 and KiDS~1444 have nearby companions at a projected separation of $\sim3$~arcsec (i.e., $\sim$17~kpc at the redshifts of the sources).

\subsection{Stellar and star formation properties}\label{sec:SFR_Mstar_prop}
We now investigate the multiwavelength properties of the three BCGs of this work. 
\subsubsection{Spectral energy distributions}\label{sec:SEDs}
Figure~\ref{fig:BCG_SEDs} shows the ultraviolet to far-infrared spectral energy distributions (SEDs) of the three sources from GAMA DR3. 
The SED modeling shown in the figure corresponds to the MAGPHYS  \citep[][]{daCunha2008} fits by \citet{Driver2018}. Photometric data include GALEX \citep{Martin2005,Morrissey2007} in the ultraviolet, SDSS  \citep{York2000} in optical, as well as VISTA Kilo-degree Infrared Galaxy Survey 
\citep[VIKING;][]{Edge2013}, WISE \citep{Wright2010}, and {\it Herschel}-ATLAS \citep{Eales2010,Valiante2016} in the near- to far-infrared.

Table~\ref{tab:BCG_properties} summarizes the BCG properties including their stellar, star formation, and dust properties from the SED fits. All three BCGs have stellar masses exceeding $10^{11}~M_\odot$, which confirms that they are very massive galaxies. Furthermore, with the only exception of KiDS~1444, the remaining two BCGs have far-infrared emission from {\it Herschel}, well modeled by a dust component. This supports the scenario that the observed infrared emission is mainly due to star formation, with minimum AGN contamination. Indeed, for the three BCGs, dust masses and luminosities are in the range $\log(M_{\rm dust}/M_\odot)\simeq8.5-8.7$ and $\log(L_{\rm dust}/L_\odot)\simeq11.5-11.8$, respectively, typical of luminous infrared galaxies (LIRGs). The SFR estimates based on the ultraviolet to far-infrared SED modeling are also high, and in the range $\sim(20-30)~M_\odot$/yr. These are consistent with, but higher, than the SFRs of main sequence (MS) galaxies of similar mass and redshifts \citep{Speagle2014}; details are provided in Table~\ref{tab:BCG_properties}.

We remind the reader that we selected the three distant star-forming BCGs from a parent sample of 684 BCGs at $z>0.3$ (Sect.~\ref{sec:BCGselection}). They represent 0.4\% of the parent sample and are among the most star-forming BCGs in KiDS. This is a rare population of BCGs that experience stellar mass assembly and are thus caught in a special phase of their evolution  \citep[see, e.g.,][for similar BCGs]{Castignani2020a,Castignani2020b}.  The selected targets are thus the {intermediate-$z$} counterparts of local star-forming BCGs ($>40~M_\odot$/yr) such as the famous Perseus~A and Cygnus~A \citep{FraserMcKelvie2014}.

%we limited ourselves to the nine BCGs at $z>0.3$ with robust 24~$\mu$m emission from WISE W4 (SNR$>3$) and we ask to observe the three with the most secure WISE W4 detections (SNR$\sim8$). Visual inspection of their RGB images (Fig.~2) confirm they are massive ellipticals surrounded by companions, possibly in the phase of merging with the BCGs in the cores of clusters, with richness based $M_{\rm 200} \sim10^{14}~M_\odot$. The BCGs also show disturbed morphology and UV emission from GALEX, possibly associated with ongoing SF activity.  Near-infrared to optical SED fits are reported in Fig.~3 and confirm the BCGs as strongly star forming ({SFR}${\gtrsim70~M_\odot}$) sources, with a total FIR 8-1000~$\mu$m luminosity in the range $L_{\rm FIR}=(6-9)10^{11}~L_\odot$ typical of LIRGs.
%They thus belong to the rare population of intermediate-$z$ star forming BCGs and we want to observe their cold gas feeding SF.

\subsubsection{Line diagnostics}\label{sec:line_diagnostics}

We further investigate the star formation properties of the BCGs using the available spectroscopy in GAMA DR3. According to \citet{Kennicutt1998}, the SFR is directly proportional to both H$\alpha$ line and [O~II] forbidden-line doublet (3726~\AA, 3729~\AA) luminosities $L_{\rm H\alpha}$ and $L_{\rm [O~II]}$, respectively. We thus use these relations to estimate the SFR, calibrated to a \citet{Chabrier2003} IMF.  We also correct for dust attenuation using the empirical relation by \citet{Garn_Best2010}, which relates dust attenuation to stellar mass as follows:

\begin{equation}
\label{eq:AHa}
 A_{\rm H\alpha} = 0.91 + 0.77 M + 0.11 M^2 - 0.09 M^3\;,
\end{equation}
where $M = \log(M_\star/M_\odot)-10$. With these conventions, the H$\alpha$- and [O~II]-based SFRs can be expressed as follows \citep[see, e.g.,][]{Gilbank2010,Zeimann2013,Old2020}:

\begin{equation}
\label{eq:SFR_Ha}
\frac{{\rm SFR}_{\rm H\alpha}}{M_\odot/{\rm yr}} = 5.0\times10^{-42}\times \frac{L_{\rm H\alpha}}{{\rm erg~s}^{-1}}\times10^{0.4\;A_{\rm H\alpha}}\;,
\end{equation}

\begin{equation}
\label{eq:SFR_OII}
 \frac{{\rm SFR}_{\rm [O~II]}}{M_\odot/{\rm yr}} =  5.0\times10^{-42}\times\frac{L_{\rm [O~II]}}{{\rm erg~s}^{-1}}\times\frac{10^{0.4\;A_{\rm H\alpha}}}{r_{\rm lines}}\;,
\end{equation}
where $r_{\rm lines}$ is the ratio  of extinguished [O~II] to H$\alpha$ fluxes.  

We found GAMA DR3 spectra for all three BCGs, taken with the AAOmega/2dF multi-fiber spectrograph on the 3.9~m Anglo-Australian Telescope. Gaussian fits to several lines are also available \citep{Gordon2017}.
Among the three BCGs, H$\alpha$ emission is detected for KiDS~0920 only, with a flux of $F_{\rm H\alpha}=(208.6\pm18.5)\times10^{-17}$~erg~s$^{-1}$~cm$^{-2}$. At a spectral resolution of $R\simeq1300$ the H$\alpha$ line is not blended with [N~II]. For the other two BCGs H$\alpha$ is redshifted to wavelengths longer than 9100~\AA, thus outside the  wavelength range of $3750~\AA\lesssim\lambda\lesssim8850~\AA$ covered by the spectrograph \citep{Hopkins2013}.

Conversely, [O~II] doublet emission is detected for all three BCGs. Similarly to \citet{Gilbank2010}, for each source we sum the fluxes of the two components of the [O~II] doublet, adding their uncertainties in quadrature. We refer to this sum simply as [O~II], as in Eq.~\ref{eq:SFR_OII}. The resulting [O~II] doublet fluxes of KiDS~0920, 1220, and 1444 are $F_{\rm [O~II]}=(188.9\pm9.3)$, $(72.6\pm15.8)$, and $(88.0\pm20.1)$, respectively, in units of  $10^{-17}$~erg~s$^{-1}$~cm$^{-2}$. 
The line ratio $r_{\rm lines}$ is equal to $r_{\rm lines}=0.9$ for KiDS~0920, while we fix it to $r_{\rm lines}=0.5$ for the other two BCGs for which we do not have H$\alpha$ measurements, as in \citet{Gilbank2010}.

Using these values, Eq.~\ref{eq:SFR_Ha} yields ${\rm SFR}_{\rm H\alpha}=(20.9\pm1.9)~M_\odot$/yr for KiDS~0920. {We obtain a similar ${\rm SFR}_{\rm H\alpha}=(27.6\pm2.6)~M_\odot$/yr if we compute the dust attenuation using the observed H$\alpha$/H$\beta$ Balmer decrement, following \citet{Dominguez2013}.} Eq.~\ref{eq:SFR_OII} yields instead ${\rm SFR}_{\rm [O~II]}=(24.1\pm5.3)$ and $(37.7\pm8.6)~M_\odot$/yr for KiDS~1220 and 1444, respectively.\footnote{We note that, for KiDS~0920 it holds ${\rm SFR}_{\rm H\alpha}={\rm SFR}_{\rm [O~II]}$ as $r_{\rm lines}=F_{\rm [O~II]}/F_{\rm H\alpha}=L_{\rm [O~II]}/L_{\rm H\alpha}$.} We find a good agreement, within the reported uncertainties, when comparing these line-based SFRs with those in Table~\ref{tab:BCG_properties} that are used throughout this work and come from the SED fits. This is remarkable, considering that discrepancies, even up to a factor of $\sim3$, are not uncommon when comparing different  estimates of the SFR \citep[e.g.,][]{Calzetti2013}.

Finally, we use the line ratios to identify the dominant mechanism of gas ionization using ratios of strong optical lines. We do it by means of the  {Baldwin, Phillips \& Terlevich (BPT)} diagram \citep{Baldwin1981,Veilleux_Osterbrock1987}. To this aim we use the [O~III]~$\lambda5007/{\rm H}\beta$, [N~II]~$\lambda6584/{\rm H}\alpha$, and [O~I]~$\lambda6300/{\rm H}\alpha$ line ratios. KiDS~0920 is the only source that has a detection in H$\alpha$, as well as in the other lines, and we can see where it falls in the BPT diagram. Using the line fluxes reported in the GAMA DR3 database we find the line ratios
$\log({\rm[O~III]}\,\lambda 5007 / {\rm H}\beta)=0.55\pm0.10$, $\log({\rm[N~II]}\,\lambda 6584 /{\rm H}\alpha)=-0.18\pm0.07$, and $\log({\rm[O~I]}\,\lambda 6300 /{\rm  H}\alpha)=-0.40\pm0.10$. These imply that KiDS~0920 falls in the AGN region of the [O~III]~$\lambda5007/{\rm H}\beta$ versus [N~II]~$\lambda6584/{\rm H}\alpha$ plane \citep{Kewley2006} as well as in the LINER region, close to that of Seyferts, of the  [O~III]~$\lambda5007/{\rm H}\beta$ versus [O~I]~$\lambda6300/{\rm H}\alpha$ plane \citep[e.g.,][]{Sarzi2010,Singh2013}. These results are not surprising, as BCGs are often associated with AGN and radio galaxies in particular \citep{Zirbel1996}. In this case the BCG is likely a Type~2 AGN, as the spectrum does not show any clear broad emission line (e.g., H$\alpha$ and H$\beta$). 

For the other two BCGs, KiDS~1220 and KiDS~1444, since we do not have H$\alpha$ we cannot locate them precisely in the BPT diagram. However, the sources have $\log({\rm[O~III]}\,\lambda 5007 / {\rm H}\beta)=-0.35\pm0.19$ and  $-0.15\pm0.18$, respectively, which are lower than the corresponding line ratio of KiDS~0920. This suggests that the two sources are more likely located in the star-forming region of the BPT diagram.

{Last, we inspected the position of our sources in the infrared color-color WISE diagram \citep{Jarrett2017}. Interestingly, KiDS~0920 is classified as an intermediate disk, while KiDS~1220 and KiDS~1444 are starbursts. These findings strengthen the selection of our three targets as star-forming galaxies and suggest that any possible contamination from a circum-nuclear dusty torus at infrared wavelengths is likely negligible, which is also supported by the good agreement between the SED- and line-based SFRs.}

\begin{figure*}[h!]\centering
\captionsetup[subfigure]{labelformat=empty}
%%%%%%%%%%%%%%%%%%%%%%%%%%%%%%%%%%%%%%%%%%%%5
\subfloat[]{\hspace{-1.0cm}\includegraphics[trim={1cm 2.5cm 3.9cm 
5.3cm},clip,width=0.37\textwidth,clip=true]{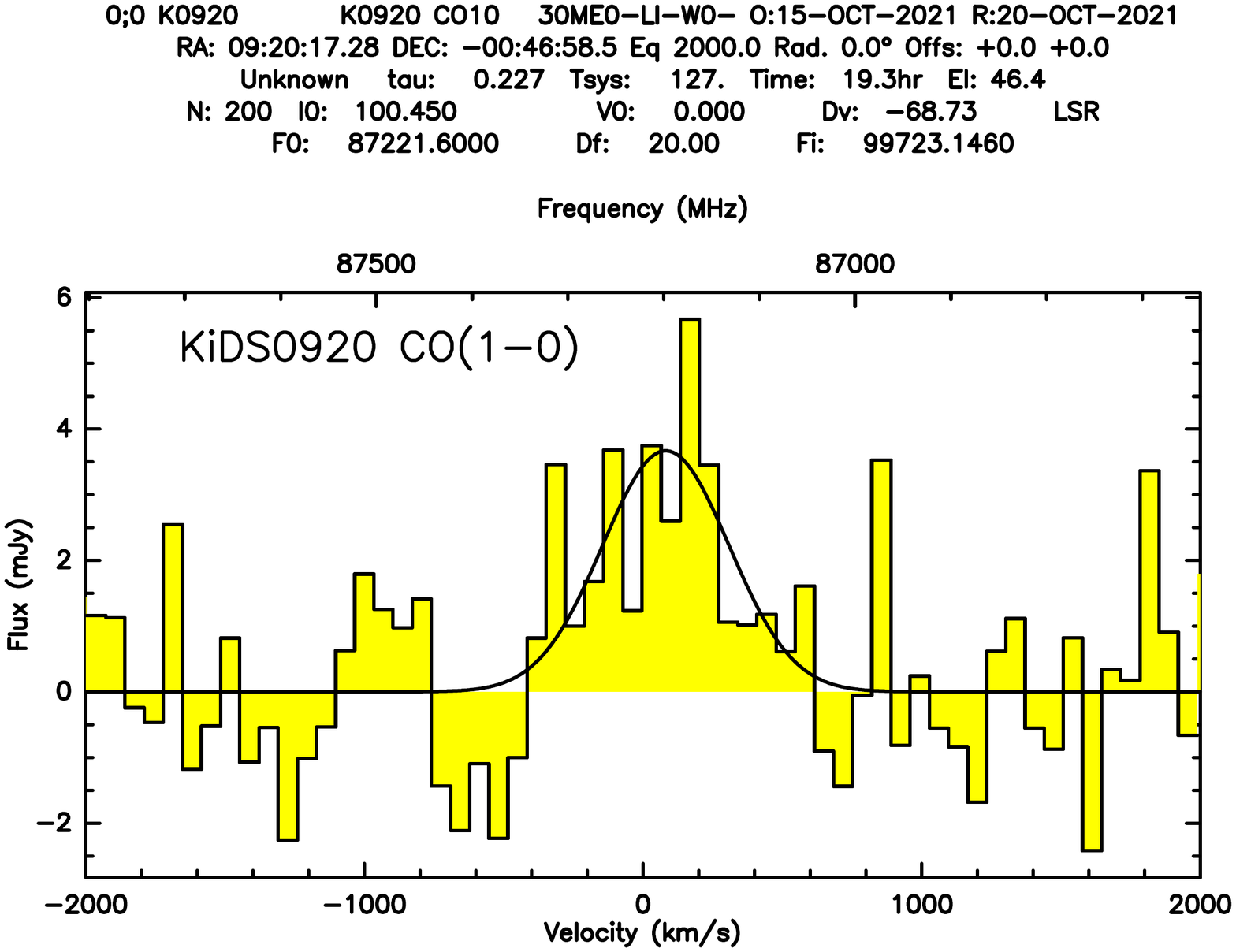}}
\subfloat[]{\includegraphics[trim={1.5cm 2.5cm 3.9cm 
5.3cm},clip,width=0.37\textwidth,clip=true]{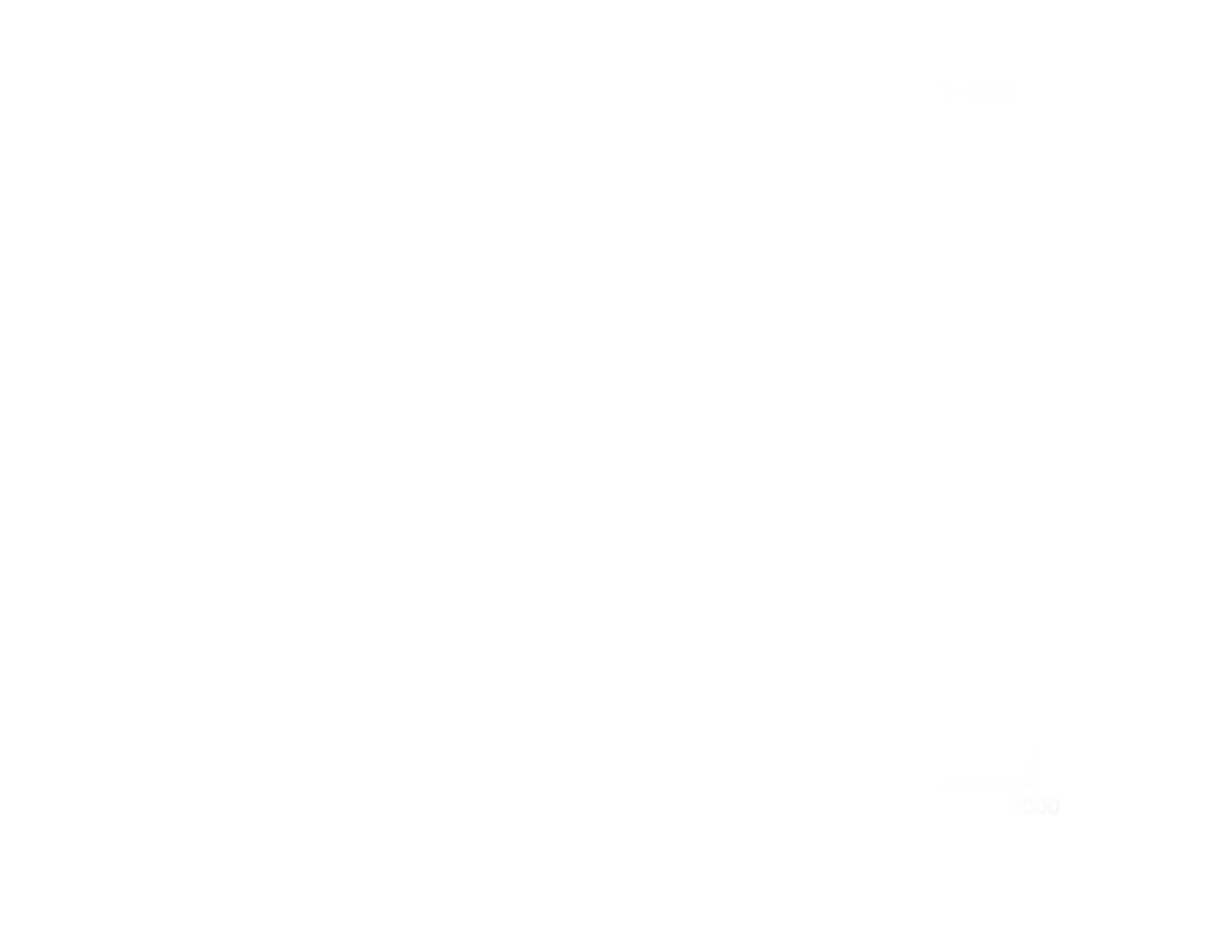}}
\subfloat[]{\includegraphics[trim={1.5cm 2.5cm 3.9cm 
5.3cm},clip,width=0.37\textwidth,clip=true]{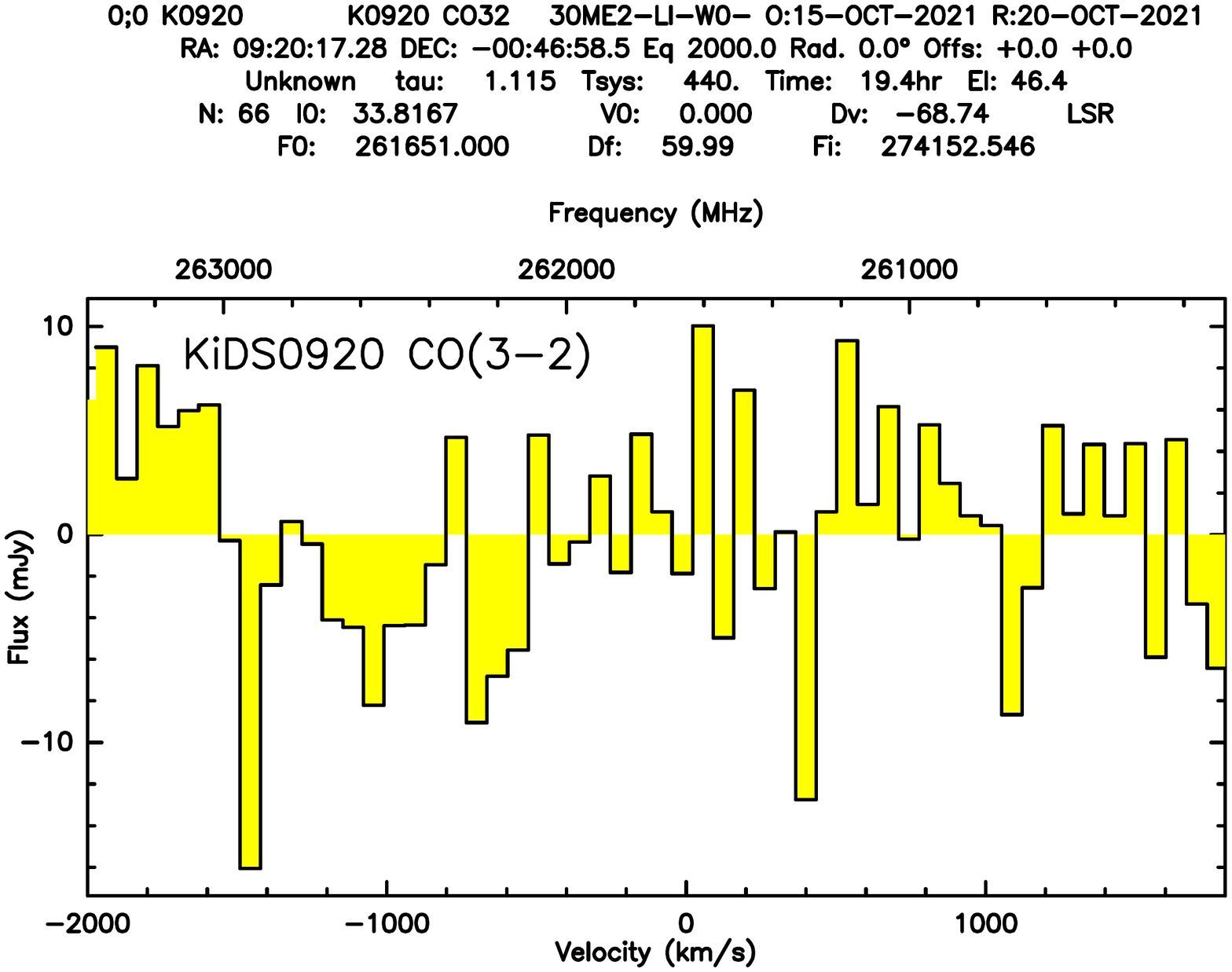}}\\
%%%%%%%%%%%%%%%%%%%%%%%%%%%%%%%%%%%%%%%%%%%%%5
\subfloat[]{\hspace{-1.0cm}\includegraphics[trim={1.5cm 2.5cm 3.9cm 
5.3cm},clip,width=0.37\textwidth,clip=true]{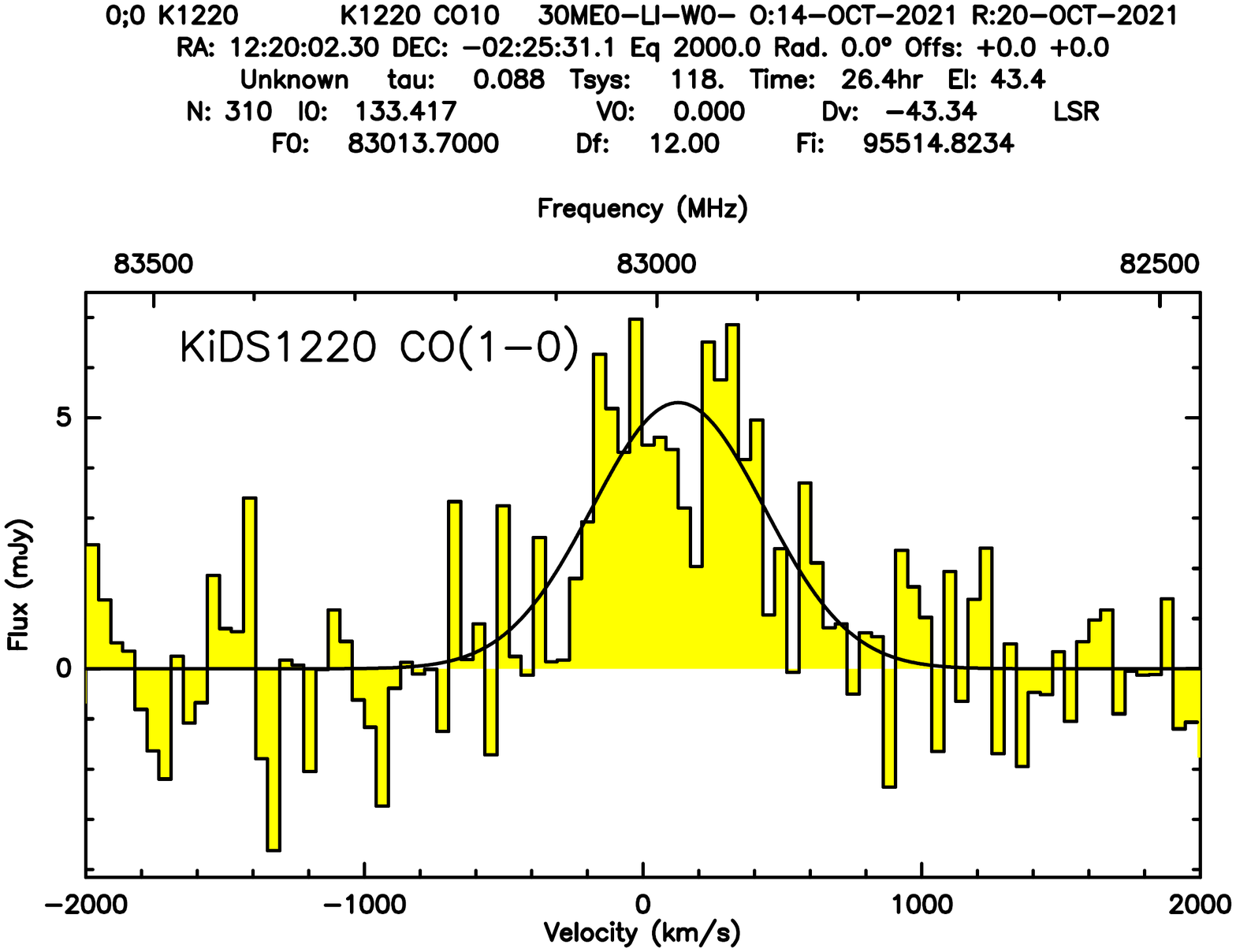}}
\subfloat[]{\includegraphics[trim={1.5cm 2.5cm 3.9cm 
5.3cm},clip,width=0.37\textwidth,clip=true]{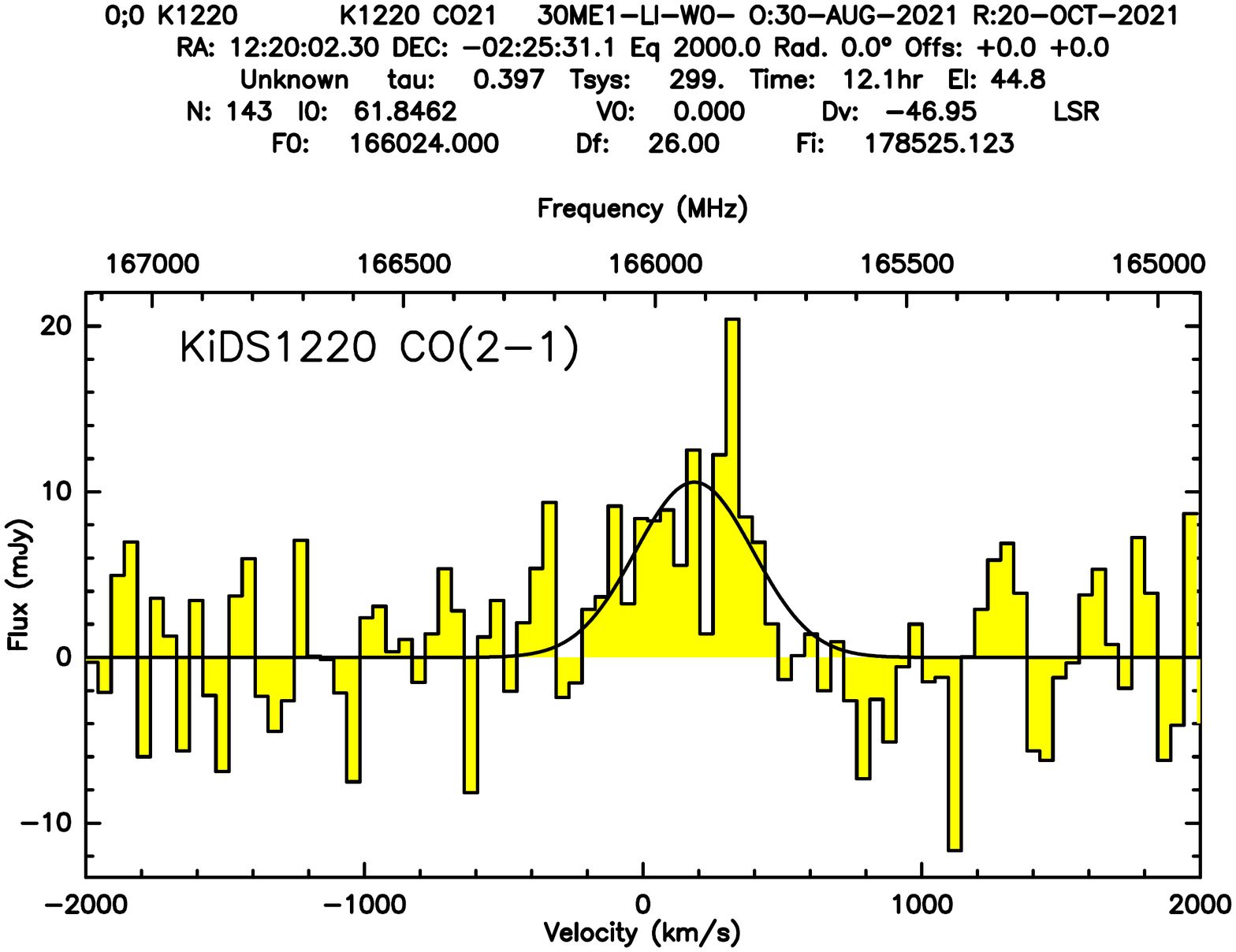}}
\subfloat[]{\includegraphics[trim={1.5cm 2.5cm 3.9cm 
5.3cm},clip,width=0.37\textwidth,clip=true]{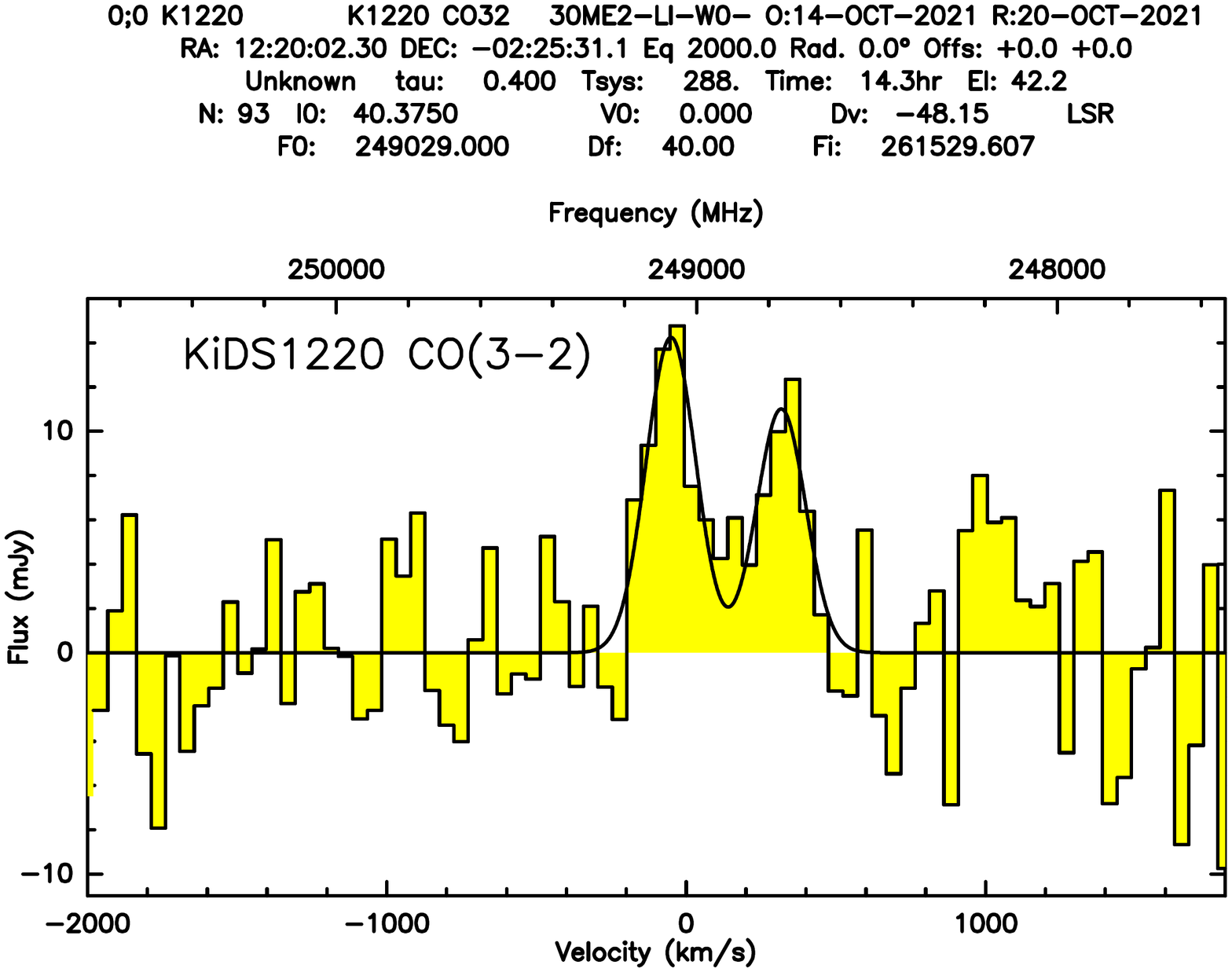}}\\
%%%%%%%%%%%%%%%%%%%%%%%%%%%%%%%%%%%%%%%%
\subfloat[]{\hspace{-1.0cm}\includegraphics[trim={1.5cm 2.5cm 3.9cm 
5.3cm},clip,width=0.37\textwidth,clip=true]{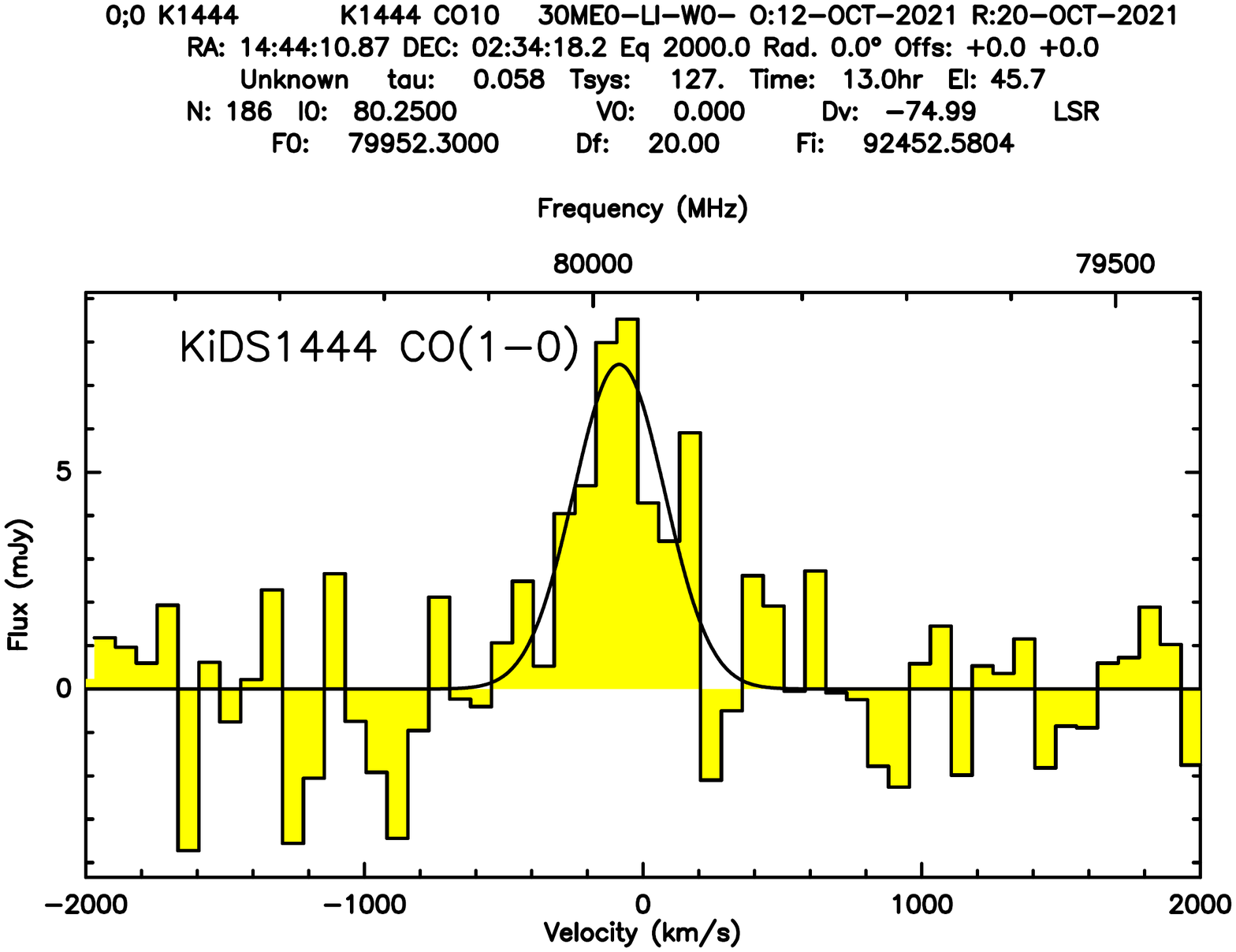}}
\subfloat[]{\includegraphics[trim={1.5cm 2.5cm 3.9cm 
5.3cm},clip,width=0.37\textwidth,clip=true]{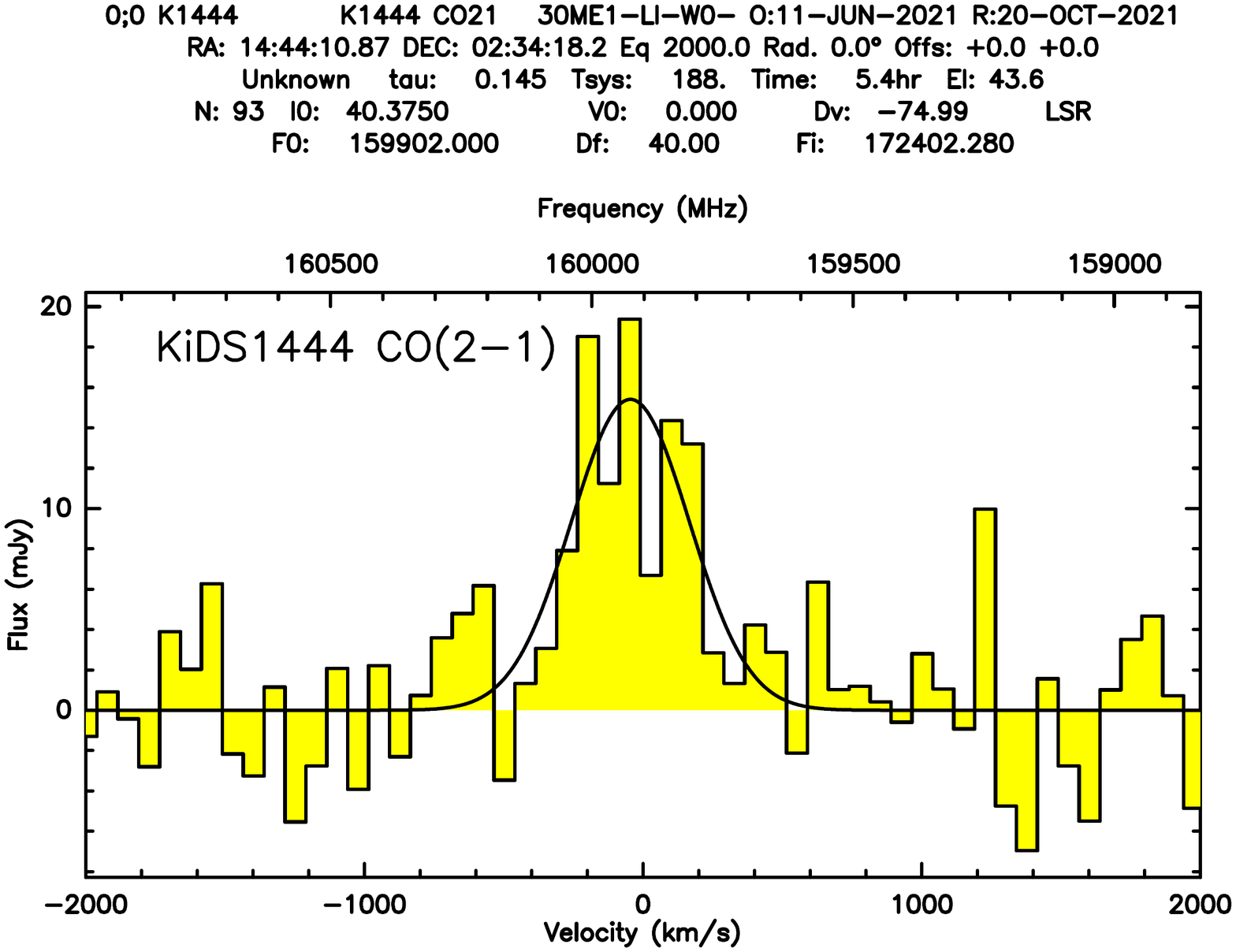}}
\subfloat[]{\includegraphics[trim={1.5cm 2.5cm 3.9cm 
5.3cm},clip,width=0.37\textwidth,clip=true]{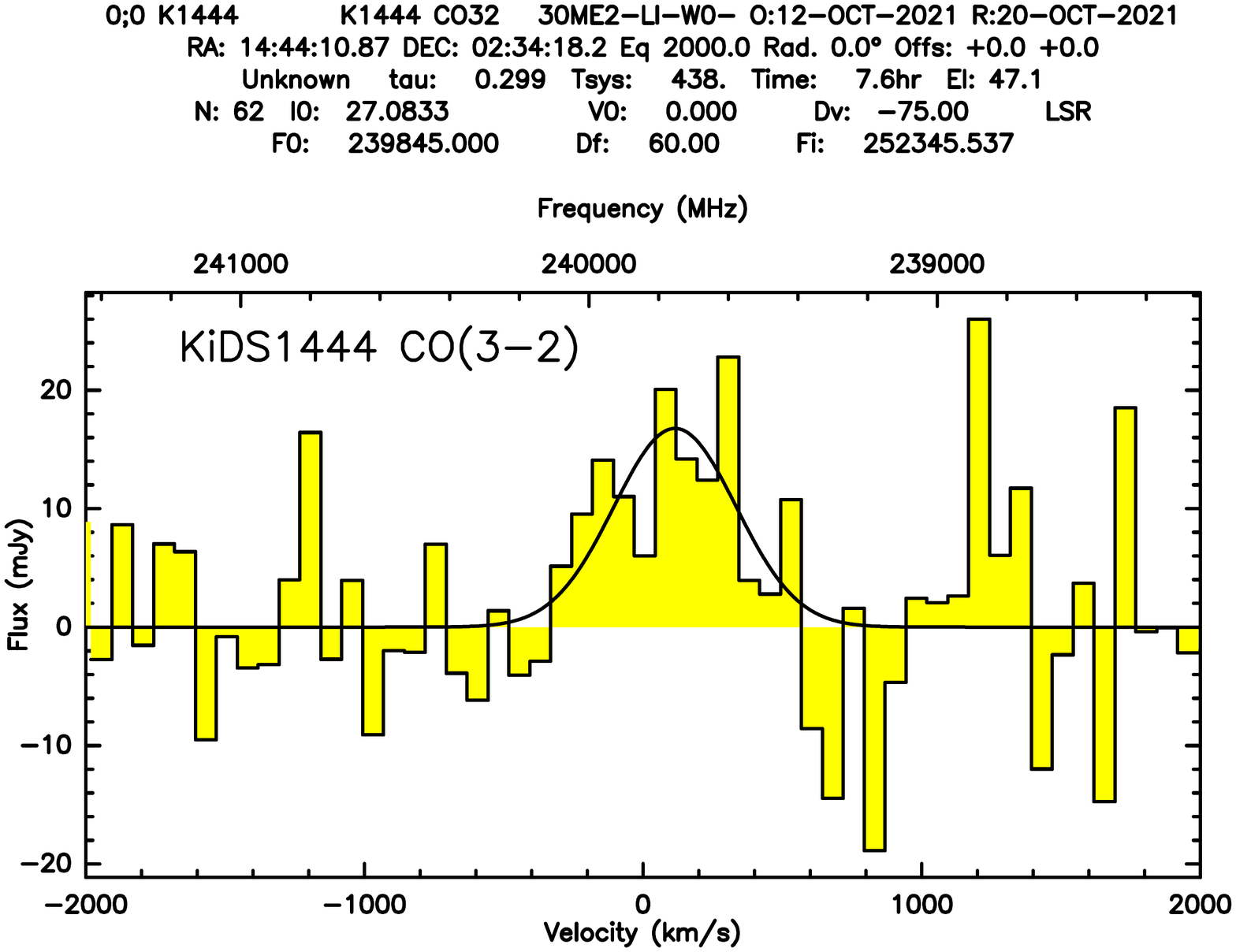}}\\
%%%%%%%%%%%%%%%%%%%%%%%%%%%
\caption{Baseline-subtracted spectra of KiDS~0920 (top row), KiDS~1220 (center row), and KiDS~1444 (bottom row). The targeted CO(J$\rightarrow$J-1) lines are reported in the top-left corner of each plot. Solid lines show the Gaussian fits to the emission lines, in the event of detection. See text for details. In all panels the flux (y axis) is plotted against the relative velocity with respect to the BCG redshift, as in Table~\ref{tab:BCG_properties} (bottom x axis) and the observer frame frequency (top x axis).}\label{fig:BCG_spectra}
\end{figure*}
%%%%%%%%%%%%%%%%%%%%%%%%%%%%%%%%%%%%%%%%%%%%5

\section{IRAM 30m observations and data reduction}\label{sec:observations_and_data_reduction}
We observed the three KiDS BCGs using the IRAM 30m telescope at Pico Veleta in Spain. The observations were carried out in June, August, and October 2021 (ID: 079-21; PI: G.~Castignani). We used the Eight Mixer Receiver (EMIR) to observe CO(J$\rightarrow$J-1) emission lines from the BCGs, where J is a positive integer denoting the total angular momentum. For each galaxy, the specific CO(J$\rightarrow$J-1) transitions were chosen to maximize the likelihood of the detection, in terms of the ratio of the predicted signal to the expected rms noise. 
%As further outlined below, when possible we also preferred low-J transitions, which are better suited for summer observations than higher-J transitions at shorter wavelengths.

The E090, E150, and E230 receivers, operating between $\sim$1-3~mm, offer 4$\times$4~GHz instantaneous bandwidth covered by the correlators. For KiDS~1220 and KiDS~1444 BCGs we used both E090 and E150 receivers, simultaneously, to observe CO(1$\rightarrow$0) and CO(2$\rightarrow$1), redshifted to 3~mm and 2~mm in the observer frame, respectively. For all three targets we also used both E090 and E230 receivers to target simultaneously CO(1$\rightarrow$0) and CO(3$\rightarrow$2) lines, redshifted to 3~mm and 1~mm, respectively. We note that we could not observe  KiDS~0920 in CO(2$\rightarrow$1) as the line is redshifted to 174.440~GHz in the observer frame, too close to the 183 GHz water vapor line, and thus prohibited because of low atmospheric transmission. 

The IRAM 30m half power beam width (HPBW) is $\sim16$~arcsec~$\frac{\lambda_{\rm obs}}{2~{\rm mm}}$ \citep{Kramer2013}, where $\lambda_{\rm obs}$ is the observer frame wavelength. All targets are thus unresolved by our observations. We refer to Sect.~\ref{sec:IRAM30m_results} for further discussion.
The wobbler-switching mode was used for all the observations to minimize the impact of atmospheric variability.
The Wideband Line Multiple Autocorrelator  was used to cover the 4$\times$4~GHz bandwidth, in each linear polarization.   We also simultaneously recorded the data with the fast Fourier transform spectrometers, as a backup, at 200 kHz resolution.

During our June and August observations, we had variable weather conditions, while in October we had good weather during our observations. Overall we had average system temperatures typical of summer season and equal to $T_{\rm sys}\simeq120-130$~K, 190-300~K, and 290-440~K at 3, 2, and 1~mm, respectively. For pointing and focus we used the following strong sources: Mars, Mercury, Venus, 3C~084, 3C~273, 3C~279, and OJ +287.

We summarize our observations in Table~\ref{tab:BCG_properties_mol_gas}.  The data reduction and analysis were performed using the {\sc CLASS} software of the {\sc GILDAS}  package\footnote{https://www.iram.fr/IRAMFR/GILDAS/}. { For each source and frequency tuning, we merged all available observations. We flagged only a few bad scans. At a velocity resolution of 60~km~s$^{-1}$ the resulting rms values of the antenna temperature (Ta$^\ast$) are similar among our three targets, at their corresponding frequency. They range between ${{\rm rms=}~0.20-0.31}$~mK at 1~mm and between ${\mathrm{rms=}~0.57-0.59}$~mK at 2~mm. At 3~mm we have instead ${{\rm rms=}~0.59}$, 0.38, and 0.96~mK for KiDS~0920, 1220, and 1444, respectively. }
Our results are presented in Sect.~\ref{sec:results}, { where we further describe the analysis of our IRAM 30m spectra.}

%NOTE: { Section updated after observations in Oct.}

%results from June observations only
%$T_{\rm sys}\simeq(180-310)$~K at 2~mm, and $T_{\rm sys}\simeq(180-310)$~K at 2~mm, and  and $T_{\rm sys}\simeq(350-610)$~K

\begin{table*}[tb]\centering
\begin{adjustwidth}{-0.9cm}{}
\begin{center}
\begin{tabular}{cccccccccccccc}
\hline
\hline
  &   &    &   &  &  &  &  &  & integration \\
 Galaxy  &  $z_{spec}$ &   CO(J$\rightarrow$J-1)  &  $\nu_{\rm obs}$ & $S_{\rm CO(J\rightarrow J-1)}\,\Delta\varv$   & S/N & FWHM & $L^\prime_{\rm CO(J\rightarrow J-1)}$ & $z_{\rm CO(J\rightarrow J-1)}$ & time \\ 
   &  & & (GHz) &  (Jy~km~s$^{-1}$) & & (km~s$^{-1}$) & ($10^{10}$~K~km~s$^{-1}$~pc$^2$) & & (hr)  \\ 
 (1) & (2) & (3) & (4) & (5) & (6) & (7) & (8) & (9) & (10) \\
 \hline
KiDS~0920 & 0.3216 & 1$\rightarrow$0 & 87.222 & $2.1\pm0.4$ & 5.4 & $535\pm104$ & $1.1\pm0.2$ & $0.3220\pm0.0002$ & 4.8 \\ %vel =  81.782 ( 51.831) km/s
          &  & 3$\rightarrow$2 & 261.651 & $<2.4$ & --- & --- & $<0.14$ & ---  & 4.9 \\
\hline
KiDS~1220 & 0.3886 & 1$\rightarrow$0 & 83.014 &   $4.1\pm0.4$  & 10.2 & $726\pm76$ & $3.2\pm0.3$ & $0.3892\pm0.0002$ & 6.6  \\ %vel = 126.730 ( 33.385)
               &  & 2$\rightarrow$1 & 166.024 &    $5.6\pm1.0$ & 5.6 & $497\pm99$ & $1.1\pm0.2$ & $0.3895\pm0.0002$ & 3.0 \\ %vel = 184.312 ( 49.234)
                  &  & 3$\rightarrow$2 & 249.029 & $5.5\pm1.0$ & 5.5 & $561\pm102$ & $0.47\pm0.09$ & $0.3890\pm0.0002$  & 3.6 \\ %vel 83.339 ( 53.391)
\hline
KiDS~1444 & 0.4417 & 1$\rightarrow$0 & 79.952 & $3.1\pm0.4$ & 7.8 & $385\pm70$ & $3.1\pm0.4$ & $0.4413\pm0.0001$ & 3.3 \\ %vel = -85.334 ( 26.931)
              &  &  2$\rightarrow$1  & 159.902 & $8.2\pm1.1$ & 7.5 & $499\pm84$ & $2.1\pm0.3$ & $0.4415\pm0.0002$ & 1.4 \\ % vel = -45.249 ( 32.534)
              &  & 3$\rightarrow$2 & 239.845 & $9.2\pm2.4$ & 3.8 & $516\pm123$ & $1.0\pm0.3$ &  $0.4423\pm0.0004$  & 1.9 \\ %vel = 115.065 ( 72.957) 
\hline
\end{tabular}
\end{center}
\caption{Results of our CO observations. Column description: (1) galaxy name;  (2) spectroscopic redshift as in Table~\ref{tab:BCG_properties}; (3-4) CO(J$\rightarrow$J-1) transition and observer frame frequency; (5) CO(J$\rightarrow$J-1) velocity integrated flux; (6) signal-to-noise ratio of the CO(J$\rightarrow$J-1) detection; (7) full width at half maximum of the CO(J$\rightarrow$J-1) line; (8) CO(J$\rightarrow$J-1) velocity integrated luminosity; (9) redshift derived from the CO(J$\rightarrow$J-1) line; (10) on-source integration time (double polar). Absent values are denoted with the symbol ---. For  KiDS~0920 the reported upper limits are at 3$\sigma$.}
\label{tab:BCG_properties_mol_gas}
\end{adjustwidth}
%%%%%%%%%%%%%%%%%%%%%%%%%%%%%%%%%%%%%%%%%%%%%%%%%%%%

%0.08, 0.08, 0.11
%0.13, 0.14, 0.17
%0.05, 0.05, 0.07

\end{table*}
%-------------------------------- Tab: CO detections  -------------------
\begin{table*}[]
\begin{center}
\begin{tabular}{ccccccccccc}
\hline\hline
Galaxy & $z_{\rm spec}$ & excitation ratio & $M_{H_2}$ & $\tau_{\rm dep}$ & $\frac{M_{H_2}}{M_\star}$ & $\frac{M_{H_2}}{M_{\rm dust}}$ & $\tau_{\rm dep, MS}$ & $\big(\frac{M_{H_2}}{M_\star}\big)_{\rm MS}$  \\
 & & & ($10^{10}M_\odot$) &  (Gyr) &  & & (Gyr) \\
 (1) & (2) & (3) & (4) & (5) & (6) & (7) & (8) & (9) \\
 \hline
%array([166, 306, 195]
%array([127, 293, 409]
%array([ 71,  95, 128]))
KiDS~0920 & 0.3216 & $r_{31}<0.13$ & $4.8\pm0.9$ & $2.2^{+0.6}_{-0.6}$ & $0.21^{+0.05}_{-0.06}$  & $166^{+127}_{-71}$ &  $1.4^{+0.2}_{-0.2}$ & $0.08^{+0.13}_{-0.05}$  \\
\hline
KiDS~1220 & 0.3886 & $r_{21}=0.34\pm0.07$ & $14.0\pm1.4$ & $4.3^{+0.6}_{-1.4}$ & $0.46^{+0.09}_{-0.12}$ & $306^{+293}_{-95}$ & $1.4^{+0.3}_{-0.2}$  & $0.08^{+0.14}_{-0.05}$  \\
 &  & $r_{31}=0.15\pm0.03$ &  &  &   &  &   \\
 \hline
KiDS~1444 & 0.4417 & $r_{21}=0.68\pm0.13$ & $13.5\pm1.7$ &  $4.0^{+1.3}_{-2.5}$ & $0.56^{+0.13}_{-0.15}$ & $195^{+409}_{-128}$ & $1.3^{+0.2}_{-0.2}$ & $0.11^{+0.17}_{-0.07}$   \\
&  & $r_{31}=0.32\pm0.11$ &  &  &   &  &   \\
 \hline
 \end{tabular}
\end{center}
 \caption{Molecular gas properties of the BCGs. Column description: (1) BCG name; (2) spectroscopic redshift as in Table~\ref{tab:BCG_properties}; (3) excitation ratio; (4) molecular gas mass $M_{H_2} =\alpha_{\rm CO} L^\prime_{\rm CO(J\rightarrow J-1)}$, where $\alpha_{\rm CO}=4.36~M_\odot\,({\rm K~km~s}^{-1}~{\rm pc}^2)^{-1}$ is assumed; (5) depletion timescale  $\tau_{\rm dep}=M_{H_2}/{\rm SFR}$; (6) molecular gas-to-stellar mass ratio; (7) molecular gas-to-dust mass ratio; (8-9) depletion timescale and molecular gas-to-stellar mass ratio predicted for MS field galaxies with redshift and stellar mass of our targets, following the prescription by \citet{Tacconi2018}, calibrated to $\alpha_{\rm CO}=4.36~M_\odot\,({\rm K~km~s}^{-1}~{\rm pc}^2)^{-1}$. For  KiDS~0920 the reported $r_{31}$ upper limit is at 3$\sigma$.} 
\label{tab:COresults}
%%%%%%%%%%%%%%%%%%%%%%%%%%%%%%%%%%%%%%
\end{table*}
%---------------------------------------------------------------------------------------

%-------------------------Fig molecular gas prop ---------------------
\begin{figure*}[h!]\centering
%%%%%%%%%%%%%%%%%%%%%%%%%%%%%%%%%%%%%%%%%%%%5
\subfloat{\hspace{0.2cm}\includegraphics[trim={1.5cm 0.cm 2.5cm 
0.5cm},clip,width=0.5\textwidth,clip=true]{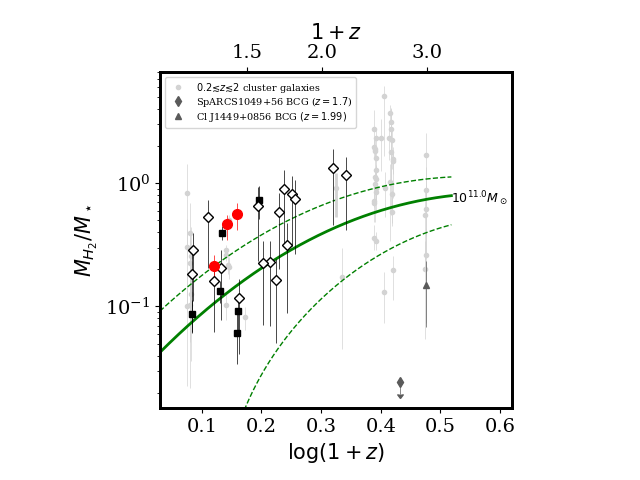}}
\subfloat{\hspace{0.2cm}\includegraphics[trim={1.5cm 0.cm 2.5cm 
0.5cm},clip,width=0.5\textwidth,clip=true]{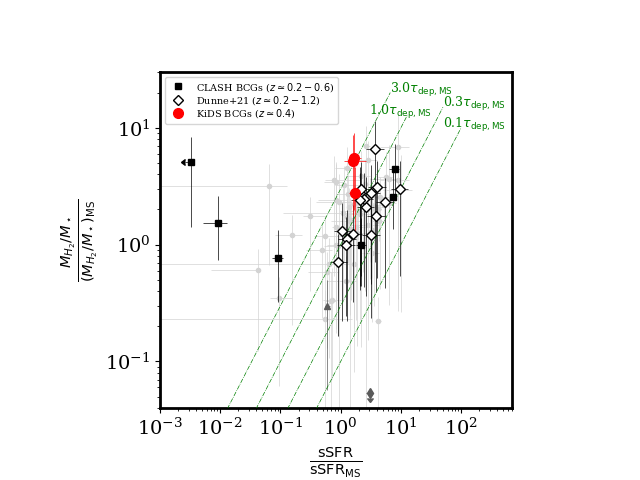}}\\
%%%%%%%%%%%%%%%%%%%%%%%%%%%
\caption{{ Molecular gas properties of distant  BCGs and cluster galaxies observed in CO}. Left: Evolution of the molecular gas-to-stellar mass ratio as a function of the redshift for cluster galaxies at $0.2\lesssim z\lesssim2$ observed in CO.  
The solid green curve is the scaling relation found  by \citet{Tacconi2018} for field galaxies in the MS and  with $\log(M_\star/M_\odot)$=11, which is the median stellar mass of all sources in the plot. The dashed green lines show the statistical 1$\sigma$ uncertainties in the model. Right: Molecular gas-to-stellar mass ratio versus the specific SFR for the cluster galaxies, both normalized to the corresponding MS values using the relations for the ratio and the SFR by \citet{Tacconi2018} and \citet{Speagle2014}, respectively. The dot-dashed green lines show different depletion times, in units of the depletion time at the MS. In both panels the points show cluster galaxies with molecular gas observations. The adopted color code is reported at the top left of each panel. The KiDS BCGs from this work are reported as red circles. Additional distant BCGs from the literature are also highlighted.}\label{fig:mol_gas1}
\subfloat{\hspace{-1.0cm}\includegraphics[trim={1.5cm 0.cm 2.5cm 
1.5cm},clip,width=0.37\textwidth,clip=true]{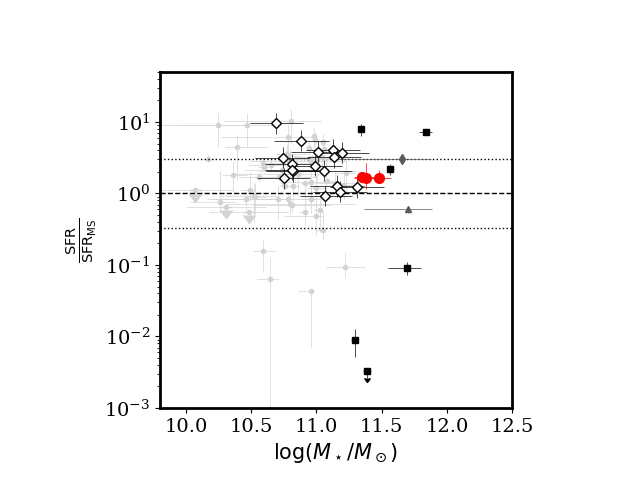}}
\subfloat{\hspace{-0.3cm}\includegraphics[trim={1.5cm 0.cm 2.5cm 
1.5cm},clip,width=0.37\textwidth,clip=true]{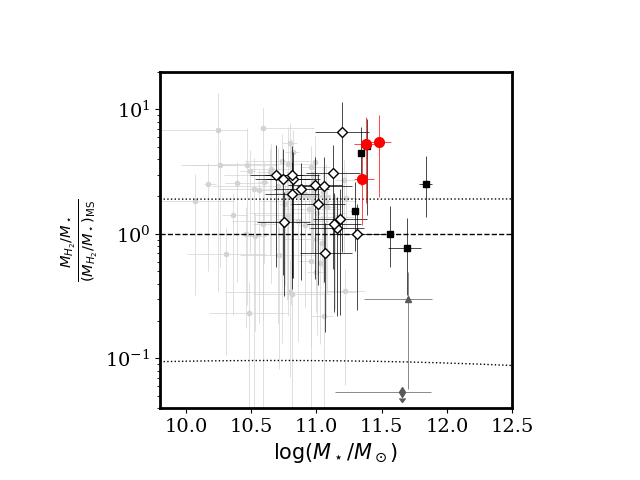}}
\subfloat{\hspace{-0.3cm}\includegraphics[trim={1.5cm 0.cm 2.5cm 
1.5cm},clip,width=0.37\textwidth,clip=true]{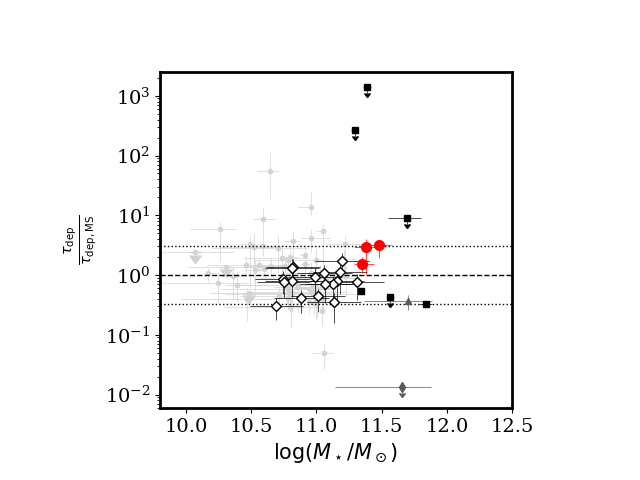}}
%%%%%%%%%%%%%%%%%%%%%%%%%%%
\caption{{ SFR (left), molecular gas-to-stellar mass ratio (center), and depletion time (right), as a function of the stellar mass for distant BCGs and cluster galaxies observed in CO.} The y-axis values are all normalized to the corresponding MS values using the relations by \citet{Speagle2014} and \citet{Tacconi2018}. The horizontal dashed lines correspond to y-axis values equal to unity, while the dotted lines denote the fiducial uncertainties associated with the MS. The uncertainty is chosen to be equal to $\pm0.48$~dex for both the left and right panels since the MS is commonly identified by $1/3<{\rm SFR}/{\rm SFR}_{\rm MS}<3$. For the central panel, the plotted uncertainties are estimated at redshift $z=0.5$. The color-coding for the data points is the same as in Fig.~\ref{fig:mol_gas1}.}\label{fig:mol_gas3}
\end{figure*}
%-------------------------------------------------------------------------

\section{The comparison samples}\label{sec:comparison_sample}
{In this section we describe samples of galaxies with available observations of their molecular gas that we use as  a comparison for our analysis.}

\subsection{Distant cluster galaxies observed in CO}\label{sec:comparison_sample_cluster_galaxies}

We want to put our galaxies in a broader context, in comparison with existing samples of distant cluster galaxies. To this aim, we consider the compilation built by \citet{Castignani2020b}, comprising 120 (proto-)cluster galaxies at $0.2\lesssim z\lesssim5.0$ with $M_\star>10^{9}~M_\odot$ and CO observations from the literature.
{ In particular, we refer to Tables~A.1, A.2 of \citet{Castignani2020b}, where stellar, star formation, and gas properties are listed.}
Since in this work we consider massive intermediate-redshift BCGs, we limit ourselves to sources in the compilation with stellar masses $M_\star>10^{10}~M_\odot$ and redshifts $z<2$ (that is, we exclude the most distant clusters and still assembling protoclusters). { As we have detected all three of our KiDS BCGs  in CO, concerning the comparison sample we similarly considered 
sources with CO detections, that is, we  excluded sources with only upper limits to the $H_2$ masses.} The only exception is represented by the SpARCS~104922.6+564032.5 BCG at $z=1.7$, further discussed below, { which is one of the few BCGs at $z>1$ with observations in CO, as well as $M_\star$ and SFR estimates.}
This selection yields 58 galaxies over 16 clusters at moderate to high redshifts: $z\sim0.2-0.5$ \citep{Geach2011,Jablonka2013,Cybulski2016};  $z\sim1-1.2$ \citep{Wagg2012,Castignani2018}; $z\sim1.5-2.0$   \citep{Aravena2012,Rudnick2017,Webb2017,Noble2017,Noble2019,Hayashi2018,Kneissl2019,Coogan2018,Castignani2020c}.  
%We refer to \citet{Castignani2020b} and references therein for details.

For our comparison,  as further discussed below,  we include additional BCGs drawn from the CLASH and SpARCS surveys. They span the redshift range $z\sim0.2-1.7$ and have been observed in CO by recent studies \citep{Fogarty2019,Castignani2020a,Castignani2020c,Dunne2021}. With the inclusion of these sources and the three BCGs presented in this work the full list comprises 84 cluster  galaxies over 42 clusters at $z\simeq0.2-2$ and stellar masses in the range $\log(M_\star/M_\odot)\simeq10-12$. {We use this sample in Sect.~\ref{sec:gas_SFR_tdep} to perform a comparison with the three BCGs of this work in terms of stellar mass, SFR, molecular gas content, and depletion time.}

%Including ${\rm SFR}\gtrsim3\times{\rm SFR}_{\rm MS}$ sources might result in biased-high molecular gas masses \citep[see also e.g.,][for discussion]{Noble2017,Castignani2018,Castignani2019}.  However, since the two most abundantly star-forming BCGs in our CLASH sample, RX1532 and M1932, have ${\rm sSFR}\gtrsim 5\times {\rm sSFR}_{\rm MS}$, we prefer not to discard high-SFR galaxies from the comparison. 

\subsection{Distant BCGs observed in CO}\label{sec:BCGs_CO_literature}
Among the distant cluster galaxies with CO observations and stellar mass estimates in the comparison sample, considered above, there are 25 BCGs, as outlined in the following.

The SpARCS~104922.6+564032.5 BCG, located at $z=1.7$.  Recent CO(1$\rightarrow$0) Jansky Very Large Array (JVLA) observations by \citet{Barfety2022} yield a large molecular gas reservoir of $M_{H_2}\sim10^{11}~M_\odot$, in agreement with the value previously found with single-dish observations by \citet{Webb2017}.  %This value is in agreement with that previously found by \citet{Webb2017} based on the CO(2$\rightarrow$1) detection, obtained with the Large Millimiter Telescope (LMT).
  However, the CO(1$\rightarrow$0) peak is offset by $\sim15$~kpc to the southeast of the BCG, which suggests that the molecular gas reservoir may be associated with cluster core BCG companions.  
  {NOrthern Extended Millimeter Array (NOEMA)} observations \citep{Castignani2020c} supported this scenario, with the detection in CO(4$\rightarrow$3) of two gas-rich BCG companions within 20~kpc from the BCG. 
  The observations allowed us to set a robust upper limit to the molecular gas reservoir $M_{H_2}<1.1\times10^{10}~M_\odot$ for the BCG { that we use in this work.}
 \smallskip\smallskip\smallskip
%%%%%%%%%%%%%%%%%%%%%%%%%%%%%%% 

ClJ1449+0856 BCG is another very distant BCG, located at $z = 1.99$. %It is constituted by a gas-rich ($M_{H_2}\gtrsim10^{10}~M_\odot$) triplet of galaxies. 
For this system, \citet{Coogan2018} detected both CO(4$\rightarrow$3) and CO(3$\rightarrow$2), possibly associated with an optically faint nearby southern component \citep{Strazzullo2018}. 

There are 17 BCGs from the recent SpARCS survey, detected in CO(2$\rightarrow$1) and S/N$\gtrsim3$ by \citet{Dunne2021}. They span the redshift range $z\sim0.2-1.2$ and have molecular gas masses in the range  $M_{H_2}\sim(10^{10}-10^{11})~M_\odot$. 

{Finally, we consider} six CLASH BCGs at intermediate redshifts $z\sim0.2-0.6$, similarly to our KiDS targets. They are M0329, A1423, M1206, M2129, RX1532, and M1932. The last two were clearly detected in CO(1$\rightarrow$0) at S/N$>6$ with the IRAM 30m \citep{Castignani2020a} and  ALMA \citep{Fogarty2019}, respectively. They are gas-rich systems, with $M_{H_2}\sim10^{11}~M_\odot$. The first four have been observed in CO(1$\rightarrow$0) or CO(2$\rightarrow$1) as part of our IRAM 30m campaign \citep{Castignani2020a}. This campaign yielded CO detections at S/N$\gtrsim3$ for the three BCGs, corresponding to molecular gas masses  in the range $M_{H_2}\sim(10^{10}-10^{11})~M_\odot$.
\smallskip\smallskip\smallskip
%%%%%%%%%%%%%%%%%%
%\item[$\bullet$] We\ also included for our comparison six additional BCG (candidates) at $z\sim0.4-1$ that were observed by \citet{Castignani2019,Castignani2020a} using the IRAM-30m as part of a large search for CO in distant star-forming BCGs, which were selected from several surveys (DES, SDSS, COSMOS, and SpARCS). These six sources are  DES-RG~399 ($z=0.39$), DES-RG~708 ($z=0.61$), COSMOS-FRI~16 ($z=0.97$), COSMOS-FRI~31 ($z=0.91$), 3C~244.1 ($z=0.43$), and  SDSS~J161112.65+550823.5 ($z=0.91$), for which we set robust upper limits to their molecular gas content in our previous studies. 

\subsection{{Galaxies with CO(1$\rightarrow$0) and CO(3$\rightarrow$2) observations}}\label{sec:r31_comparison}
In addition to cluster galaxies with observations in CO, we considered a second compilation of { sources} with molecular gas observations in the first and third CO transitions. { We stress that the number of such sources is limited in the literature as detecting galaxies in both these transitions is expensive in terms of observational resources.} We use these galaxies in Sect.~\ref{sec:excitation_ratios} to investigate the relation between the excitation ratios and both star formation and gas properties, in comparison with the KiDS BCGs. We are particularly interested in CO(1$\rightarrow$0) and CO(3$\rightarrow$2) as we observed all three BCGs of this work at these transitions. { The galaxies considered for the comparison, outlined below, are massive, $\log(M_\star/M_\odot)\geq10$, similarly to our BCGs. They also cover  broad ranges in redshift and SFR, which allow us to effectively search for possible relations and trends.}

First, {we considered two samples of local galaxies with stellar, gas, and star formation properties from \citet{Lamperti2020}. These are 25 star-forming galaxies at $z\sim0.02-0.05$ from the xCOLD GASS survey \citep{Saintonge2017}, with $\log(M_\star/M_\odot)=10.0-11.0$ and ${\rm SFR}=(0.9-35)~M_\odot$/yr. In addition, we include 36 AGN  at $z\lesssim0.04$ from the  BAT AGN Spectroscopic Survey (BASS)\footnote{http://www.bass-survey.com/}, with $\log(M_\star/M_\odot)=9.7-11.0$ and ${\rm SFR}\simeq(0.2-56)~M_\odot$/yr.   These two samples have CO(1$\rightarrow$0) and CO(3$\rightarrow$2) observations, taken with the IRAM 30m and JCMT telescopes, respectively \citep{Saintonge2017,Lamperti2020}.  In particular, for both BASS and xCOLD GASS samples, in this work we use the $H_2$ gas masses and CO(3$\rightarrow$2) / CO(1$\rightarrow$0) excitation ratios, both corrected for beam, as reported in \citet{Lamperti2020}.} 
 
 {We additionally included five $2.5<z<2.7$ (ultra)LIRGs with infrared luminosities in the range $\sim(11.6-12.9)~L_\odot$, stellar masses $\log(M_\star/M_\odot)=10.4-11.1$, and ${\rm SFR}\simeq(35-347)~M_\odot$/yr. Both CO(1$\rightarrow$0) and CO(3$\rightarrow$2) observations are available for these sources, as part of the ALMA Spectroscopic Survey in the {\it Hubble} Ultra Deep Field  \citep[ASPECS;][]{Boogaard2020}.}

\section{Results}\label{sec:results}

\subsection{Molecular gas properties}\label{sec:IRAM30m_results}
With our observations we clearly detect all three targeted BCGs in multiple transitions, with a signal-to-noise ratio (S/N) in the range S/N$\simeq(3.8-10.2)$, as outlined in the following. Our observations thus increase the still limited sample of rare distant star-forming BCGs with detections of their molecular gas. Two out of the three targeted BCGs, namely KiDS~1220 and KiDS~1444 were detected { in CO(1$\rightarrow$0), CO(2$\rightarrow$1), and CO(3$\rightarrow$2).} The third one, KiDS~0920, was detected in  CO(1$\rightarrow$0), while we set an upper limit to the CO(3$\rightarrow$2) emission. The corresponding spectra are shown in Fig.~\ref{fig:BCG_spectra}. 

These results imply that we double the still limited sample of distant {($z>0.2$)} BCGs with clear detections in two distinct CO lines. To the best of our knowledge there are in fact only other two distant BCGs detected in multiple CO transitions. These are the intermediate-redshift ($z\sim0.4$) BCGs RX1532 and M1932 from CLASH \citep{Postman2012}. RX1532 was detected both in CO(1$\rightarrow$0) and CO(2$\rightarrow$1) with our recent CLASH IRAM 30m campaign \citep{Castignani2020a}. M1932 was instead detected in CO(1$\rightarrow$0), CO(3$\rightarrow$2), and CO(4$\rightarrow$3) with ALMA by \citet{Fogarty2019}.

%{THE FOLLOWING PART SHOULD BE REVISED}

% { what follows is speculative, put it in the discussion  --} {It is thus possible that these two BCGs are in the phase of, or close to merging with the companions, which may explain the observed disturbed morphology for KiDS~1444 and its tail, possibly originated by tidal interactions. { up to here ---}}

%Therefore, we cannot firmly exclude the possibility that they contribute to the total observed CO emission. This is also suggested by the large full-width at half maximum (FWHM) in the range $\sim~385-726$~km~s$^{-1}$ that is measured for the detected CO lines, thus larger than FWHM$\sim300$~km$^{-1}$, which is the typical value for massive galaxies in the literature. The possible interaction of the BCGs with their companions may also replenish the BCGs gas reservoirs, thus ultimately contributing to substain the high level of star formation activity observed in the targeted BCGs.

%{UP TO HERE}

Based on the optical morphologies illustrated in Fig.~\ref{fig:BCG_images} the three targeted KiDS BCGs are unresolved by our observations, assuming a CO-to-optical size ratio $\sim0.5$ \citep{Young1995} and given that the IRAM 30m beam is $\sim16$~arcsec at 2~mm \citep[][see our Sect.~\ref{sec:observations_and_data_reduction}]{Kramer2013}. 
We detect all targeted lines except CO(3$\rightarrow$2) of KiDS~0920, this yields an exceptional success rate of $88\%$, that is, seven detections out of eight targeted lines, higher than that of recent surveys of distant BCGs \citep{Castignani2020a,Castignani2020b,Dunne2021}. 
We then fit each detected CO line with a Gaussian curve, after removing the baseline with a linear fit. We used instead a combination of two Gaussian curves to fit the CO(3$\rightarrow$2) line of { KiDS~1220}, as the spectrum shows a clear double-horn profile. Hints of a double-peaked line are also present in the CO(1$\rightarrow$0) and CO(2$\rightarrow$1) spectra.

We believe that the observed double-peaked CO emission is originated from the KiDS~1220 BCG only and we further discuss the underlying structural parameters in Sect.~\ref{sec:double_horn_modeling}. A possible alternative is that the nearby companion of KiDS~1220 BCG contributes to the observed CO(3$\rightarrow$2) emission. However, this is less likely as the source is at a projected separation of 2.7~arcsec, so that with a HPBW of 9.6~arcsec \citep{Kramer2013} the efficiency is suppressed down to 80\% the maximum.  
%power = power_max*exp(-0.5*(sep/sigma)^2)
%sep = 2.7 arcsec
%HPBW = FWHM = 2.355 * sigma
%HPBW = 9.6
With a similar argument, we believe that the CO emission observed for KiDS~1444 comes from the BCG only, with null or negligible contamination from the nearby companion within the beam. However, it is possible that the KiDS~1220 and KiDS~1444 BCGs are interacting with their companions, which may favor the replenishment of the BCGs' gas reservoirs, thus ultimately contributing to sustain the high level of star formation activity observed in the targeted BCGs.

We detect KiDS~0920 in CO(1$\rightarrow$0), and it is likely that the detected molecular gas comes from the elongated disk-like structure seen in Fig.~\ref{fig:BCG_images}. However, we do not find any evidence of CO(3$\rightarrow$2) emission, so that we first removed the baseline in the CO(3$\rightarrow$2) spectrum with a linear fit. We then estimated the rms noise level for the antenna temperature (Ta$^\ast$). Using  standard conversions, outlined below, we converted the rms into a 3$\sigma$ upper limit to the integrated CO flux, at a resolution of 300~km/s, which corresponds to the typical full width at half maximum for massive galaxies and BCGs in particular \citep{Edge2001,Dunne2021}.

In Table~\ref{tab:BCG_properties_mol_gas}, we report the results of our observations, where we apply standard efficiency corrections to convert i) Ta$^\ast$ into the main beam temperature $T_{\rm mb}$ and then ii) $T_{\rm mb}$ into the corresponding CO line flux, with the conversion factor of 5~Jy/K. In particular, we adopt the following efficiency  corrections $T_{\rm mb}/T{\rm a}^\ast = 1.2$, $1.4$ and $2.0$ for $\sim$3, 2, and 1~mm observations, respectively.\footnote{https://www.iram.es/IRAMES/mainWiki/Iram30mEfficiencies} 

We then derive the CO(J$\rightarrow$J-1) luminosity $L^{\prime}_{\rm CO(J\rightarrow J-1)}$, in units K~km~s$^{-1}$~pc$^2$, from the velocity integrated CO(J$\rightarrow$J-1) flux $S_{\rm CO(J\rightarrow J-1)}\,\Delta\varv\ $, in units Jy~km~s$^{-1}$.
To this aim we used Eq.~(3) from \citet{Solomon_VandenBout2005}:
\begin{equation}
\label{eq:LpCO}
 L^{\prime}_{\rm CO(J\rightarrow J-1)}=3.25\times10^7\,S_{\rm CO(J\rightarrow J-1)}\,\Delta\varv\,\nu_{\rm obs}^{-2}\,D_L^2\,(1+z)^{-3}\,,
\end{equation}
where $\nu_{\rm obs}$ is the observer frame frequency, in GHz, of the CO(J$\rightarrow$J-1) transition, $D_L$ is the luminosity distance in Mpc, and $z$ is the redshift of the BCG.

{All three BCGs are detected in CO(1$\rightarrow$0): this enables us to derive the total $H_2$ gas mass as} $M_{H_2}=\alpha_{\rm CO}L^{\prime}_{\rm CO(1\rightarrow0)}$. The use of the CO(1$\rightarrow$0) transition is advantageous with respect to higher-J ones as it does not require any assumption on the excitation ratio
$r_{\rm J1}= L^{\prime}_{\rm CO(J\rightarrow J-1)}/L^{\prime}_{\rm CO(1\rightarrow0)}$.

The three KiDS BCGs have similar specific star-formation rates,  ${\rm sSFR}\simeq1.6\times {\rm sSFR}_{\rm MS}$ (see Table~\ref{tab:BCG_properties}). They are all less than the value of $3\times {\rm sSFR}_{\rm MS}$, below which the galaxy sSFR is within the typical scatter of the MS. To estimate $H_2$ gas masses we therefore assume a Galactic CO-to-$H_2$ conversion factor of $\alpha_{\rm CO}=4.36~M_\odot\,({\rm K~km~s}^{-1}~{\rm pc}^2)^{-1}$, commonly adopted for star-forming galaxies at the MS. The use of a single $\alpha_{\rm CO}$ factor also allows us to do a homogeneous comparison in terms of gas content.

As all three BCGs are detected in CO(1$\rightarrow$0), we use both LO and LI sidebands to search for the CN(N=1$\rightarrow$0) doublet, which is redshifted to $\sim$(79-86)~GHz in the observer frame, close to the CO(1$\rightarrow$0) line. However, none of the three BCGs shows evidence for any of the CN(N=1$\rightarrow$0) J=1/2$\rightarrow$1/2 and CN(N=1$\rightarrow$0) J=3/2$\rightarrow$1/2 lines of the doublet. This is not surprising, since they are usually much fainter than CO(1$\rightarrow$0). In our recent study we performed a similar search for the CN(N=1$\rightarrow$0) doublet in the BCG RX1532 at $z\sim0.4$, which also led to a non-detection \citep{Castignani2020a}. However, thanks to the long integration times of several hours at 3~mm (Table~\ref{tab:BCG_properties_mol_gas}) we reach low rms values of 1.7, 1.3, and 2.3~mJy for the KiDS~0920, 1220, and 1444 BCGs, respectively, at a resolution of 60~km/s.

We do not attempt to set any upper limit to the continuum emission of the BCGs by using the available total 15~GHz (LI, LO, UI, UO) bandwidths,  for each polarization. In fact, the BCGs are faint and there is significant intrinsic atmospheric instability at millimeter wavelengths that prevents us from robustly determining the continuum levels or estimating upper limits.

We also estimate the depletion timescale $\tau_{\rm dep}=M_{H_2}/{\rm SFR}$, which is the characteristic time to deplete the molecular gas reservoirs. To compute $\tau_{\rm dep}$ we use the SED-based SFR estimates reported in Table~\ref{tab:BCG_properties}. 
Similarly, we  also estimate the molecular-gas-to-stellar-mass ratio, $M_{H_2}/M_\star$, and the ratio between the molecular gas and the dust masses.  For a comparison we compute the depletion time $\tau_{\rm dep, MS}$ and the molecular gas to stellar mass ratio $\big(\frac{M_{H_2}}{M_\star}\big)_{\rm MS}$ for MS galaxies in the field with redshift and stellar mass equal to those of our BCGs, by using empirical prescriptions by \citet{Tacconi2018}, calibrated to the CO-to-$H_2$ conversion factor of $\alpha_{\rm CO}=4.36~M_\odot\,({\rm K~km~s}^{-1}~{\rm pc}^2)^{-1}$ used in this work. 

In Table~\ref{tab:COresults} we summarize the molecular gas properties for the three KiDS BCGs. As both KiDS~1220 and KiDS~1444 are detected in multiple CO transitions, in the Table we also list their $r_{21}$ and $r_{31}$ excitation ratios, while for KiDS~0920 we report the upper limit to $r_{31}$.

%{ GC: comment on Mgas, gas-to-Mstar ratio, MH2/Mdust, Mdust, etc.. here or below}

%-------------------------Fig excitation ratios  ---------------
\begin{figure*}[th!]\centering
\captionsetup[subfigure]{labelformat=empty}
%%%%%%%%%%%%%%%%%%%%%%%%%%%%%%%%%%%%%%%%%%%%5
\subfloat[]{\hspace{0.cm}\includegraphics[trim={2.2cm 0.1cm 3.2cm 
1.5cm},clip,width=0.4\textwidth,clip=true]{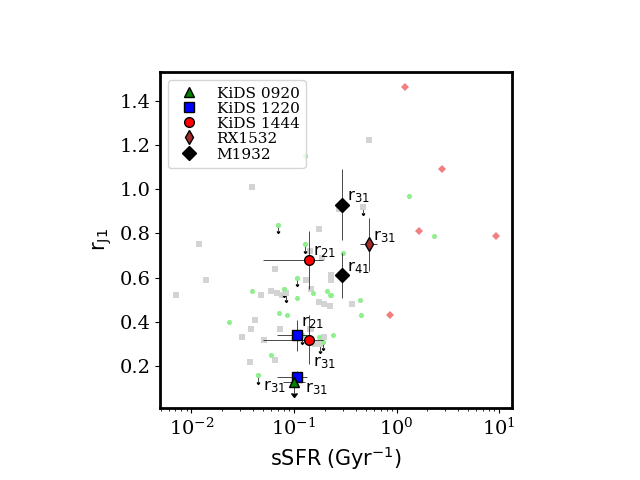}}
\subfloat[]{\hspace{-0.cm}\includegraphics[trim={2.2cm 0.1cm 3.2cm 
1.5cm},clip,width=0.4\textwidth,clip=true]{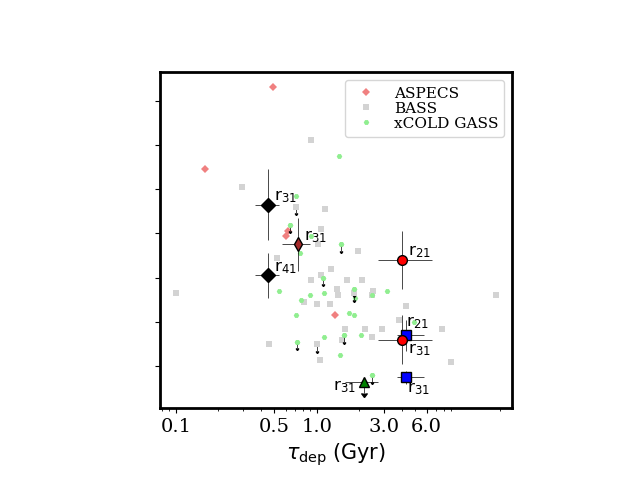}}
%%%%%%%%%%%%%%%%%%%%%%%%%%%
\caption{Excitation ratios $r_{\rm J1}$ plotted against the {sSFR} (left) and the depletion time (right) for intermediate-redshift BCGs with multiple CO(J$\rightarrow$J-1) detections. {Excitation ratios $r_{\rm 31}$ of ASPECS, BASS, and xCOLD GASS sources are reported for a comparison.}}\label{fig:excitation}
\end{figure*}
%%%%%%%%%%%%%%%%%%%%%%%%%%%%%%%%%%%%%%%%%%%%5

%-------------------------Fig color magnitude plots  ---------------
\begin{figure}[htb!]\centering
\captionsetup[subfigure]{labelformat=empty}
%%%%%%%%%%%%%%%%%%%%%%%%%%%%%%%%%%%%%%%%%%%%5
\subfloat[]{\hspace{0.cm}\includegraphics[trim={0cm 0cm 0cm 
0cm},clip,trim = {2.5cm 0.5cm 3cm 1.7cm}, width=0.36\textwidth,clip=true]{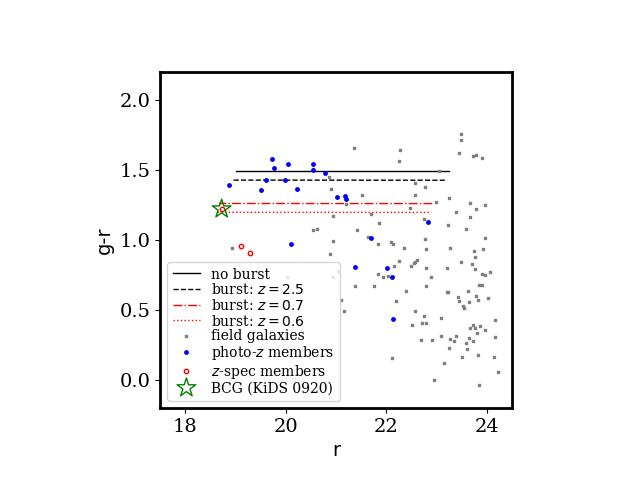}}
\hspace{0.1cm}\subfloat[]{\hspace{0.cm}\includegraphics[trim={0cm 0cm 0cm 0cm},clip,trim = {2.5cm 0.5cm 3cm 1.7cm},width=0.36\textwidth,clip=true]{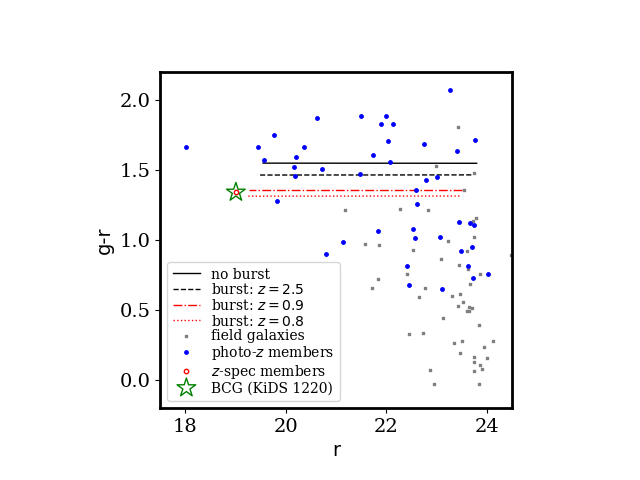}}
\hspace{0.1cm}\subfloat[]{\hspace{0.cm}\includegraphics[trim={0cm 0cm 0cm 
0cm},clip,trim = {2.5cm 0.5cm 3cm 1.7cm},width=0.36\textwidth,clip=true]{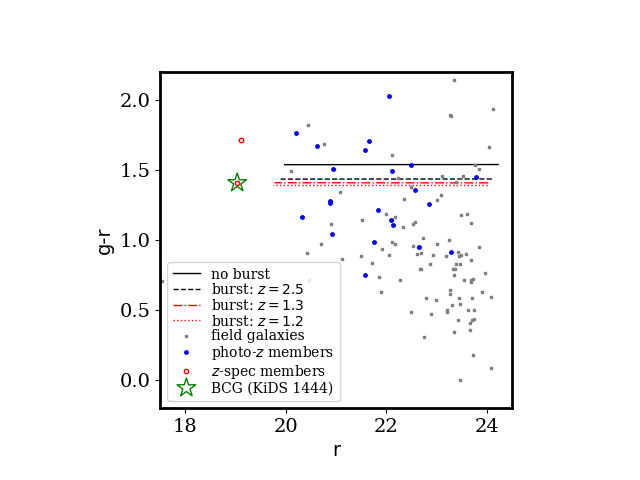}}
%%%%%%%%%%%%%%%%%%%%%%%%%%%
\caption{Color magnitude plots for the sources in the field of the BCGs KiDS~0920 (top), KiDS~1220 (center), and KiDS~1444 (bottom). Field galaxies (gray points), photometrically (blue points) and spectroscopically (red points) selected cluster members, and the BCGs (green stars) are highlighted. Red-sequence models are shown as horizontal lines. See the legend and text for further details.}\label{fig:CM_plots}
\end{figure}
%%%%%%%%%%%%%%%%%%%%%%%%%%%%%%%%%%%%%%%%%%%%5

\subsection{Gas fraction, star formation, and depletion time}\label{sec:gas_SFR_tdep}

We now exploit the compilation of cluster galaxies with available $H_2$ gas masses, SFR, and $M_\star$ estimates {presented in Sect.~\ref{sec:comparison_sample_cluster_galaxies}.}
Figure~\ref{fig:mol_gas1} displays the ratio of molecular gas to stellar mass  $M_{H_2}/M_\star$ as a function of both redshift and specific SFR for the compilation of cluster galaxies with 
observations in CO and stellar mass estimates.

In Fig.~\ref{fig:mol_gas3} we show instead the SFR, $M_{H_2}/M_\star$, and the depletion time ($\tau_{\rm dep})$, all normalized to their MS values, as a function of the stellar mass. {In both  figures, to allow a homogeneous comparison, we have used $\alpha_{\rm CO}=4.36~M_\odot\,({\rm K~km~s}^{-1}~{\rm pc}^2)^{-1}$ for all data points and the overplotted \citet{Tacconi2018} relations. We also highlight the} KiDS BCGs (red dots) and BCGs from the literature, while the remaining cluster galaxies in our comparison sample are shown in background (gray dots). By looking at Fig.~\ref{fig:mol_gas3} it is striking that our BCGs have high stellar masses, $\log(M_\star/M_\odot)\sim11.4$, that are systematically higher than the SpARCS BCGs in \citet{Dunne2021}. This result  strengthens the BCG association for our targets, further discussed in Sect.~\ref{sec:color_magnitude_plots}.

Interestingly, the figures also show that all three KiDS BCGs are gas-rich, with $H_2$ gas masses largely exceeding $10^{10}~M_\odot$ and $H_2$-to-stellar mass ratios in the range $\sim0.2-0.6$, which are values similar to those found by previous studies of star-forming BCGs at intermediate redshifts \citep[SpARCS and CLASH survey;][]{Fogarty2019,Castignani2020a,Dunne2021}, highlighted in the figures. 

Furthermore, while our KiDS BCGs have $M_{H_2}/M_\star$ ratios higher than those of MS galaxies, they also have MS levels of star formation activity (Fig.~\ref{fig:mol_gas3} left).  These results imply that our targeted BCGs convert the molecular gas into stars  quite inefficiently, having depletion timescales $\tau_{\rm dep} = M_{H_2}/{\rm SFR}$ in the range $\sim(2-4)$~Gyr. As illustrated in Fig.~\ref{fig:mol_gas3} (right) these values are still consistent with the depletion timescales of MS galaxies, given the uncertainties, but higher than those found for star-forming CLASH and SpARCS BCGs. This result is intriguing and may suggest that the gas fueling in the three KiDS BCGs differs somehow from that of previously studied star-forming BCGs at intermediate redshifts. Similarly, as outlined in Table~\ref{tab:COresults}, our three KiDS BCGs have $H_2$-to-dust mass ratios in the range $\sim170-300$, thus exceeding  the ratio of $\sim100$, typical of distant star-forming galaxies \citep{Berta2016,Scoville2014,Scoville2016}.

To explain both the high $\tau_{\rm dep}$ and the high  $M_{H_2}/M_{\rm dust}$, a possibility is that a substantial amount of the $H_2$ gas has been recently accreted, but still not efficiently converted into stars. This accretion may have occurred via filaments (as it may be the case of KiDS~0920, which shows an elongated morphology), or via interaction with the companions (for KiDS~1220 and 1444). {Alternatively, adopting a lower $\alpha_{\rm CO}$ conversion factor than the Galactic one, as usually done for starbursts, would lead to lower $M_{H_2}$ and $\tau_{\rm dep}$.} Motivated by these results, in the following Section we further investigate the differences of the selected BCGs with respect those studied in previous works by means of the excitation ratio. 

%In Sect.~?? we instead study the mass assembly and star formation of the BGCs by means of color-magnitude plots. Finally, in Sect.~?? we perform a 

%{GC: we can expand here or directly in the discussion as we did for the CLASH paper. The BCGs are gas rich, but normal in SFR. This implies high tdep and low SFE. It may be that a fraction of the observed H2 is originated from recently infalling pristine gas, possibly in the form of accretion from filaments, or accretion induced by the interaction with the companions.\\ We find that our BCGs are systematically more massive than Dunne+21 BCGs, this strenghen the BCG association for our targets.}

\subsection{Excitation ratios}\label{sec:excitation_ratios}
To the best of our knowledge there exist only a few distant BCGs with CO detections in different transitions,  as outlined in Sect.~\ref{sec:IRAM30m_results}. These are RX1532 
\citep[$z=0.361$,][]{Castignani2020a} and M1932 \citep[$z=0.353$,][]{Fogarty2019}, in addition to the KiDS BCGs of this work. 
As the sources are all detected in CO(1$\rightarrow$0) as well as in CO(2$\rightarrow$1),  CO(3$\rightarrow$2), or CO(4$\rightarrow$3) we consider the $r_{21}$,  $r_{32}$, and  $r_{43}$ excitation ratios, whenever available. 

As pointed out above, all these intermediate-$z$ BCGs are star-forming and span a quite broad range in the ${\rm SFR}\simeq(20-200)~M_\odot/{\rm yr}$. On the other hand the BCGs are gas-rich, with similar gas content in terms of $M_{H_2}\simeq(0.5-1.4)\times10^{11}~M_\odot$ and $M_{H_2}/M_\star\simeq(0.1-0.6)$, which are less scattered than the SFR. Consistently with this result, we do not find any trend when plotting the excitation ratios $r_{\rm J1}$ against the $M_{H_2}$ or $M_{H_2}/M_\star$.  

Figure~\ref{fig:excitation} displays the excitation ratios $r_{\rm J1}$ of the BCGs against the {sSFR} (left) and the depletion time (right). {For the latter we used $\alpha_{\rm CO}=4.36~M_\odot\,({\rm K~km~s}^{-1}~{\rm pc}^2)^{-1}$ for all sources in the plot, to allow a homogeneous comparison.}
The plots show a clear dichotomy between the CLASH BCGs (M1932 and RX1532) that show the highest ${\rm SFR}\gtrsim100~M_\odot/{\rm yr}$ and the KiDS BCGs 
considered in this work, with lower ${\rm SFR}\simeq(20-30)~M_\odot$/yr.  The latter not only have higher depletion times, as discussed in Sect.~\ref{sec:gas_SFR_tdep}, but also lower excitation ratios. 

{As all BCGs have observations in both CO(1$\rightarrow$0) and CO(3$\rightarrow$2) we include in Fig.~\ref{fig:excitation} the ASPECS, BASS, and xCOLD GASS sources presented in Sect.~\ref{sec:r31_comparison}, for a comparison.
When considering the $r_{31}$ values of the KiDS BCGs and the comparison galaxies altogether we find that the excitation ratio turns out to be correlated with sSFR and anticorrelated with the depletion time, at  a level of 3.1$\sigma$ (${\rm p-value}=1.81\times10^{-3}$) and 4.7$\sigma$~ (${\rm p-value}=2.73\times10^{-6}$), as found with the Spearman test, respectively.} We also note that, at fixed BCG, it holds $r_{41}<r_{32}<r_{21}$ as indeed higher-J CO transitions are associated with denser gas, more difficult to excite.

These findings suggest that highly excited gas is preferentially found in highly star-forming and cool-core BCGs such as RX1532 and M1932, {as well as in star-forming galaxies of the comparison samples (e.g., ASPECS).}  As further discussed in \citet{Castignani2020a} the cool-core cluster environments likely
favor the condensation and the inflow of gas onto the BCGs
themselves. The present work further supports this scenario, while suggesting meanwhile that SFRs lower than $\sim100~M_\odot$/yr appear to be not sufficient to excite the gas to the highest levels. {The latter is true for the KiDS BCGs, as well as for lower mass xCOLD GASS and BASS galaxies of the comparison sample. Interestingly, trends of the excitation ratios as a function of star formation and gas properties were also reported in previous studies of local galaxies \citep{Lamperti2020,Yao2003,Leech2010}, while in this work we find that they are also valid for star-forming and distant BCGs.}

{Although the excitation ratios of the BCGs nicely follow the expected trends as a function of the sSFR and the depletion time, we note that the $r_{\rm J1}$ values of the KiDS BCGs are lower than those of distant star-forming galaxies, for which $r_{31}\simeq0.6$, on average \citep{Carilli_Walter2013,Bothwell2013,Genzel2015,Boogaard2020}.
We remind that the IRAM 30m HPBW is $\sim$24, 16, and 10~arcsec for the CO(1$\rightarrow$0), CO(2$\rightarrow$1), and CO(3$\rightarrow$2) transitions, respectively. 
As all KiDS BCGs have nearby companions within 10~arcsec in projection, the latter ones might contribute to the observed CO(1$\rightarrow$0) and CO(2$\rightarrow$1) emission, while the contamination is likely negligible for CO(3$\rightarrow$2), as already discussed in Sect.~\ref{sec:IRAM30m_results}.
The relatively low $r_{\rm J1}$ values of the KiDS BCGs might therefore be explained if the BCG companions are rich in gas. High resolution interferometric observations are however needed to have a definitive answer, as in the case of SpARCS~104922.6+564032.5 BCG at $z = 1.7$ \citep[][see our Sect.~\ref{sec:BCGs_CO_literature}]{Castignani2020c}.}

%- to the best our our knownledge there are only a few BCGs with multiple CO detections in the distant universe, see also earlier section \\
%- therefore we double the number of these kind of sources \\
%- we investigate the excitation ratios as a function of the star formation and gas properties \\
%- all sources with multiple CO detections are gas-rich, both in Mgas and Mgas/Mstar, with similar values of the two\\
%- however, we find  a correlation in excitation vs tau and excitation vs SFR\\
%- there is a clear dicotomy between low SFR and high SFR BCGs. the former have also lower excitation ratios \\
%- the correlation is tested with the spearman test, and the highest significant is found when only r31 is considered \\
%- at a given source, $r41<r31<r21$, as it should be, however highly excited gas is found only when $SFR>100~Mo/yr$. \\

\subsection{Color-magnitude plots}\label{sec:color_magnitude_plots}
We now study the cluster galaxy population of the three KiDS clusters of this work  by means of color-magnitude (CM) plots, with a particular focus on the BCGs. To this aim we select sources in the KiDS DR3 within a separation of 2~arcmin from each of the three targeted BCGs. 

In Fig.~\ref{fig:CM_plots} we report the corresponding \textsf{g-r} versus \textsf{r} plots, where \textsf{g} and \textsf{r} are AB apparent magnitudes, {obtained with the OmegaCAM camera on the VST (see Sect.~\ref{sec:BCGselection}).}
%which are part of  the KiDS DR3 photometric data release, and obtained with the OmegaCAM camera on the VLT Survey Telescope (VST). 
At the redshifts of our targeted BCGs the rest frame 4000~\AA~ break is redshifted to $\sim(5300-5800)$~\AA, between the \textsf{g} and \textsf{r} filters, {whose  effective wavelengths are 4800~\AA~ and 6250~\AA, respectively.} The chosen filters are thus optimal {to detect  the red sequence in the considered clusters.}

%to catch the presence, if any, of the red sequence for the cluster galaxy populations of the three clusters considered.

In each CM plot we highlight the BCGs (green stars). We also distinguish field galaxies (gray dots) from fiducial cluster members, which have been selected as follows. Photometrically selected cluster members (filled blue dots) are those with AMICO cluster membership probabilities greater than $50\%$, which implies that the galaxies are formally more likely to belong to the clusters than the field. The selected threshold potentially allows us to reach a completeness level of $\gtrsim90\%$ and at the same time limit the contamination from the field \cite[e.g.,][]{George2011,Castignani_Benoist2016}. Similarly to previous studies \citep[e.g.,][]{Rozo2015}, we also selected  spectroscopic cluster members (open red symbols) as those with a line-of-sight velocity separation $\delta \varv<2000$~km~s$^{-1}$ with respect to the BCG, where $\delta \varv = c\,\frac{|z_{\rm spec} - z_{\rm BCG}|}{ 1 + z_{\rm BCG}}$, $c$ is the speed of light, $z_{\rm BCG}$ is the BCG redshift, and $z_{\rm spec}$ is the spectroscopic redshift of the fiducial cluster member. 

The CM plots show that our targeted BCGs are indeed the brightest (and the most massive) among the selected cluster members. There is only one exception, which is a photometrically selected member with \textsf{r}=18.0 and a projected separation of 63~arcsec {($\sim$0.3~Mpc)} in projection from the KiDS~1220 BCG. The galaxy is thus much ($\sim1$~mag) brighter than BCG and we suspect it is a foreground source, despite its photometric redshift of $0.46\pm0.07$.

In all CM plots we also appreciate a clustering of red sources at \textsf{g-r}$\sim1.5$. { However, the observed clustering is mitigated} by the photometric redshift selection of cluster members. Interestingly, the targeted BCGs are systematically bluer. These findings suggest that the considered cluster cores are sufficiently mature with their galaxy populations well evolved in terms of color properties. On the other hand, the bluer colors of the BCGs are consistent with the fact that we selected them to be star-forming so that they are caught in a rare phase of their evolution. These results motivated us to perform a red-sequence modeling of the photometric data, similarly to previous studies \citep{Kotyla2016,Castignani2019}, as described in the following.

To model the red sequence, we used the Galaxy Evolutionary Synthesis Models (GALEV) tool\footnote{http://www.galev.org} \citep{Kotulla2009}.  We adopt chemically consistent models of elliptical galaxies, according to the GALEV prescriptions. We impose a galaxy formation at high redshift ($z=8$), consistent with formation scenarios of massive ellipticals \citep{Partridge_Peebles1967,Larson1975,DeLucia2006} We verified that changing the formation redshift between $z=6.5-20$ does not significantly change the red-sequence model in the CM plot.

We also assume that the total initial mass of the galaxy is between $1\times10^{10}~M_\odot$ and $5\times10^{11}~M_\odot$. For our red-sequence models we further rely on a \citet{Kroupa2001} IMF between 0.1 and 100~$M_\odot$, which agrees well with the \citet{Chabrier2003} IMF. We note that the SED modeling of the BCGs (Sect.~\ref{sec:SEDs}) relies on the \citet{Chabrier2003} IMF, which is, however, not included in the GALEV library.

With the above assumptions we performed different red-sequence models, which we report  as horizontal lines in Fig.~\ref{fig:CM_plots}, as further outlined in the following. For all three CM plots we show a model assuming no burst, which yields the reddest red sequence possible at the cluster redshift. We also report  different models assuming a single strong burst of star formation, exponentially declining with time. For these models we assume a burst duration of 2~Gyr (e-folding time) and a burst strength of 100\%, which is the fraction of the available gas at the onset of burst that is transformed into stars during the burst. We report different bursty red-sequence models, corresponding to single bursts of star formation occurring between $z=0.6-2.5$, as shown in the legend of Fig~\ref{fig:CM_plots}.

Interestingly, models with a star formation burst at $z=2.5$ (dashed lines in Fig~\ref{fig:CM_plots}), that is, approximately at the peak of the cosmic SFR density \citep{Madau_Dickinson2014}, yield colors that are quite similar to those inferred from models with no burst of star formation (solid lines in Fig~\ref{fig:CM_plots}).  Furthermore, while these two models reproduce fairly well the colors of red cluster members, those of the star-forming BCGs are bluer. 
%This is particularly true for KiDS~0920 and 1220 at the lowest redshifts.
This analysis motivates us to perform additional {evolutionary} models. We indeed find that recent bursts at $z\sim0.6-0.7$, $z\sim0.8-0.9$, and $z\sim1.2-1.3$  are  needed to better reproduce the observed \textsf{g-r} colors, and also the observed \textsf{r}-band magnitudes of  the three KiDS~0920, 1220, and 1444 BCGs, respectively. For KiDS~1444 BCG at the highest redshift, a burst at $z=2.5$ still reproduces fairly well the BCG \textsf{g-r} color. However, a later burst at $z\sim1.2-1.3$ appears to be favored as it implies both a {bluer} color and a brighter magnitude, more similar to those of the BCG.  
However, we note that neither of the adopted red-sequence models is able to reach the low \textsf{r}-band magnitude of KiDS~1444 BCG, despite the high $5\times10^{11}~M_\odot$ initial mass adopted in the modeling. This possibly suggests that a faster mass growth (e.g., via mergers) than that predicted by the close-box GALEV evolutionary models would be a viable alternative for the BCG.\\

Overall, our analysis confirms that the three considered clusters have some evidence of a red sequence, well modeled assuming either no burst {or} an old burst of star formation at cosmic noon ($z=2.5$). The associated BCGs show instead bluer colors, which are better reproduced assuming a recent burst of star formation. These results agree well with the fact that the BCGs are star-forming and gas-rich. With our observations and the modeling we are not able to explain precisely the physical origin for the observed high star formation, the blue colors, and the large gas reservoirs. However, recently accreted pristine gas may explain the observed properties, a scenario that we further discussed in Sect.~\ref{sec:gas_SFR_tdep}.

%-------------------------Fig double horn modeling  ---------------
\begin{figure}[]\centering
\captionsetup[subfigure]{labelformat=empty}
%%%%%%%%%%%%%%%%%%%%%%%%%%%%%%%%%%%%%%%%%%%%5
\subfloat[]{\hspace{0.cm}\includegraphics[clip,trim = {0cm 0cm 0cm 0cm}, width=0.5\textwidth,clip=true]{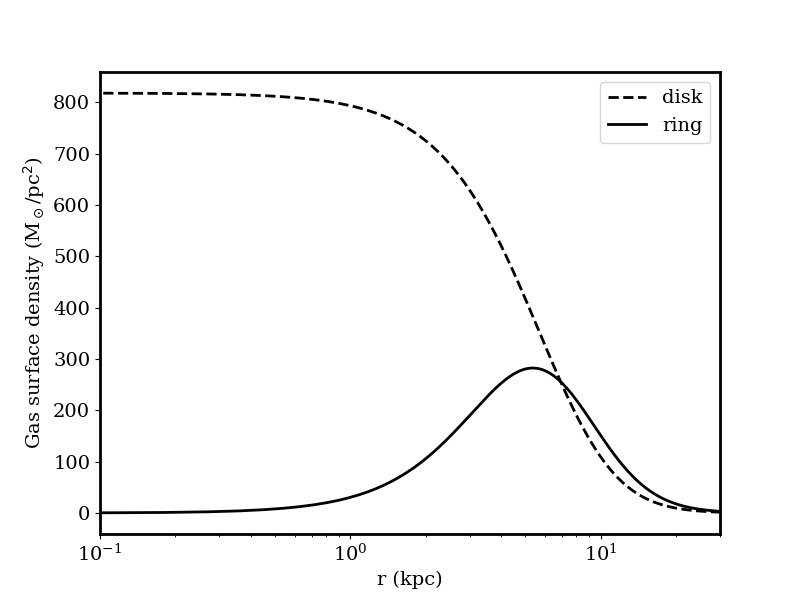}}
\hspace{0.1cm}\subfloat[]{\hspace{0.cm}\includegraphics[clip,trim={0cm 0cm 0cm 
0cm},width=0.5\textwidth,clip=true]{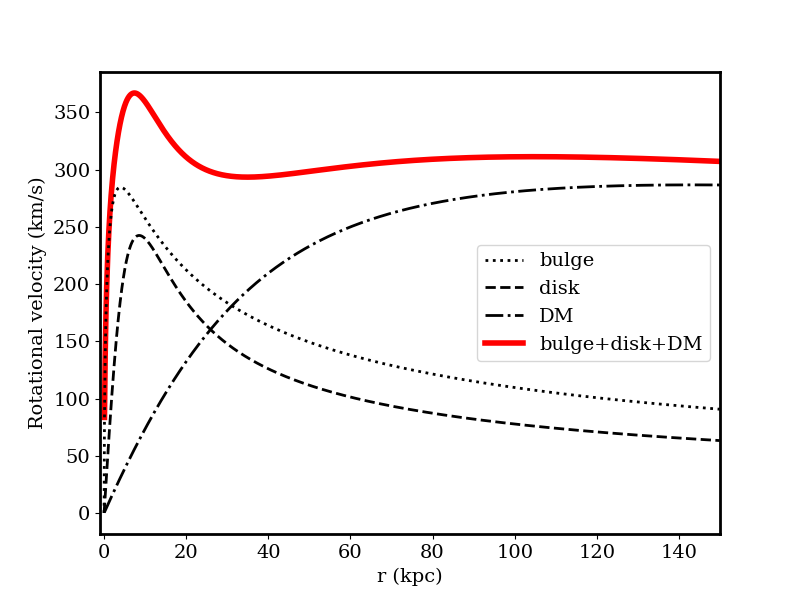}}
\hspace{0.1cm}\subfloat[]{\hspace{0.cm}\includegraphics[trim={0cm 0cm 0cm 
0cm},clip,trim={0cm 0cm 0cm 
0cm},width=0.5\textwidth,clip=true]{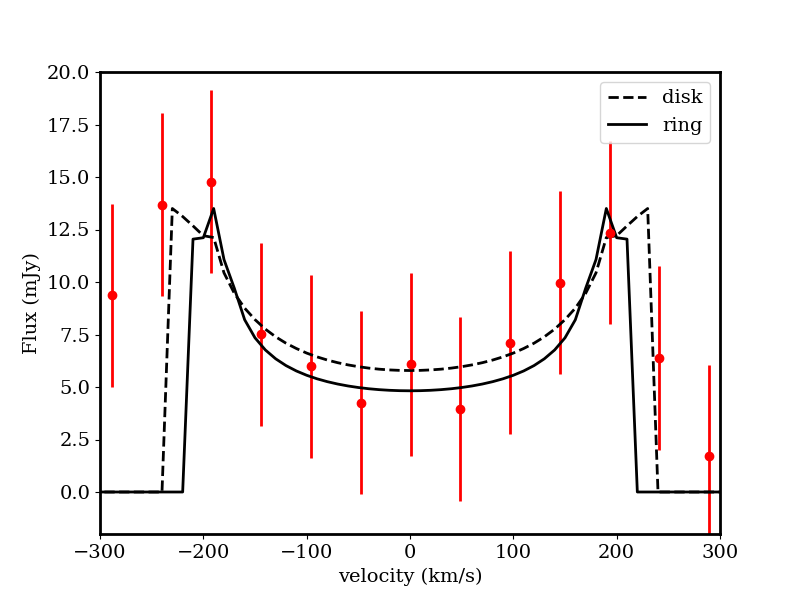}}
%%%%%%%%%%%%%%%%%%%%%%%%%%%
\caption{Modeling of the double-horn CO(3$\rightarrow$2) emission line for KiDS~1220. Top: $H_2$ gas density profile. Center: Rotational velocity curve. Bottom: Observed spectrum and modeling. See the legend and text for further details.}\label{fig:double_horn_modeling}
\end{figure}

\subsection{Double-horn modeling for KiDS~1220. Structural parameters}\label{sec:double_horn_modeling}

We now specifically investigate the structural properties of KiDS~1220. As discussed in Sect.~\ref{sec:IRAM30m_results} this BCG shows evidence for a double-horn profile in the CO(3$\rightarrow$2) emission. This is indicative of a low concentration of the molecular gas toward the center of the galaxy, which results in the observed flux dimming at small velocities in Fig.~\ref{fig:BCG_spectra}. Asymmetric spectra as the one observed for the KiDS~1220 BCG are commonly found in HI in many galaxies 
\citep{Richter_Sancisi1994,Haynes1998,Matthews1998,Yu2020}, as the atomic gas reservoirs are more extended and fragile than the molecular gas. Indeed the latter is usually more centrally peaked, with an exponentially decaying profile  
\citep{Nishiyama2001,Regan2001,Leroy2008}.

{However, asymmetric CO spectra similar to the one found in KiDS~1220 are also reported in previous studies, mostly for low-$z$ galaxies \citep{Wiklind1997,Combes2007,Castignani2022b}, including BCGs \citep{Salome2015}.} Following \citet{Wiklind1997} and \citet{Salome2015} we attempt to link the observed double-horn spectrum to the structural parameters of the gas reservoir, as follows. 

The observed velocity integrated $S_{\rm CO(3\rightarrow2)}\Delta\varv$ spectrum 
can be related to the $H_2$ gas density profile $n(r)$ and the rotational velocity $V_{\rm rot}(r)$ via Eq.~4 of \citet{Wiklind1997}:
\begin{equation}
\label{eq:model_double_horn}
 S_{\rm CO(3\rightarrow2)}\Delta\varv \propto \int\int \frac{n(r) r}{V_{\rm rot}(r)\sin i \sqrt{1-\bigg(\frac{\varv}{V_{\rm rot}(r)\sin i}\bigg)^2}}~dr~d\varv\,,
\end{equation}
where $r$ is the projected radius from galaxy center and $i$ is the inclination with respect to the line of sight ($i = 0$~deg is face–on). We then parameterize the gas density profile with a \citet{Toomre1963} disk of order $n=2$, which corresponds to a flat Plummer distribution:
\begin{equation}
 n(r)=n_0\Bigg(1+\frac{r^2}{d^2}\Bigg)^{-5/2}\,.
\end{equation}
The above expression implies a quadratic contribution to the rotational velocity equal to $V_{\rm rot,gas}^2(r)=V_{\rm rot,d}^2(r)$, as in  Eq.~8 of \citet{Wiklind1997}. Here $d$ is the characteristic scale of the disk, and $n_0=\frac{3 M_{\rm H_2}}{2\pi d^2}$
is the central gas density, whose value is fixed by matching the surface integral of $n(r)$ to the total gas mass $M_{H_2}=1.4\times10^{11}~M_\odot$.

As an alternative to the flat disk we also consider an even less centrally concentrated distribution of $H_2$ as described by a ring. We model a ring-like distribution as the difference of two gaseous disks with scales $d_1>d_2$:  
\begin{equation}
 n(r)=n_0\Bigg[\Bigg(1+\frac{r^2}{d_1^2}\Bigg)^{-5/2}-\Bigg(1+\frac{r^2}{d_2^2}\Bigg)^{-5/2}\Bigg]\,,
\end{equation}
where $n_0=\frac{3 M_{\rm H_2}}{2\pi (d_1^2-d_2^2)}$ is fixed by the surface integral, as in the case of a disk, while the quadratic contribution to the rotational velocity becomes equal to $V_{\rm rot,gas}^2(r)=V_{\rm rot,d_1}^2(r)-V_{\rm rot,d_2}^2(r)$. 

We then express the total rotational velocity as the quadratic sum of the contributions from the gaseous component (disk or ring), the stellar bulge, and the dark matter (DM) halo, as follows:
\begin{equation}
 V_{\rm rot}(r)^2=V_{\rm rot,gas}(r)^2+ V_{\rm rot,bulge}(r)^2+V_{\rm rot,DM}(r)^2\,.
\end{equation}
Following \citet{Wiklind1997}, the stellar bulge is given by
\begin{equation}
V_{\rm rot,bulge}(r)^2 =\frac{GM_\star r}{(r+\frac{r_{\rm eff}}{1.8153})^2}\,.
\end{equation}
Here $M_\star=3.02\times10^{11}~M_\odot$ (Table~\ref{tab:BCG_properties}) and $r_{\rm  eff}=7.28$~kpc are the stellar mass  and half-light radius of KiDS~1220, respectively. We estimated the latter in \textsf{r}-band
assuming a fixed S\'{e}rsic index $n_{\textrm{S\'ersic}}=4$, typical of ellipticals. To this aim we used the {\sc SourceXtractor++} (v 0.16) package\footnote{\url{https://astrorama.github.io/SourceXtractorPlusPlus}}, which is the successor to the {\sc SeXtractor} tool \citep{Bertin_Arnouts1996}.  A  position-dependent model of the point spread function (PSF) was  derived with the {{\sc PSFEx}} tool\footnote{\url{www.astromatic.net}} and used as input in 
{\sc SourceXtractor++} to fit the galaxy with a S\'{e}rsic profile plus the model PSF.

We then express the DM halo contribution as in Eq.~5 of \citet{Castignani2012}, which corresponds to the  universal rotation curve halo model by \citet{Salucci2007}:
\begin{equation}
 V_{\rm rot,DM}(r)^2=6.4\frac{\rho_0 r_0^3}{r}\Bigg[\log\Bigg(1+\frac{r}{r_0}\Bigg)-\arctan{\frac{r}{r_0}}+\frac{1}{2}\log\Bigg(1+\frac{r^2}{r_0^2}\Bigg)\Bigg]\,.
\end{equation}
According to the model, both the scale length and the central density can be related to the DM mass \citep{Salucci2007,Memola2011}, and consequently $V_{\rm rot,DM}(r)^2$. In order to reproduce a flattening of the rotational velocity profile at a large distance $r\sim100$~kpc, as it is the case of massive galaxies such as KiDS~1220, we assume a DM mass of $M_{\rm DM}=5\times10^{12}~M_\odot$, which implies $\rho_0 = 2.40\times10^{-25}$~g~cm$^{-3}$, $r_0 = 44.2$~kpc,  a stellar to DM halo mass ratio of $M_\star/M_{\rm DM}=6\%$ and a flattening of the rotation velocity to a value of $\sim300$~km~s$^{-1}$. 

To determine the structural parameters of the gaseous reservoir we then vary the disk scales d, d$_1$, and d$_2$ from 50~pc up to 20~kpc. From the integration in Eq.~\ref{eq:model_double_horn} over all radii and over the velocity support of the CO(3$\rightarrow$2) line we can relate the observed spectrum to the one inferred from our modeling. 
The overall normalization is fixed by the height of the line. Increasing the inclination angle $i$ has the net effect of increasing the velocity separation between the two peaks in the spectrum \citep{Wiklind1997}. As $M_{H_2}$ is fixed, decreasing the disk scales $d$, $d_1$ has the net effect of increasing the central density $n_0$, and thus the concentration, which ultimately reduces the observed depth in the spectrum at small velocities, with respect to the peaks. Given these relations, we find that both a ring and a disk with an inclination of $i=40$~deg reproduce the observed spectrum well. The inferred sizes are $d=d_1=9$~kpc and $d_1-d_2=1$~kpc. With these parameters we get  peak $H_2$-gas-density values of 800 and 300~$M_\odot$/pc$^2$ in the case of a disk and a ring, respectively, which are typical values found for LIRGs, like our BCGs,  along the Kennicutt-Schmidt relation. 

Figure~\ref{fig:double_horn_modeling} (top) displays the gas surface density profiles. In the central panel we report the rotational velocity profile in the case of a $d=9$~kpc disk, while similar curves are obtained in the case of a ring with the above parameters. In the bottom panel of Fig.~\ref{fig:double_horn_modeling} we show both the observed spectrum and our modeling in both cases of a gaseous disk or a ring. As the central density scales as the square of the inverse of disk scale, we stress that by adopting a smaller radius for the molecular gas reservoir we significantly increase the central density up to high values $\gg1000~M_\odot$/pc$^2$, which are less likely. At the same time we also increase the concentration and we are not able anymore to reproduce well the flux dimming at small velocities in the spectrum, even assuming an extreme edge-on orientation ($i=90$~deg).

Overall, our results thus yield an extended molecular gas reservoir that is reminiscent of the extended disk phase described in \citet{Olivares2022} for local BCGs. During this phase the star formation and so the molecular gas is more likely to be detected. Similar to what was suggested by these authors, the angular momentum may prevent the cold gas from being rapidly accreted, which would explain the low concentration and the large extent of molecular gas in KiDS~1220. 
Following the evolutionary scenario proposed by \citet{Olivares2022}, the BCG may then experience a rapid compaction phase, similarly to that previously suggested for distant star-forming galaxies \citep{Barro2013,Barro2017}, including BCGs \citep{Castignani2020b}.

\section{Summary and conclusions}\label{sec:conclusions}
%, and studied their interplay with their mega-parsec environments. This work is part of a larger campaign targeting galaxies in and around distant clusters in CO, and in particular BCGs, with the final goal to evaluate the impact of dense mega-parsec scale environemnts in processing the molecular gas reservoirs that feed the star formation \citep{Castignani2018,Castignani2019,Castignani2020a,Castignani2020b,Castignani2020c,Castignani2020d}.

In this work we have investigated the molecular gas and star formation properties of three star-forming BCGs at intermediate redshifts, drawn from the \cite{Radovich2020} sample of BCGs within the third data release of KiDS \citep[][]{deJong2017,Kuijken2019}. We have selected three targets from a large sample of 684 spectroscopically confirmed $z>0.3$ BCGs in the  equatorial KiDS-N field, namely KiDS~0920 ($z=0.3216$), KiDS~1220 ($z=0.3886$), and KiDS~1444 ($z=0.4417$), that show significant infrared emission{ (Sect.~\ref{sec:BCGselection})}. Our sources thus represent a small fraction (0.4\%) of the parent sample and are among the most star-forming BCGs in KiDS. They have line-based (H$\alpha$, [O~II]) and SED-based ${\rm SFR}\simeq(20-30)~M_\odot$/yr, as well as dust masses and luminosities  typical of LIRGs { (Sects.~\ref{sec:SEDs} and \ref{sec:line_diagnostics})}. By performing BPT line diagnostics, we find  KiDS~0920 is classified as a dusty AGN, while the other two BCGs likely fall in the star-forming region of the BPT diagram{ (Sect.~\ref{sec:line_diagnostics})}.
The selected BCGs thus constitute a rare subpopulation of star-forming BCGs that are experiencing stellar mass assembly and are thus caught in a special phase of their evolution, similar to other distant star-forming BCGs \citep[e.g.,][]{Castignani2020a,Castignani2020b}.

%. By looking at available far-infrared to ultravioled spectral energy distibution (SED) modeling \citep{Driver2018} imply dust masses and luminosities are in the range $\log(M_{\rm dust}/M_\odot)\simeq8.5-8.7$ and $\log(L_{\rm dust}/L_\odot)\simeq11.5-11.8$, respectively, typical of luminous infrared galaxies (LIRGs). 

%With such star formation and dust properties, 
%the three BCGs of this work are the high-$z$ counterparts of local star forming BCGs ($>40~M_\odot$/yr) such as the famous Perseus~A and Cygnus~A \citep{FraserMcKelvie2014}. 

%{ I ARRIVED HERE}
In order to investigate the mass growth and the fueling of gas feeding star formation, we observed the three BCGs with the IRAM 30m telescope, targeting the first three CO transitions { (Sect.~\ref{sec:observations_and_data_reduction})}. We clearly detect at ${\rm S/N}=(3.8 - 10.2)$ all three sources in multiple transitions. Our campaign thus yields a high success rate of {88$\%$,} much higher than that of previous studies that targeted distant BCGs in CO{ (Sect.~\ref{sec:IRAM30m_results})}. With the present work we also double the number of distant BCGs that are clearly detected in{ at least} two CO transitions.

We then compared the molecular gas properties of the KiDS BCGs with those of a sample of distant cluster galaxies{ (Sect.~\ref{sec:comparison_sample})} with molecular gas observations from the literature as well as with scaling relations of MS field galaxies. Our BCGs appear to be rich in gas, with $M_{H_2}\simeq(0.5-1.4)\times10^{11}~M_\odot$ and $M_{H_2}/M_\star\simeq(0.2-0.6)$. These results place them among the most gas-rich BCGs at intermediate redshifts.

The KiDS BCGs of this work have a high content of molecular gas, well above the MS. Conversely, they also have star formation activities typical of MS galaxies { (Sect.~\ref{sec:gas_SFR_tdep})}. This implies that our targeted BCGs accumulate large molecular gas reservoirs that are not efficiently converted into stars, having depletion timescales $\tau_{\rm dep}\simeq(2-4)$~Gyr, higher than those previously found in distant star-forming BCGs from the CLASH and SpARCS surveys \citep{Fogarty2019,Castignani2020a,Dunne2021}.

All these results are intriguing and suggest that the gas fueling the three KiDS BCGs differs somehow from that of previously studied star-forming BCGs at intermediate redshifts. 
Consistent with this interpretation, we find a correlation between the excitation ratio and the star formation activity of the BCGs{ (Sect.~\ref{sec:excitation_ratios})}, in the sense that the highest excitation ratios are found only in highly star-forming BCGs (${\rm SFR}>100~M_\odot$/yr), which also have short depletion times of $\sim(0.5-0.7)$~Gyr \citep{Fogarty2019,Castignani2020a}.
Our KiDS BCGs instead have lower excitation ratios, lower ${\rm SFRs}\simeq(20-30)~M_\odot$/yr, and longer $\tau_{\rm dep}$.

We further investigated the star formation properties of the KiDS BCGs by performing a CM analysis{ (Sect.~\ref{sec:color_magnitude_plots})}. We find that the BCGs are bluer and brighter than photometrically selected cluster members. Red-sequence modeling suggests that  recent bursts of star formation are needed to better reproduce both the observed \textsf{g-r} colors and the \textsf{r}-band magnitudes of the three BCGs.

One possible explanation for {the observed phenomenology} is that the $H_2$ gas, or a substantial amount of it, was recently accreted by the KiDS BCGs but still not efficiently converted into stars. The infall of pristine gas may also explain the $M_{H_2}/M_{\rm dust}\simeq170-300$ ratios that we find for our KiDS BCGs. This accretion may have occurred via filaments (as may be the case of KiDS~0920, which shows an elongated morphology) or via interaction with companions, which are in fact found for KiDS~1220 and 1444.

Interestingly, for KiDS~1220 we have additionally found a double-horn emission in CO(3$\rightarrow$2), which implies low gas concentration toward the galaxy center. The CO(3$\rightarrow$2) emission is well modeled with an extended molecular gas reservoir of $\sim9$~kpc { (Sect.~\ref{sec:double_horn_modeling})}, which is reminiscent of a mature extended disk phase observed in local BCGs by \citet{Olivares2022}. This phase may precede a compaction phase \citep{Barro2013,Barro2017,Castignani2020b}, and later the quenching.

\begin{acknowledgements}
{ We thank the anonymous referee for helpful comments which contributed to improving the paper.}
This work is based on observations carried out under project numbers 079-21 with the IRAM 30m telescope. GC thanks IRAM staff at Granada for their help with these observations. IRAM is supported by INSU/CNRS (France), MPG (Germany) and IGN (Spain).
{GC, LM, CG, and FM  acknowledge the support from the grant ASI n.2018-23-HH.0. LM and CG also acknowledge the support from the grant PRIN-MIUR 2017 WSCC32. MS acknowledges financial contribution from contracts ASI-INAF n.2017-14-H.0 and INAF mainstream project 1.05.01.86.10.}
The work uses GAMA data-products. GAMA is a joint European-Australasian project based around a spectroscopic campaign using the Anglo-Australian Telescope. The GAMA input catalogue is based on data taken from the Sloan Digital Sky Survey and the UKIRT Infrared Deep Sky Survey. Complementary imaging of the GAMA regions is being obtained by a number of independent survey programmes including GALEX MIS, VST KiDS, VISTA VIKING, WISE, Herschel-ATLAS, GMRT and ASKAP providing UV to radio coverage. GAMA is funded by the STFC (UK), the ARC (Australia), the AAO, and the participating institutions. The GAMA website is \href{http://www.gama-survey.org/}{http://www.gama-survey.org/}. The work is based on data products from observations made with ESO Telescopes at the La Silla Paranal Observatory under programme IDs 177.A-3016, 177.A-3017 and 177.A-3018, and on data products produced by Target/OmegaCEN, INAF-OACN, INAF-OAPD and the KiDS production team, on behalf of the KiDS consortium. OmegaCEN and the KiDS production team acknowledge support by NOVA and NWO-M grants. Members of INAF-OAPD and INAF-OACN also acknowledge the support from the Department of Physics \& Astronomy of the University of Padova, and of the Department of Physics of Univ. Federico II (Naples). 
\end{acknowledgements}
%\newpage


\begin{thebibliography}{1000}
%\bibitem[Adams(1977)]{Adams1977} Adams, T.~F. 1977, ApJS, 33, 19
%\bibitem[Annunziatella et al.(2014)]{Annunziatella2014} Annunziatella, M., Biviano, A., Mercurio, A. et al. 2014, A\&A, 571, 80
%\bibitem[Annunziatella et al.(2016)]{Annunziatella2016} Annunziatella, M., Mercurio, A., Biviano, A. et al. 2016, A\&A, 585, 160
\bibitem[Aravena et al.(2012)]{Aravena2012} Aravena, M., Carilli, C.~L., Salvato, M. et al. 2012, MNRAS, 426, 258 
\bibitem[Baldry et al.(2010)]{Baldry2010} Baldry, I.~K., Robotham, A.~S.~G., Hill, D.~T. et al. 2010, MNRAS, 404, 86
\bibitem[Baldry et al.(2018)]{Baldry2018} Baldry, I.~K., Liske, J., Brown, M.~J.~I. et al. 2018, MNRAS, 474, 3875
\bibitem[Baldwin et al.(1981)]{Baldwin1981} Baldwin, J.~A., Phillips, M.~M., \& Terlevich, R. 1981, PASP, 93, 5
\bibitem[Barfety et al.(2022)]{Barfety2022} Barfety, C., Valin, F.-A., Webb, T.~M.~A et al. 2022, ApJ, 930, 25
\bibitem[Barro et al.(2013)]{Barro2013} Barro, G., Faber, S.~M. et al. 2013, ApJ, 765, 104
\bibitem[Barro et al.(2017)]{Barro2017} Barro, G., Faber, S.~M., Koo, D.~C. et al. 2017, ApJ, 840, 47
\bibitem[Bellagamba et al.(2018)]{Bellagamba2018} Bellagamba, F., Roncarelli, M., Maturi, M. et al. 2018, MNRAS, 473, 5221
\bibitem[Bellagamba et al.(2019)]{Bellagamba2019}  Bellagamba, F., Sereno, M., Roncarelli, M. et al. 2019, MNRAS, 484, 1598
\bibitem[Benítez(2000)]{Benitez2000} Benítez, N. 2000, ApJ, 536, 571
\bibitem[Berta et al.(2016)]{Berta2016} Berta S., Lutz D., Genzel R. et al. 2016, A\&A, 587, A73
\bibitem[Bertin \& Arnouts(1996)]{Bertin_Arnouts1996} Bertin, E. \& Arnouts, S. 1996, A\&AS, 117, 393
%\bibitem[Bezanson et al.(2019)]{Bezanson2019} { Bezanson, R., Spilker, J.,  Williams, C.~C. et al. 2019, ApJ, 873, 19}
%\bibitem[Bigiel et al.(2008)]{Bigiel2008} Bigiel F., Leroy A., Walter F. et al. 2008, AJ, 136, 2846
%\bibitem[Bolatto et al.(2013)]{Bolatto2013} { Bolatto, A.~T., Wolfire, M., \& Leroy, A.~K 2013, ARA\&A, 51, 207}
\bibitem[Bonaventura et al.(2017)]{Bonaventura2017} Bonaventura, N.~R., Webb, T.~M.~A.,  Muzzin, A. et al. 2017, MNRAS, 469, 1259
\bibitem[Boogaard et al.(2020)]{Boogaard2020} Boogaard, L.~A., van der Werf, P., Weiss, A. et al. 2020, ApJ, 902, 109
\bibitem[Bothwell et al.(2013)]{Bothwell2013} Bothwell, M. S., Smail, I., Chapman, S. C., et al. 2013, MNRAS, 429, 3047
%\bibitem[Burke et al.(2015)]{Burke2015} Burke, C., Hilton, M., Collins, C. et al. 2015, 2015, MNRAS, 449, 2353
%\bibitem[Calzetti et al.(2000)]{Calzetti2000} Calzetti, D., Armus, L., Bohlin, R. C., et al. 2000, ApJ, 533, 682
\bibitem[Calzetti(2013)]{Calzetti2013} Calzetti D., 2013, Star Formation Rate Indicators,  Cambridge University Press, Cambridge, UK, p.~419
\bibitem[Carilli \& Walter(2013)]{Carilli_Walter2013}  Carilli, C.~L. \& Walter, F. 2013, ARA\&A, 51, 105
%\bibitem[Caminha et al.(2019)]{Caminha2019} Caminha, G.~B., Rosati, P., Grillo, C. et al. 2019, A\&A, 632, 36
%\bibitem[Carilli \& Walter(2013)]{Carilli_Walter2013} Carilli, C.~L., \& Walter, F. 2013, ARA\&A, 51, 105
\bibitem[Castignani \& Benoist(2016)]{Castignani_Benoist2016} Castignani, G., \& Benoist, C.,  2016,  A\&A, 595, 111
\bibitem[Castignani et al.(2012)]{Castignani2012} Castignani, G., Frusciante, N., Vernieri, D. et al. 2012, Natural Science, 4, 265-270
\bibitem[Castignani et al.(2018)]{Castignani2018} Castignani, G., Combes, F., Salom\'{e}, P. et al. 2018,  A\&A, 617, 103
\bibitem[Castignani et al.(2019)]{Castignani2019} Castignani, G., Combes, F., Salom\'{e}, P. et al. 2019, A\&A, 623, 48
\bibitem[Castignani et al.(2020a)]{Castignani2020a} Castignani, G., Pandey-Pommier, M., Hamer, S.~L. et al. 2020a, A\&A, 640, 65
\bibitem[Castignani et al.(2020b)]{Castignani2020b} Castignani, G., Combes, F., Salom\'{e}, P. et al. 2020b, A\&A, 635, 32 
\bibitem[Castignani et al.(2020c)]{Castignani2020c} Castignani, G., Combes, F., \& Salom\'{e}, P., 2020c,  A\&A 635, L10 %BCG z=1.7
\bibitem[Castignani et al.(2020d)]{Castignani2020d} Castignani, G., Jablonka, P., Combes, F. et al. 2020d, A\&A, 640, 64
\bibitem[Castignani et al.(2022a)]{Castignani2022a} Castignani, G., Meyer, E., Chiaberge, M. et al. 2022a, A\&AL, 661, 2
\bibitem[Castignani et al.(2022b)]{Castignani2022b} Castignani, G., Combes, F., Jablonka, P. et al. 2022b, A\&A, 657, 9
\bibitem[Chabrier(2003)]{Chabrier2003} Chabrier G., 2003, PASP, 115, 763
%\bibitem[Chen \& Hwang(2017)]{Chen_Hwang2017} Chen, Y.-C., \& Hwang, C.-Y., 2017, Ap\&SS, 362, 230
\bibitem[Collins et al.(2009)]{Collins2009} Collins, C.~A., Stott, J.~P.,  Hilton, M. et al. 2009, Nature, 458, 603
\bibitem[Combes et al.(2007)]{Combes2007} Combes, F., Young, L.~M., \& Bureau, M., 2007, MNRAS, 377, 1795
\bibitem[Coogan et al.(2018)]{Coogan2018} { Coogan, R.~T., Daddi, E., Sargent, M.~T. et al. 2018, MNRAS, 479, 703}
%\bibitem[Cooke et al.(2016)]{Cooke2016} Cooke, K.~C., O'Dea, C.~P., Baum, S.~A. et al. 2016, ApJ, 833, 224
\bibitem[Cybulski et al.(2016)]{Cybulski2016} Cybulski, R., Yun, M.~S., Erickson, N. et al. 2016, MNRAS, 459, 3287
%\bibitem[Daddi et al.(2015)]{Daddi2015} Daddi, E., Dannerbauer, H., Liu, D. et al. 2015, A\&A, 577, 46
\bibitem[D'Amato et al.(2020)]{Damato2020} D'Amato, Q., Gilli, R., Prandoni, I. et al. 2020 A\&A, 641, 6D
\bibitem[da Cunha et al.(2008)]{daCunha2008} da Cunha, E., Charlot, S., \& Elbaz, D. 2008, MNRAS, 388, 1595
\bibitem[de Jong et al.(2017)]{deJong2017} de Jong, J.~T.~A., Verdoes Kleijn, G.~A., Erben, T. et al. 2017, A\&A, 604, 134
\bibitem[De Lucia et al.(2006)]{DeLucia2006} De Lucia, G., Springel, V., White, S.~D.~M. et al. 2006, MNRAS, 366, 499
\bibitem[De Lucia \& Blaizot(2007)]{DeLucia_Blaizot2007} De Lucia, G., \& Blaizot, J. 2007, MNRAS, 375, 2
%\bibitem[DeMaio et al.(2019)]{DeMaio2019} DeMaio, T., Gonzalez, A.~H., Zabludoff, A. et al. 2020, MNRAS, 491, 3751
%\bibitem[de Souza et al.(2016)]{deSouza2016} de Souza, R.~S., Dantas, M.~L.~L., Krone-Martins, A. et al. 2016, MNRAS, 461, 2115
%\bibitem[Devereux et al.(1994)]{Devereux1994} Devereux, N., Taniguchi, Y., Sanders, D. B., et al. 1994, AJ, 107, 2006
\bibitem[Domínguez et al.(2013)]{Dominguez2013} Domínguez, A., Siana, B., Henry, A.~L. et al. 2013, ApJ, 763, 145
%\bibitem[Donahue et al.(2014)]{Donahue2014} Donahue, M., Voit, G. M., Mahdavi, A., et al. 2014, ApJ, 794, 136
%\bibitem[Donahue et al.(2015)]{Donahue2015}Donahue, M., Connor, T., Fogarty, K., et al. 2015, ApJ, 805, 177
%\bibitem[Donahue et al.(2016)]{Donahue2016} Donahue, M., Ettori, S., Rasia, E. et al. 2016, ApJ, 819, 36, 2016 
\bibitem[Driver et al.(2018)]{Driver2018} Driver, S.~P., Andrews, S.~K., da Cunha, E. et al. 2018, MNRAS, 475, 2891
\bibitem[Dunne et al.(2021)]{Dunne2021} Dunne, D.~A., Webb, T.~M.~A., Noble, A. et al. 2021, ApJ, 909, 29
%\bibitem[Durret et al.(2019)]{Durret2019} Durret, F., Tarricq, Y., M\'{a}rquez, I. et al. 2019, A\&A, 622, 78
\bibitem[Eales et al.(2010)]{Eales2010} Eales, S., Dunne, L., Clements, D. et al. 2010, PASP, 122, 499
\bibitem[Edge et al.(2013)]{Edge2013} Edge, A., Sutherland, W., Kuijken, K., et al. 2013, The Messenger, 154, 32
\bibitem[Edge(2001)]{Edge2001} Edge, A., 2001, MNRAS, 328, 762
\bibitem[Fabian(1994)]{Fabian1994} Fabian, A.~C. 1994, ARA\&A,32, 277
\bibitem[Fabian(2012)]{Fabian2012} Fabian, A.~C. 2012, ARA\&A, 50, 455
%\bibitem[Fogarty et al.(2015)]{Fogarty2015} Fogarty, K., Postman, M., Connor, T. et al. 2015, ApJ, 813, 117
%\bibitem[Fogarty et al.(2017)]{Fogarty2017} Fogarty, K., Postman, M., Larson, R. et al. 2017, ApJ, 846, 103
\bibitem[Fogarty et al.(2019)]{Fogarty2019} Fogarty, K., Postman, M., Li, Y. et al. 2019, ApJ, 879, 103
\bibitem[Fraser-McKelvie et al.(2014)]{FraserMcKelvie2014} Fraser-McKelvie, A., Brown, M.~J.~I., \& Pimbblet, K.~A., 2014, MNRAS, 444, 63
%\bibitem[Freundlich et al.(2019)]{Freundlich2019} Freundlich, J., Combes, F., Tacconi, L. J. et al., 2019, A\&A, 622, 105
\bibitem[Garn \& Best(2010)]{Garn_Best2010} Garn, T., \& Best, P. N. 2010, MNRAS, 409, 421
%\bibitem[Gaspari et al.(2012)]{Gaspari2012} Gaspari, M., Ruszkowski, M., \& Sharma, P. 2012, ApJ, 746, 94
\bibitem[Geach et al.(2011)]{Geach2011} Geach, J.~E., Smail, I., Moran, S.~M. et al. 2011, ApJL, 730, 19
\bibitem[Genzel et al.(2015)]{Genzel2015} Genzel, R., Tacconi, L. J., Lutz, D. et al. 2015, ApJ, 800, 20
\bibitem[George et al.(2011)]{George2011} George, M.~R., Leauthaud, A., Bundy, K. et al. 2011, ApJ, 742, 125
\bibitem[Gilbank et al.(2010)]{Gilbank2010} Gilbank, D.~G., Baldry, I.~K., Balogh, M.~L. et al. 2010, MNRAS, 405, 2594
%\bibitem[Gobat et al.(2018)]{Gobat2018} { Gobat, R., Daddi, E., Magdis, G., et al. 2018, Nature Astronomy, 2, 239}
\bibitem[Gordon et al.(2017)]{Gordon2017} Gordon, Y.~A., Owers, M.~S., Pimbblet, K.~A. et al. 2017, MNRAS, 465, 267
%\bibitem[Hamer et al.(2012)]{Hamer2012} Hamer, S.~L.,  Edge, A.~C., Swinbank, A.~M. et al. 2012, MNRAS, 421, 3409 %The relation between line emission and brightest cluster galaxies in three exceptional clusters: evidence for gas cooling from the intracluster medium
\bibitem[Hausman \& Ostriker(1978)]{Hausman_Ostriker1978} Hausman, M.~A. \& Ostriker, J.~P., 1978. , ApJ, 224, 320
\bibitem[Hayashi et al.(2018)]{Hayashi2018} { Hayashi, M., Tadaki, K., Kodama, T. et al. 2018, ApJ, 856, 118}
\bibitem[Haynes et al.(1998)]{Haynes1998} Haynes, M.~P., Hogg, D.~E., Maddalena, R.~J. et al. 1998, AJ, 115, 62
%\bibitem[Heckman et al.(1978)]{Heckman1978} Heckman, T.~M., Balick, B., Sullivan, W.~T., 1978,  ApJ, 224, 745
%\bibitem[Hicks et al.(2010)]{Hicks2010} Hicks, A.~K., Mushotzky, R., \& Donahue, M. 2010, ApJ,719, 1844
\bibitem[Hildebrandt et al.(2021)]{Hildebrandt2021} Hildebrandt, H., van den Busch, J.~L., Wright, A.~H. et al. 2021, A\&A, 647, 124
\bibitem[Hopkins et al.(2013)]{Hopkins2013} Hopkins, A.~M., Driver, S.~P., Brough, S. et al. 2013, MNRAS, 430, 2047
%\bibitem[Huang \& Gu(2009)]{Huang_Gu2009} Huang, S., \& Gu, Q.-S., 2009, MNRAS, 398, 1651
\bibitem[Jablonka et al.(2013)]{Jablonka2013}  Jablonka, P., Combes, F., Rines, K. et al. 2013, A\&A, 557, 103
\bibitem[Jarrett et al.(2017)]{Jarrett2017} Jarrett, T.~H., Cluver, M.~E., Magoulas, C. et al. 2017, ApJ, 836, 182
\bibitem[Kennicutt(1998)]{Kennicutt1998} Kennicutt, R.~C.~J. 1998, ARAA, 36, 189
\bibitem[Kewley et al.(2006)]{Kewley2006} Kewley, L.~J., Groves, B., Kauffmann, G. et al. 2006, MNRAS, 372, 961
%\bibitem[Kewley et al.(2013)]{Kewley2013} Kewley, L.~J., Maier, C., Yabe, K. et al. 2013, ApJ, 774, 10
%\bibitem[Kiuchi et al.(2009)]{Kiuchi2009} Kiuchi, G., Ohta, K., \& Akiyama, M. 2009,  ApJ, 696, 1051
\bibitem[Kneissl et al.(2019)]{Kneissl2019}  Kneissl, R., del Carmen Polletta, M., Martinache, C. et al. 2019, A\&A, 625, 96
\bibitem[Kotulla et al.(2009)]{Kotulla2009} Kotulla, R., Fritze, U., Weilbacher, P. et al. 2009, MNRAS 396, 462 
\bibitem[Kotyla et al.(2016)]{Kotyla2016} Kotyla, J.~P., Chiaberge, M., Baum, S. et al. 2016, ApJ, 826, 46 
\bibitem[Kramer et al.(2013)]{Kramer2013}  Kramer, C.,  Pe\~{n}alver, J. \& Greve A., 2013, ''Improvement of the IRAM 30m telescope beam pattern'', www.iram-institute.org$/$medias$/$uploads$/$eb2013-v8.2.pdf
%\bibitem[Krist (1995)]{Krist1995} Krist, J. 1995, ASPC, 77, 349
%\bibitem[Krist et al.(2011)]{Krist2011} Krist, J.~E., Hook, R.~N. \& Stoehr, F. 2011, SPIE, 8127
\bibitem[Kroupa(2001)]{Kroupa2001} Kroupa, P. 2001, MNRAS, 322, 231
\bibitem[Kuijken et al. (2019)]{Kuijken2019} Kuijken, K., Heymans, C., Dvornik, A. et al. 2019, A\&A, 625, 2
\bibitem[Lamperti et al.(2020)]{Lamperti2020} Lamperti, I., Saintonge, A., Koss, M. et al. 2020, ApJ, 889, 103
\bibitem[Larson (1975)]{Larson1975} Larson R.~B., 1975, MNRAS, 173, 671
\bibitem[Larson et al.(1980)]{Larson1980} Larson, R.~B., Tinsley, B.~M., \& Caldwell, C.~N., 1980, ApJ, 237, 692
\bibitem[Lauer et al.(2014)]{Lauer2014} Lauer, T.~R., Postman, M., Strauss, M.~A. et al., 2014, ApJ, 797, 82
\bibitem[Lavoie et al.(2016)]{Lavoie2016} Lavoie, S., Willis, J. P., Démoclès, J. et al. 2016, MNRAS, 462, 4141
%\bibitem[Lee et al.(2017)]{Lee2017}  Lee, M.~M., Tanaka, I., Kawabe, R. et al. 2017,  ApJ, 842, 55
\bibitem[Leech et al.(2010)]{Leech2010} Leech, J.; Isaak, K.~G.; Papadopoulos, P.~P. et al. 2010, MNRAS, 406, 1364
\bibitem[Leroy et al.(2008)]{Leroy2008} Leroy, A.~K., Walter, F., Brinks, E., et al. 2008, AJ, 136, 2782
%\bibitem[Leroy et al.(2013)]{Leroy2013} Leroy A.~K., Walter F., Sandstrom K. et al. 2013, AJ, 146, 19
%\bibitem[Li \& Bryan(2014)]{Li_Bryan2014} Li, Y., \& Bryan, G. L. 2014, ApJ, 789, 153
%\bibitem[Li et al.(2015)]{Li2015} Li, Y., Bryan, G. L., Ruszkowski, M., et al. 2015, ApJ, 811, 72
\bibitem[Lidman et al.(2012)]{Lidman2012} Lidman, C., Suherli, J., Muzzin, A., et al. 2012, MNRAS, 427, 550
\bibitem[Madau \& Dickinson(2014)]{Madau_Dickinson2014} Madau, P. \& Dickinson, M., 2014, ARA\&A, 52, 415
\bibitem[Martin et al.(2005)]{Martin2005} Martin, D.~C., Fanson, J., Schiminovich, D. et al. 2005, ApJ, 619, 1
\bibitem[Matthews et al.(1998)]{Matthews1998} Matthews, L.~D., van Driel, W., \& Gallagher, J.~S., III, 1998, AJ, 116, 1169
\bibitem[Maturi et al(2019)]{Maturi2019} Maturi, M., Bellagamba, F., Radovich, M. et al. 2019, MNRAS, 485, 498
%\bibitem[McDonald et al.(2013)]{McDonald2013} McDonald, M., Benson, B., Veilleux, S. et al. 2013, ApJL, 765, 37
\bibitem[McDonald et al.(2014)]{McDonald2014} McDonald, M., Swinbank, M., Edge, A.~C. et al. 2014, ApJ, 784, 18
\bibitem[McDonald et al.(2016)]{McDonald2016} McDonald, M., Stalder, B., Bayliss, M., et al. 2016, ApJ, 817, 86
\bibitem[McNamara et al.(2000)]{McNamara2000} McNamara, B.~R., Wise, M., Nulsen, P. E. J., et al. 2000, ApJL, 534, 135
\bibitem[McNamara \& Nulsen(2012)]{McNamara_Nulsen2012} McNamara, B.~R., \& Nulsen, P.~E.~J. 2012, NJPh, 14, 055023
%\bibitem[McNamara et al.(2014)]{McNamara2014} McNamara, B.~R., Russell, H.~R., Nulsen, P.~E.~J. et al. 2014, ApJ, 785, 44 %A 1010 Solar Mass Flow of Molecular Gas in the A1835 Brightest Cluster Galaxy
\bibitem[Memola et al.(2011)]{Memola2011} Memola, E., Salucci, P., \& Babić, A., 2011, A\&A, 534, 50
\bibitem[Moore et al.(1999)]{Moore1999} Moore, B., Lake, G., Quinn, T. et al. 1999, MNRAS, 304, 465
\bibitem[Morrissey et al.(2007)]{Morrissey2007} Morrissey, P., Conrow, T., Barlow, T.~A. et al. 2007, ApJS, 173, 682
\bibitem[Nishiyama et al.(2001)]{Nishiyama2001} Nishiyama, K., Nakai, N., \& Kuno, N. 2001, PASJ, 53, 757
\bibitem[Noble et al.(2017)]{Noble2017} Noble, A.~G., McDonald, M., Muzzin, A., et al. 2017, ApJ, 842, 21
\bibitem[Noble et al.(2019)]{Noble2019} { Noble, A.~G., Muzzin, A., McDonald, M. et al. 2019, ApJ, 870, 56}
\bibitem[Old et al.(2020)]{Old2020} Old, L.~J., Balogh, M.~L., van der Burg, R.~F.~J. et al. 2020, MNRAS, 493, 5987
\bibitem[Olivares et al.(2019)]{Olivares2019}  Olivares, V., Salom\'{e}, P., Combes, F., et al. 2019, A\&A, 631, 22O 
\bibitem[Olivares et al.(2022)]{Olivares2022} Olivares, V., Salome, P., Hamer, S. L. et al. 2022, arXiv:220107838
\bibitem[O'Sullivan et al.(2021)]{OSullivan2021} O'Sullivan, E., Combes, F., Babul, A. et al. 2021, MNRAS, 508, 379
%\bibitem[Papadopoulos et al.(2000)]{Papadopoulos2000} Papadopoulos, P.~P., R\"{o}ttgering, H.~J.~A., van der Werf, P.~P. et al. 2000, ApJ, 528, 626
%\bibitem[Papadopoulos et al.(2012)]{Papadopoulos2012} Papadopoulos, P.~P., van der Werf, P.~P., Xilouris, E.~M. et al. 2012, MNRAS, 426, 2601
\bibitem[Partridge \& Peebles (1967)]{Partridge_Peebles1967} Partridge R.~B., \& Peebles P.~J.~E., 1967, ApJ, 147, 86
%\bibitem[Peng et al.(2002)]{Peng2002} { Peng, C.~Y., Ho, L.~C., Impey, C.~D. et al. 2002, AJ, 124, 266}
%\bibitem[Peng et al.(2010)]{Peng2010} { Peng, C.~Y., Ho, L.~C., Impey, C.~D. et al. 2010, AJ, 139, 2097}
\bibitem[Peterson et al.(2003)]{Peterson2003} Peterson, J.~R., Kahn, S.~M., Paerels, F.~B.~S., et al. 2003, ApJ, 590, 207
\bibitem[Peterson \& Fabian(2006)]{Peterson_Fabian2006} Peterson, J.~R., \& Fabian, A.~C. 2006, PhR, 427, 1
%\bibitem[Planck Collaboration(2018)]{PlanckCollaborationVI2018} Planck Collaboration results VI, 2018, arXiv:180706209
\bibitem[Postman et al.(2012)]{Postman2012} Postman, M., Coe, D., Ben\'{i}tez, N., et al. 2012, ApJS, 199, 25
%\bibitem[Postman et al.(2012b)]{Postman2012b} Postman, M., Lauer, T.~R., Donahue, M., et al. 2012b, ApJ, 756, 159 
\bibitem[Radovich et al.(2020)]{Radovich2020} Radovich, M., Tortora, C., Bellagamba, F. et al. 2020, MNRAS, 498, 4303
%\bibitem[Rawle et al.(2012)]{Rawle2012} Rawle, T.~D., Edge, A.~C., Egami, E., et al. 2012, ApJ, 747, 29
\bibitem[Regan et al.(2001)]{Regan2001} Regan, M.~W., Thornley, M.~D., Helfer, T.~T., et al. 2001, ApJ, 561, 218
%\bibitem[Riechers et al.(2011)]{Riechers2011} Riechers D.~A., Carilli C.~L., Maddalena R.~J et al. 2011, ApJL, 739, 32
\bibitem[Richter \& Sancisi(1994)]{Richter_Sancisi1994} Richter, O.-G., Sancisi, R.,  1994, A\&A, 290, 9
%\bibitem[Riess et al.(2019)]{Riess2019} Riess, A.~G., Casertano, S., Yuan, W. et al. 2019, ApJ, 876, 85
%\bibitem[Riess(2019)]{Riess2019b} Riess, A.~G., 2019 Nature Reviews Physics, 2, 10
\bibitem[Rozo et al.(2015)]{Rozo2015} Rozo, E., Rykoff, E.~S., Becker, M. et al. 2015, MNRAS, 453, 38
\bibitem[Rudnick et al.(2017)]{Rudnick2017} Rudnick, G., Hodge, J., Walter, F. et al. 2017, ApJ, 849, 27
%\bibitem[Russell et al.(2014)]{Russell2014} Russell, H.~R., McNamara, B.~R.,  Edge, A.~C. et al. 2014,  ApJ, 784, 78 %Massive Molecular Gas Flows in the A1664 Brightest Cluster Galaxy
\bibitem[Russell et al.(2017)]{Russell2017} Russell, H.~R., McDonald, M., McNamara, B.~R. et al. 2017, ApJ, 836, 130
%\bibitem[Russell et al.(2019)]{Russell2019} Russell, H.~R., McNamara, B.~R., Fabian, A.~C. et al. 2019, MNRAS, 490, 3025
\bibitem[Rykoff et al.(2014)]{Rykoff2014} Rykoff, E.~S., Rozo, E., Busha, M.~T. et al. 2014, ApJ, 785, 104
\bibitem[Saintonge et al.(2017)]{Saintonge2017} Saintonge, A.,  Catinella, B., Tacconi, L.~J. et al. 2017, ApJS, 233, 22
\bibitem[Salom\'{e} \& Combes(2003)]{Salome_Combes2003} Salom\'{e} P. \&  Combes, F. 2003, A\&A, 412, 657
\bibitem[Salom\'{e} et al.(2006)]{Salome2006} Salom\'{e} P., Combes, F., Edge, A.~C. et al. 2006, A\&A, 454, 437
\bibitem[Salomé et al.(2015)]{Salome2015} Salomé, Q.; Salomé, P.; Combes, F. 
%\bibitem[Salpeter(1955)]{Salpeter1955} Salpeter, E.~E. 1955,ApJ,121, 161
\bibitem[Salucci et al.(2007)]{Salucci2007} Salucci, P., Lapi, A., Tonini, C., et al. 2007, MNRAS, 378, 41-47
%\bibitem[Sargent et al.(2015)]{Sargent2015} Sargent, M. T., Daddi, E., Bournaud, F., et al. 2015, ApJL, 806, L20
\bibitem[Sarzi et al.(2010)]{Sarzi2010} Sarzi, M., Shields, J.~C., Schawinski, K. et al. 2010, MNRAS, 402, 2187
%\bibitem[Schruba et al.(2011)]{Schruba2011} Schruba A., Leroy A.~K., Walter F. et al. 2011, AJ, 142, 37  
\bibitem[Scoville et al.(2014)]{Scoville2014} Scoville, N., Aussel, Sheth, K. et al. 2014, ApJ, 783, 
\bibitem[Scoville et al.(2016)]{Scoville2016} Scoville, N.,  Sheth, K.,  Aussel, H. et al. 2016, ApJ,820, 83, 2016
\bibitem[Singh et al.(2013)]{Singh2013} Singh, R., van de Ven, G., Jahnke, K. 
et al. 2013, A\&A, 558, 43
%\bibitem[Solomon et al.(1997)]{Solomon1997} { Solomon, P.~M., Downes, D., Radford, S.~J.~E. et al. 1997, ApJ, 478, 144}
\bibitem[Solomon \& Vanden Bout(2005)]{Solomon_VandenBout2005} Solomon, P.~M. \& Vanden Bout, P.~A., 2005, ARA\&A, 43, 677
\bibitem[Speagle et al.(2014)]{Speagle2014} { Speagle, J.~S., Steinhardt, C.~L., Capak, P.~L. et al. 2014, ApJS, 214, 15}
\bibitem[Stott et al.(2011)]{Stott2011} Stott, J.~P., Collins, C.~A., Burke, C. et al., 2011, MNRAS, 414, 445
\bibitem[Stott et al.(2011)]{Stott2011} Stott, J.~P., Collins, C.~A., Burke, C. et al., 2011, MNRAS, 414, 445
\bibitem[Stott et al.(2012)]{Stott2012} Stott, J.~P., Hickox, R.~C., Edge, A.~C. et al., 2012, MNRAS, 422, 2213
\bibitem[Strazzullo et al.(2018)]{Strazzullo2018} Strazzullo, V., Coogan, R.~T., Daddi, E. et al. 2018, ApJ, 862, 64
\bibitem[Tacconi et al.(2018)]{Tacconi2018} Tacconi, L.~J., Genzel, R., Saintonge, A., et al. 2018, ApJ, 853, 179
%\bibitem[Tadaki et al.(2019)]{Tadaki2019} Tadaki, K., Kodama, T., Hayashi, M. et al. 2019, PASJ, 71, 40
\bibitem[Toomre(1963)]{Toomre1963} Toomre, A. 1963, ApJ, 138, 385
\bibitem[Tremblay et al.(2016)]{Tremblay2016} Tremblay, G.~R., Oonk, J.~B.~R., Combes, F. et al. 2016, Nature, 534, 218
%\bibitem[Trudeau et al.(2019)]{Trudeau2019} Trudeau, A., Webb, T., Hlavacek-Larrondo, J. et al. 2019, MNRAS, 487, 1210
\bibitem[Valiante et al.(2016)]{Valiante2016} Valiante, E., Smith, M.~W.~L., Eales, S. et al. 2016, MNRAS, 462, 3146
%\bibitem[van der Wel et al.(2014)]{vanderWel2014} { van der Wel, A., Franx, M., van Dokkum, P.~G. et al. 2014, ApJ, 788, 28}
\bibitem[Veilleux \& Osterbrock(1987)]{Veilleux_Osterbrock1987} Veilleux S., \& Osterbrock D.~E., 1987, ApJS, 63, 295
%\bibitem[Voit(2005)]{Voit2005} Voit, G. Mark, 2005, RvMP, 77, 207
%\bibitem[Voit \& Donahue(2015)]{Voit_Donahue2015} Voit, G.~M., \& Donahue, M. 2015, ApJL, 799, L1
%\bibitem[Voit et al.(2017)]{Voit2017} Voit, G.~M., Meece, G., Li, Y., et al. 2017, ApJ, 845, 80
\bibitem[Wagg et al.(2012)]{Wagg2012} Wagg, J., Pope, A., Alberts, S. et al. 2012, ApJ, 752, 91 
\bibitem[Webb et al.(2015a)]{Webb2015a} Webb, T., Muzzin, A., Noble, A. et al. 2015a, ApJ, 814, 96 
\bibitem[Webb et al.(2015b)]{Webb2015b} Webb, T., Noble, A., De Groot, A. et al. 2015b,  ApJ, 809, 173
\bibitem[Webb et al.(2017)]{Webb2017} Webb, T.~M.~A., Lowenthal, J., Yun, M. et al. 2017, ApJ, 844, 17
\bibitem[Wen et al.(2012)]{Wen2012} Wen, Z.~L., Han, J.~L., and Liu, F.~S. 2012, ApJS, 199, 34
\bibitem[White(1976)]{White1976} White, S.~D.~M., 1976, MNRAS, 177, 717
\bibitem[Wiklind et al.(1997)]{Wiklind1997} Wiklind, T., Combes, F., Henkel, C., \& Wyrowski, F. 1997, A\&A, 323, 727
\bibitem[Wright et al.(2010)]{Wright2010} Wright, E.~L., Eisenhardt, P.~R.~M., Mainzer, A.~K. et al. 2010, AJ, 140, 1868
\bibitem[Yao et al.(2003)]{Yao2003} Yao, L., Seaquist, E.~R., Kuno, N. et al. 2003, ApJ, 588, 771
\bibitem[York et al.(2000)]{York2000} York, D.~G., Adelman, J., Anderson, J.~E. Jr. et al. 2000, AJ, 120, 1579
\bibitem[Young et al.(1995)]{Young1995} Young, J.~S., Xie, S., Tacconi, L. et al. 1995, ApJS, 98, 219
%\bibitem[Yu et al.(2018)]{Yu2018} Yu, H., Tozzi, P., van Weeren, R. et al. 2018, ApJ, 853, 100
\bibitem[Yu et al.(2020)]{Yu2020} Yu, N., Ho, L.~C., \& Wang, J. 2020,  ApJ, 898, 102
\bibitem[Zeimann et al.(2013)]{Zeimann2013} Zeimann, G.~R., Stanford, S.~A., Brodwin, M. et al. 2013, ApJ, 779, 137
\bibitem[Zirbel(1996)]{Zirbel1996} Zirbel, E.~L. 1996, ApJ, 473, 713
\end{thebibliography}
\end{document}